\documentclass{aa}  
\usepackage{graphicx}
\usepackage{multirow}
\usepackage{subfigure}
\usepackage{xcolor}
\usepackage{tablefootnote}

\usepackage{txfonts}
\newcommand{\gaia}{$Gaia$~EDR3}
\newcommand{\cluster}{NGC\,2244}
\newcommand{\ha}{H${\alpha}$}

\begin{document} 

   \title{Stellar population of the Rosette Nebula and NGC\,2244\thanks{Table \ref{tab:spt} is only available in electronic form
at the CDS via anonymous ftp to cdsarc.u-strasbg.fr (130.79.128.5)
or via \protect\url{http://cdsweb.u-strasbg.fr/cgi-bin/qcat?J/A+A/}}}

   \subtitle{Application of the probabilistic random forest}

   \author{K. Mu\v{z}i\'c\inst{1}, 
   V. Almendros-Abad\inst{1}, 
   H. Bouy\inst{2}, 
   K. Kubiak\inst{1},
   K. Pe\~na Ramírez\inst{3},
   A. Krone-Martins\inst{4},\\
   A. Moitinho\inst{1}
   \and
   M. Concei\c{c}\~{a}o\inst{5} 
          }

   \institute{CENTRA, Faculdade de Ci\^{e}ncias, Universidade de Lisboa, Ed. C8, Campo Grande, P-1749-016 Lisboa, Portugal\\
              \email{kmuzic@sim.ul.pt}
         \and
           Laboratoire d’Astrophysique de Bordeaux, Univ. Bordeaux, CNRS, B18N, All\'ee Geoffroy Saint-Hillaire, 33615 Pessac, France
        \and Centro de Astronomía (CITEVA), Universidad de Antofagasta, Av. Angamos 601, Antofagasta, Chile
        \and Donald Bren School of Information and Computer Sciences, University of California, Irvine, CA 92697, USA
        \and Instituto de Astrofísica e Ci\^{e}ncias do Espaço - Faculdade de Ci\^{e}ncias, Universidade de Lisboa, Ed. C8, Campo Grande, P-1749-016 Lisboa, Portugal }

   \date{Received; accepted}
 
  \abstract
   {Measurements of internal dynamics of young clusters and star-forming regions are crucial to fully understand the process of their formation. A basic prerequisite for this is a well-established and robust list of probable members. } 
   {In this work, we study the $2.8\degr \times 2.6\degr$ region in the emblematic Rosette Nebula, centred in the young cluster \cluster, with the aim of constructing the most reliable candidate member list to date. Using the obtained catalogue, we can determine various structural and kinematic parameters, which can help to draw conclusions about the past and the future of the region.}
   {We constructed a catalogue containing optical to mid-infrared photometry, as
   well as accurate positions and proper motions from $Gaia$\,EDR3 for the sources in the field of the Rosette Nebula. We applied the probabilistic random forest algorithm to derive the membership probability for each source within our field of view. Based on the list of almost 3000 probable members, of which about a third are concentrated within the radius of 20$'$ from the centre of \cluster, we identified various clustered sources and stellar concentrations in the region, and estimated the average distance to the entire region at
    $1489\pm37$\,pc,  $1440\pm32$\,pc to \cluster, and $1525\pm36$\,pc to NGC\,2237. The masses, extinction, and ages were derived by fitting the spectral energy distribution to the atmosphere and evolutionary models, and the internal dynamic was assessed via proper motions relative to the mean proper motion of \cluster.}
   {\cluster~is showing a clear expansion pattern, with an expansion velocity that increases with radius. Its initial mass function (IMF) is well represented by two power laws ($dN/dM\propto M^{-\alpha}$), with slopes $\alpha = 1.05 \pm 0.02$ for the mass range 0.2 - 1.5\,M$_\odot$ and $\alpha = 2.3 \pm 0.3$ for the mass range 1.5 - 20\,M$_\odot$, and it is in agreement with slopes detected in other star-forming regions. The mean age of the region, derived from the HR diagram, is $\sim2\,$Myr. We find evidence for the difference in ages between \cluster~and the region associated with the molecular cloud, which appears slightly younger. The velocity dispersion of \cluster~is well above the virial velocity dispersion derived from the total mass ($1000\pm70$\,M$_\odot$) and half-mass radius ($3.4\pm0.2$\,pc). From the comparison to other clusters and to numerical simulations, we conclude that \cluster~may be unbound and that it possibly may have even formed in a super-virial state.}
   {}


   \titlerunning{Stellar population of the Rosette Nebula}
    \authorrunning{K. Mu\v{z}i\'c et al.}
    
   \maketitle
%

\section{Introduction}

For the last three decades,
it has generally been accepted that most (70-90\%) of star formation in the Milky Way occurs in embedded clusters, with loose OB associations emerging as remnants of their dissolution  \citep{1991ASPC...13....3L, lada&lada03}. In this view, often referred to as the 'monolithic scenario', embedded clusters emerge as initially bound structures, whose expansion leads to the unbound associations we observe today. This expansion is often thought to be driven by the expulsion of residual gas because of stellar feedback \citep{goodwin06,baumgardt07,banerjee15}; although, other processes may contribute as well \citep{gieles12}.  

However, a contemporary alternative view of star formation has regarded it as a hierarchical process operating over a large range of scales \citep{1979SvAL....5...12E,1983MNRAS.203...31E,1994ASPC...65..125L,1996ApJ...466..802E}. In this 'hierarchical picture', star formation proceeds on multiple scales, from kiloparsec-sized super-complexes, to OB associations and star clusters on smaller scales, down to multiple stellar groupings and single stars.
With stars forming across a continuous distribution of gas densities, unbound associations may form in situ, with similar low densities as observed today. In the last years, access to wide-field observations, along with recent theoretical results have contributed to consolidate this view of the in situ formation of unbound associations in the Milky Way and beyond \citep{2010MNRAS.409L..54B, gouliermis18,grudic18,ward20}.

Historically, the two views of star formation proceeded more or less separately on different scales, which was motivated by observational circumstances: the works discussing the hierarchical picture (e.g. \citealt{1983MNRAS.203...31E, 1996ApJ...466..802E}) looked typically at large, galactic-scale fields, whereas the monolithic scenario emerged from studies over small field-of-view observations of star-forming regions possible with infrared arrays available in the 1990s. Although the two views of star formation are sometimes presented as being opposed to each other, they should better be viewed as being complementary and operating on different scales. It may as well be that there exist both types of OB associations, those that formed in situ as well as the expanded ones. After all, having $70-90\%$ of stars 
forming in embedded clusters, as claimed by \citet{lada&lada03}, is still compatible with a fair number forming 'on the loose'; although, the exact percentages may be revisited in the future. 



Detailed studies of young stellar object (YSO) populations in both embedded and non-embedded regions have revealed that most star-forming regions appear to be clumpy, with their populations distributed in several subclusters \citep{kuhn14,sills18,kuhn21}. Measurements of internal kinematics of these subclusters are needed to understand how they relate to each other, that is to say whether the tendency is towards merging into larger clusters, or a dispersal of individual subclusters once the molecular gas is gone from the system. 
However, detailed studies of internal kinematics and structural properties of young star clusters have been challenging due to instrumental issues (the need for high-precision astrometry or, alternatively, radial velocities for a large number of stars), as well as due to those related to the determination of membership. Recently, both issues have seen major improvements thanks to \textit{Gaia} \citep{Gaiamission}, which led to precise measurements of velocity dispersion and the detection of expansion patterns in a substantial number of clusters younger than 5 Myr \citep{kuhn19}. 
As of the membership selections, those based on optical and near-infrared (NIR) photometry typically suffer from significant contamination by field objects, while X-ray and mid-infrared-excess selections, though quite robust, come with a bias towards objects with a particular set of properties, and they are unable to uncover entire populations. Spectroscopy, on the other hand, becomes unfeasible for regions spanning large areas, and consequently containing large numbers of objects. Precise proper motions, especially when combined with parallaxes for systems relatively close to the Sun, provide a crucial piece of information for membership determination, and can be combined with photometric criteria to uncover reliable sets of pre-main sequence (PMS) candidate members.

Recently, various machine learning (ML) techniques, both supervised and unsupervised, have been used to aid in this process, and have provided membership lists in a large number of regions, from the youngest ones with ages of $\sim$1\,Myr to the oldest known open clusters \citep[e.g.][]{gao18, marton19,miret19,melton20,cantatgaudin18,cantatgaudin20,galli21,penaramirez21}. While the unsupervised techniques can efficiently deal with a large number of clusters and thus help to improve the statistics on the overall properties of young Galactic populations, they may still come with some non-negligible amount of contamination. On the other hand, supervised ML techniques allow us to use all the prior knowledge on the properties of cluster population, potentially resulting in cleaner candidate member lists. This, of course, brings the challenge of properly selecting a training set, since all biases and the contamination of this training set will be propagated towards the final classification and thus to the sample that will be used for further astrophysical studies.

Most traditional ML methods are often not designed to deal with measurement uncertainties, which can influence their overall performance on a typical low- or moderate-signal-to-noise ratio astronomical dataset \citep{baron19}. Several approaches to this problem have been explored and presented in astronomical literature (e.g. \citealt{upmask,castro18,naul18,shy21}), but there are in general very few existing tools that take dataset uncertainties into account during the model construction. One of these tools is the probabilistic random forest (PRF) classifier \citep{reis19}, a modified version of the original random forest algorithm, in which both the
features and the labels are treated as probability distribution functions (PDFs), rather than deterministic values. Each random variable is represented by a PDF, with a mean corresponding to a measurement and a variance to the measurement uncertainty. In this work, we apply the PRF algorithm to uncover the stellar population in a well-known young region of star formation, the Rosette Nebula.

The Rosette Nebula is the most prominent and active star-forming region in the Mon OB2 association. In its centre, it hosts a young star cluster \cluster, whose massive stars are thought to be responsible for the evacuation of much of the interstellar material from the centre of the HII bubble \citep{romanzunigalada08}. Numerous studies of the \cluster~population have revealed a presence of several tens of OB stars, along with a large population of low-mass stars and substellar candidates \citep{park02, balog07, romanzunigalada08, wang08,bonatto09,muzic19,lim21}. Studies of the wider region of the Rosette nebula witness that the star-forming activity is by no means limited to its most prominent cluster, revealing several smaller groups associated with the molecular cloud, as well as more extended regions containing young stars \citep{phelps&lada97, romanzuniga08, poulton08, cambresy13, ybarra13, kuhn14}. 
Previous estimates of the distance to \cluster~vary between 1400 and 1700\,pc \citep{ogura&ishida81, perez87,hensberge00,park02,lombardi11,martins12,bell13,kharchenko13,kuhn19,muzic19,lim21}, with the most recent estimates from $Gaia$ data yielding distances around 1500\,pc. Most authors agree on the age of $\sim 2\,$Myr \citep{perez87,lim21}. In this paper, we reassess both the age and the distance to the region, using the $Gaia$ EDR3 parallaxes for the latter \citep{GaiaEDR3}. 

Previous member selections in the Rosette Nebula have been made based on X-ray \citep{wang08}, NIR \citep{romanzuniga08}, and mid-infrared (MIR; \citealt{balog07}) properties, as well as proper motions from $Gaia$ (alone or in combination with photometric selection; \citealt{muzic19,kuhn19,cantatgaudin20, lim21}). Some of these studies deal with the wide region spanned by the nebula, while others only concentrate on the central cluster \cluster. In this work, we apply the PRF algorithm to the sources detected in the $2.8\degr \times2.6\degr$  region in the direction of the Rosette Nebula, accompanied with a carefully selected training set, with a goal to obtain the most reliable and unbiased view of the stellar population in the region to date, and study its structure, mass, age, and kinematics.

This paper is structured as follows. In Section~\ref{sec:dataset}, we present the details of the dataset used in this work, including photometric and astrometric catalogues, as well as the spectroscopic data. The membership selection method using the PRF algorithm is given in Section~\ref{sec:PRF}, followed by the analysis of spatial and kinematic properties of the selected members (Section~\ref{sec:kinematics}), as well as their further physical characterisation (Section~\ref{sec:physical_character}).
Section~\ref{sec:ngc2244_properties} is centred around the properties of \cluster, including its mass function, mass segregation, and a discussion on its dynamical state. Finally, Section~\ref{sec:summary} summarises the work, and lists the main conclusions.

\section{Dataset}
\label{sec:dataset}
\subsection{Photometric and astrometric catalogue}
We used the methodology developed
as part of the Dynamical Analysis of Nearby Clusters (DANCe) project
\citep{cosmicdance} to gather optical and near-infrared photometry as well as accurate positions and proper motions for the sources in the field of the Rosette Nebula, centred at the cluster \cluster.

The field of view in this study satisfies the following conditions:
96.6\degr $\leq \alpha \leq$ 99.4\degr and 3.6\degr $\leq \delta \leq$ 6.2\degr.
This field has been chosen such to include the region containing the main bulk of CO emission of the Rosette cloud (\citealt{heyer06}; see Fig. 8 in \citealt{romanzuniga08}). It also fully encompasses several structures identified and studied in previous works on the stellar populations of \cluster~and the rest of the nebula \citep[e.g.][]{wang08, kuhn14, meng17}.
We searched for wide field images covering this field in the following archives: the European Southern Observatory (ESO), the National Optical-Infrared Astronomy Research Laboratory Archive (NOIRLab), the Canada France Hawaii Telescope (CFHT) archive hosted at the Canadian Astronomy Data Centre (CADC), the Isaac Newton Group (ING), and the WFCAM Science (WSA). 
The data found in these public archives were complemented with our own observations using the ESO Very Large Telescope and its near-infrared camera HAWK-I \citep{hawki}, under the programme ID 0104.C-0369. 
Table~\ref{tab:observations} gives an overview of the various instruments used for this study.

In all cases except for CFHT/MegaCam, DECam, WIRCam and UKIRT images, the raw  data and associated calibration frames were downloaded and processed following standard procedures  using an updated version of \emph{Alambic} \citep{alambic}, a software suite developed and optimised for the processing of large multi-CCD imagers. In the case of MegaCam and WIRCam at CFHT, the images processed and calibrated with the \emph{Elixir} \citep{Elixir} and the  `I`iwi\footnote{\url{https://www.cfht.hawaii.edu/Instruments/Imaging/WIRCam/IiwiVersion2Doc.html}} pipelines, respectively, were retrieved from the CADC archive. In the case of DECam, the images  were retrieved from the NOIRLab public archive and processed with the community pipeline \citep{DECampipeline}. UKIRT images processed by the Cambridge Astronomical Survey Unit were retrieved from the WFCAM Science Archive.
After a rejection of problematic images (mostly due to loss of guiding, tracking or read-out problems) and of problematic pixels using {\sc MaxiTrack} and {\sc MaxiMask} \citep{maximask}, the dataset included 2\,187 individual images originating from ten cameras. 

\subsubsection{Astrometric calibration}

The astrometric calibration was performed as described in \citet{olivares19} using \textit{Gaia} EDR3 \citep{GaiaEDR3} as astrometric reference. The final average internal and external 1-$\sigma$ residuals were estimated to be $\sim$15~mas for high signal-to-noise (photon noise limited) sources. 
The most important steps are briefly outlined here.

\begin{itemize}
\item PSF modelling: a non-parametric spatially variable model of the PSF was computed using \texttt{PSFex} \citep{bertin2013}.

\item Source detection: sources with more than three pixels above 1.5 standard deviations of the local background were detected and their instrumental fluxes and positions measured using the empirical PSF with \texttt{SExtractor} \citep{sextractor}.
We note that the PSF modelling has been used for astrometry only, while for the photometry measurements using an automatic adaptive aperture were preferred.
\item The global astrometric solution is computed iteratively by \texttt{Scamp} \citep{bertin2006} by minimising the quadratic sum of differences in position between overlapping detections from pairs of catalogues. Positions are tied to the \textit{Gaia} EDR3 release. 
\end{itemize}

\begin{figure}
   \centering
   \includegraphics[width=0.45\textwidth]{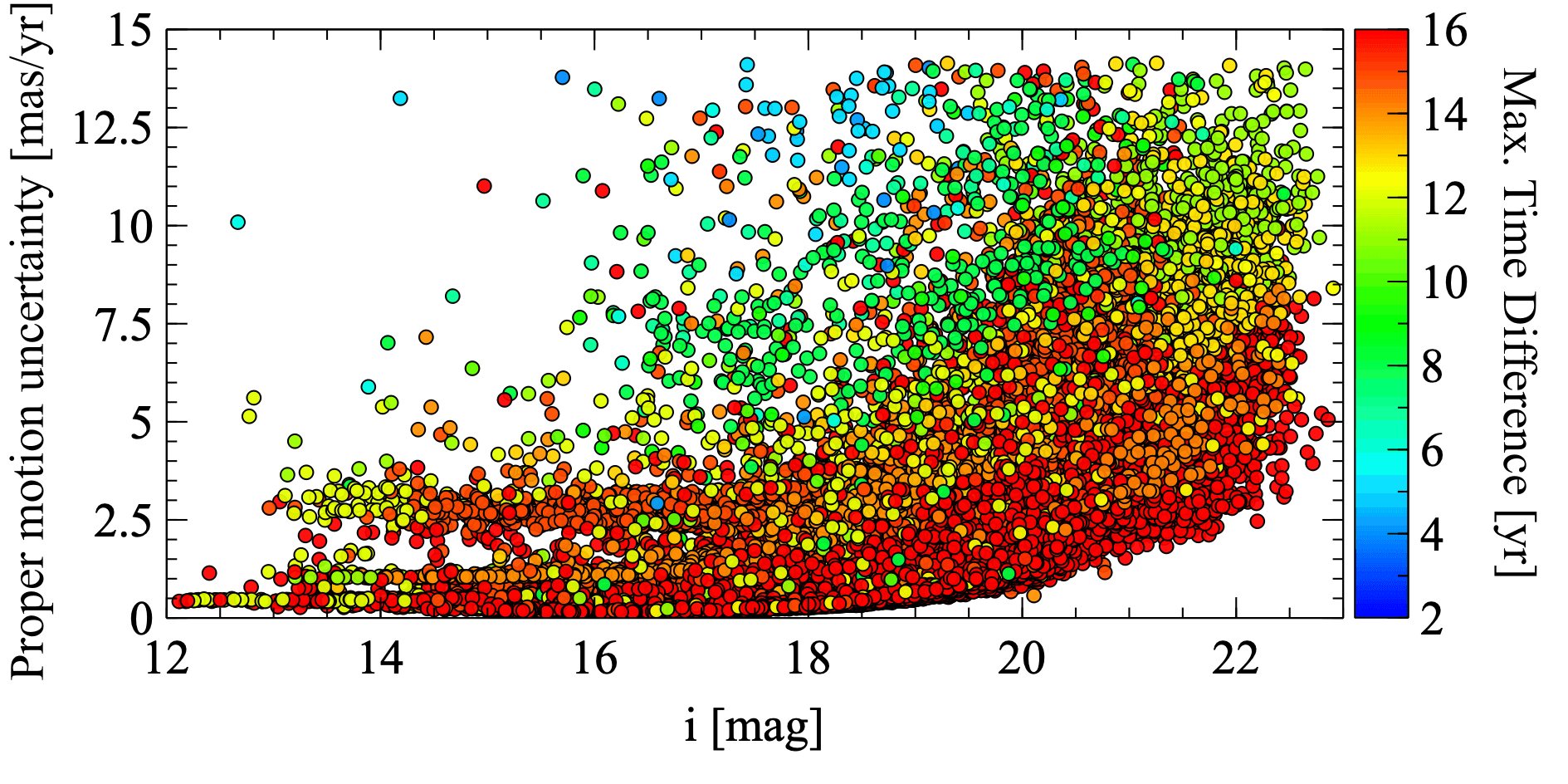}
      \caption{Proper motion uncertainty as a function of $i$-band magnitude and the maximum time difference between epoch used in proper motion calculation for each source.}
         \label{fig:ppm_err}
   \end{figure}

As explained in \citet{cosmicdance}, the computed proper motions are relative to one another and display an offset with respect to the geocentric celestial reference system. We estimate the offsets by computing the median of the difference between our proper motions and those of the \textit{Gaia} EDR3 after rejecting outliers using the modified Z-score. The typical proper motion uncertainties of our catalogue are shown in Fig.~\ref{fig:ppm_err}.

\subsubsection{Photometric calibration}
The photometric calibration was performed only for the $H\alpha$, $u$, $g$, $r$, $i$, $z$, $y$ and $J$, $H$, $K_S$ images, and was not attempted for the other various Str\"omgren or narrow band images (except $H\alpha$). The photometric zero-point of all individual images obtained in the $u$, $g$, $r$, $i$, $z$, $y$, $H\alpha$ and $J$, $H$, $K_S$ filters was computed by direct comparison of the instrumental {\sc SExtractor} \citep{sextractor} \verb|MAG_AUTO| magnitudes with an external catalogue: 
\begin{itemize}
\item $u$ images were tied to the SDSS catalogue \citep{sdss};
\item $g$, $r$, $i$, $z$, $y$ images were tied to the Pan-STARRS PS1 catalogue \citep{pan-starrs};
\item $H\alpha$ images were  tied to the IPHAS DR2 photometry \citep{iphas}; 
\item $B$, $V$ images were tied to APASS \citep{apass};
\item $R$, $I$ (VIMOS/VLT; \citealt{vimos}) images were calibrated on the Johnson-Kron-Cousins system using standard fields obtained on the same night;
\item $Y$ images were tied to the UKIDSS photometric system \citep{ukidss};
\item $J, H, K_S$ images were tied to the 2MASS catalogue \citep{twomass}.
\end{itemize}

\begin{figure}
   \centering
   \includegraphics[width=0.45\textwidth]{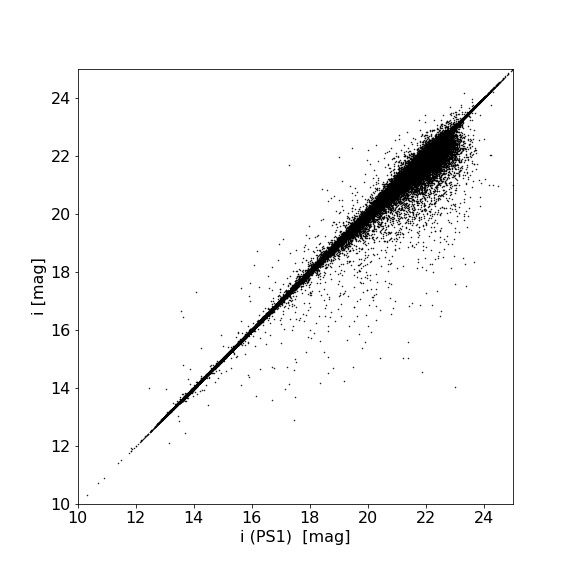}
      \caption{Comparison between our final $i$-band magnitude and the $i$-band magnitude reported in Pan-STARRS.}
         \label{fig:iband}
   \end{figure}

The individual zero-points were computed as follows.
First, clean point-like sources were selected using {\sc SExtractor}  $|$\verb|SPREAD_MODEL|$|\le$0.02. The \verb|SPREAD_MODEL| is a morphometric linear discriminant parameter obtained when fitting a Sérsic model, which is useful to classify point-sources from extended sources \citep[see e.g.][]{cosmicdance}. Second, a difference between the Kron and PSF instrumental magnitudes smaller than 0.02~mag ($|$\verb|MAG_AUTO|-\verb|MAG_PSF|$|\le$0.02~mag), used as an additional test to reject extended or problematic sources. Third,
{\sc SExtractor} \verb|FLAG|=0, to avoid any problem related to blending, saturation, or truncation. The closest match within 1\arcsec\, to these clean point-like sources was then found in the reference catalogue. The zero-point was computed as the median of the difference between the reference and instrumental (\verb|MAG_AUTO|) magnitudes. The median absolute deviation are typically of the order of 0.01$\sim$0.09~mag depending on the filter. Figure~\ref{fig:iband} shows a comparison between the final $i$-band magnitudes of our catalogue and that reported in the Pan-STARRS catalogue.

\verb|MAG_AUTO| magnitudes correspond to the total flux obtained by integrating pixel values within an adaptively scaled aperture computed using Kron’s first moment algorithm \citep{kron1980}. They are preferred over \verb|MAG_PSF| magnitudes obtained using PSF fitting because the latter give very poor measurements for extended sources (galaxies), leading to biased zeropoints when comparing to the above-mentioned reference catalogues.

\subsubsection{Deep stacks}
We median-combined all the photometrically and astrometrically registered images obtained with the same camera and in the same filter to obtain a deep stack and extract the corresponding photometry. These stacks made of many epochs were not used for the proper motion measurements but allowed us to significantly improve the detection sensitivity in all filters, and recover or improve the photometry of faint sources obtained in the individual images. These measurements have a higher signal-to-noise and are therefore preferred over the measurements obtained from the individual images, whenever available.

\subsubsection{External astrometry and photometry}
As in \citet{cosmicdance}, the photometry and astrometry extracted from the images are complemented by that reported in external catalogues. The recent \textit{Gaia} EDR3 catalogue provides accurate parallaxes and proper motions for sources down to $G\sim$20~mag. We retrieved all the sources reported in the \textit{Gaia} EDR3 catalogue over the area covered by our images and cross-matched with our catalogue using the closest match within a 1\arcsec\, radius. Because \textit{Gaia}'s proper motion measurements are much more accurate and reliable than our ground-based measurements, we always prefer them over our own measurements, and keep our proper motion measurements only for sources without any counterpart in the \textit{Gaia} EDR3 catalogue. As our catalogue was constructed using $Gaia$ EDR3 as the primary astrometric reference, it is already tied to the Gaia Celestial Reference Frame.

Given the heterogeneous spatial coverage of the various datasets, the DANCe catalogue described above (and including $H\alpha$, $u$, $g$, $r$, $i$, $z$, $y$, $J$, $H$, and $K_S$) is complemented with \textit{Gaia} EDR3 astrometry and photometry ($G$, $BP$, and $RP$), as well as Pan-STARRS ($g$, $r$, $i$, $z$, and $y$; \citealt{pan-starrs}), 2MASS ($J$, $H$, and $K_S$; \citealt{twomass}), and AllWISE \citep[W1-W4,][]{WISE} photometry.
These additional photometric values were directly included in our catalogue when no measurement was available for the corresponding source in our data; otherwise, the weighted average of all -- of our and external catalogues' -- measurements is adopted. In some cases, significant differences exist among the filters of the same photometric band. These systematic differences contribute to the noise of the photometric dispersion in the colour-magnitude and colour-colour diagrams used for our analysis. Moreover, recent variability studies of young clusters found typical amplitudes of 0.03~mag \citep[e.g.][]{rebull16, stauffer16}. Since our aim is to identify cluster members, and to take into account both the systematic differences among filters of the same band and the intrinsic photometric variability, we conservatively add, in quadrature, 0.05~mag to all the photometric uncertainties in our catalogue.

\begin{table*}
\centering
\caption{Instruments used in this study \label{tab:observations}}

\begin{tabular}{lccccccclc}\hline\hline
Observatory   & Instrument        & Filters              & Platescale   & Field of view & Epoch(s)  \\
                      &                          &                         & [\arcsec pixel$^{-1}$]    &                  &                     &                           &      \\
\hline
CTIO (Blanco)  & DECam   & $g,r,z$  & 0.26 & 1.1\degr radius & 2018  \\
CFHT   & MegaCam   & $u,g,r$  & 0.18 & 1\degr$\times$1\degr & 2005   \\
CFHT  & CFH$12$K & $B,V,R,I$ & $0\farcs21 $ & $42\arcmin\times28\arcmin$        & 1999 \\
CFHT & WIRCam  & $H2$  & $0\farcs3  $ & $20\arcmin\times20\arcmin$        & 2011   \\
INT & WFC & many & 0.33 & 34\arcmin$\times$34\arcmin & 2002-2014  \\
UKIRT &  WFCAM & $J,H,K_S,H2$ & 0.4 & 40\arcmin$\times$40\arcmin & 2005-2011\\
Kitt Peak Mayall & Flamingos & $J,H,K_S$ & 0.32 & 11\arcmin$\times$11\arcmin & 2006 \\
NTT & SofI  & $K_S$ & 0.29 & 4.95\arcmin$\times$4.95\arcmin & 2001  \\
ESO VLT & VIMOS         & $R,I$         & $0\farcs205$ & $28\arcmin\times32\arcmin$        & 2007   \\
ESO VLT & HAWK-I        & $Y,J,H,K$   & $0\farcs106$ & $7.5\arcmin\times7.5\arcmin$      & 2019-2020 \\
\hline
\end{tabular}
\end{table*}

\subsubsection{Completeness of our catalogue}

\begin{table}
\centering
\caption{Photometric completeness intervals \label{tab:completeness}}

\begin{tabular}{lcc}\hline\hline
Filter   & Bright end [mag]        & Faint end [mag]\\
\hline
$G$ & 6.0  & 20.3        \\
$g$ & 13.0 & 21.5 \\
$r$ & 13.0 & 20.6 \\
$i$ & 13.0 & 19.7 \\
$z$  & 13.0 &  19.2  \\
$y$  & 12.5 & 18.6   \\
$J$ & 8.0 & 16.5 \\
$H$ & 6.0 & 15.8 \\
$K_S$& 6.0 & 15.4\\
$H\alpha$& 13.0 & 19.6 \\
\hline
\end{tabular}
\end{table}

\begin{figure}
   \centering
   \includegraphics[width=0.45\textwidth]{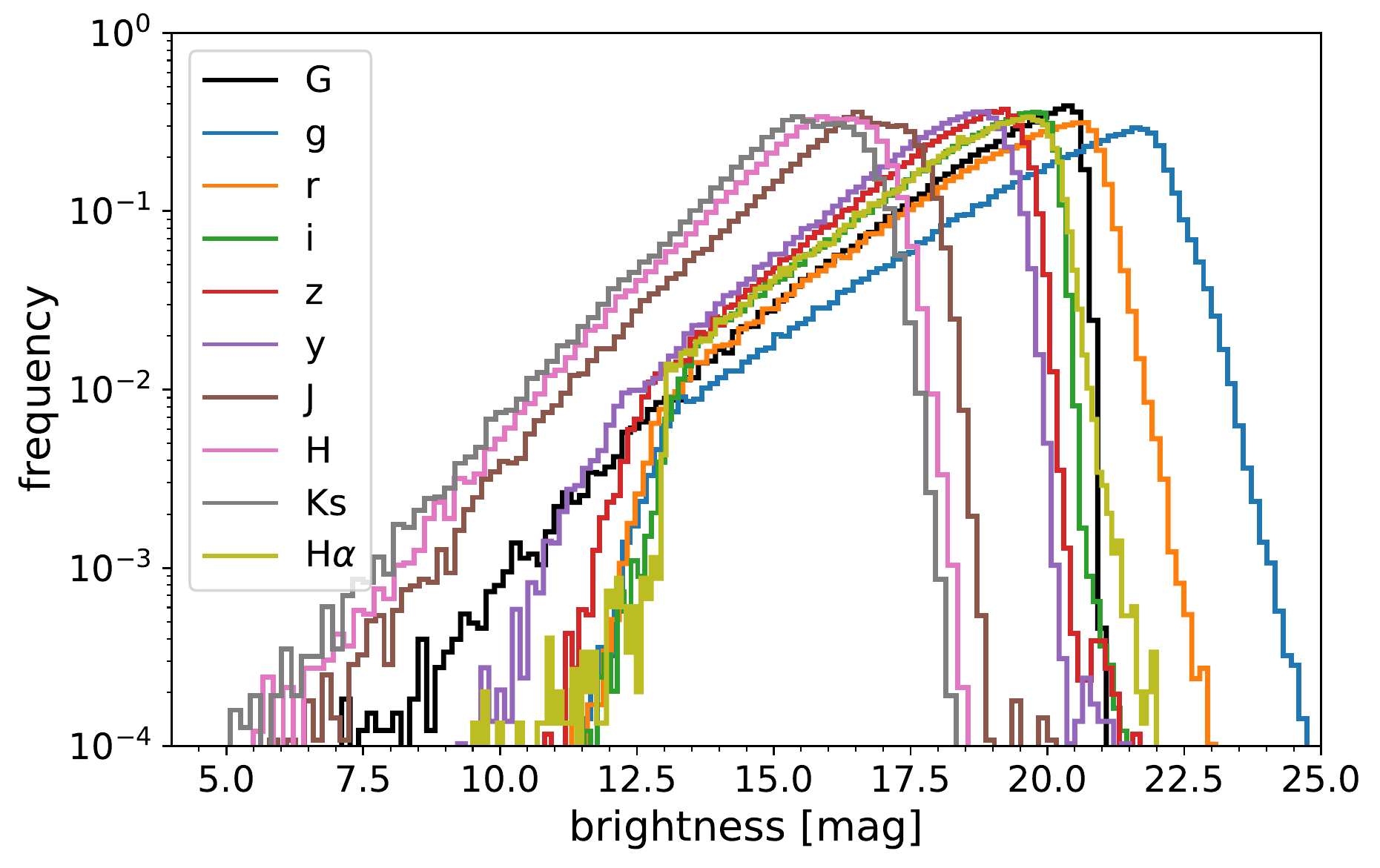}
      \caption{Density of sources with proper motion measurements from $Gaia$ EDR3 as a function of magnitude.}
         \label{fig:completeness}
   \end{figure}
   
The goal of this work is to obtain a detailed study of the stellar population of the Rosette Nebula and the cluster \cluster. To have a uniform data quality over the entire field, we limit ourselves to 
the photometric bands extending over the entire field of view (listed in Table~\ref{tab:completeness}), together with the astrometry from \textit{Gaia}\,EDR3. This selection includes $\sim200\,000$ sources, or $\sim40\%$ of the original catalogue. The remaining parts of our compiled catalogue described in previous sections are left for a future publication.

Given the heterogeneous origin of our dataset, its photometric completeness may be a complex distribution. To calculate the completeness for each photometric band that will be used in the rest of the work, we first restrict the catalogue to the objects that have proper motion measurement from $Gaia$ EDR3. We
define the completeness limit of each photometric band as the magnitude at which the density of sources reaches a maximum value (Fig.~\ref{fig:completeness}). The completeness intervals are reported in Table~\ref{tab:completeness}. The faint end for most of the filters is equivalent to masses of $0.1 - 0.2\,M_\odot$ for the age of 2 Myr, distance of 1500\,pc and the average extinction of \cluster~(A$_V$ = 1.5\,mag; \citealt{muzic19}), according to BT-Settl models \citep{allard11}. We therefore adopt $0.2\,M_\odot$ as the low-mass completeness limit later in the paper. 
For the bright end completeness limits, there is no simple way to establish a mass equivalence, since the iso-mass contours in Hertzsprung-Russell (HR) diagrams run nearly parallel to the temperature axis (see Section~\ref{sec:massage}).

\subsection{Spectroscopy}
\label{sec:vimos}

\subsubsection{Observations and data reduction}
Optical spectra have been obtained using VIsible MultiObject Spectrograph (VIMOS; \citealt{vimos}) at the ESO's Very Large Telescope, in the MOS mode. They have been obtained as a part of the programme 080.C-0697 (PI King) and retrieved from the ESO archive.  VIMOS contains four CCDs arranged in a $2\times2$ array, each  covering a field-of-view of $7'\times8'$ with a pixel resolution of 0\farcs205. The four quadrants are separated by gaps of approximately $2'$. The spectra were obtained using the medium-resolution grism (MR), covering the wavelength range of 5000 -- 10000\,\AA, with a dispersion of 2.3\AA\,pixel$^{-1}$. This, in combination with the slit of the $1''$ width, results in a spectral resolution of R$\sim 580$. The observations cover 16 different pointings, with a total on-source time per pointing between 40\,s and 9200\,s. Since the data have been taken prior to the detector upgrade in 2009, they suffer from strong fringing at wavelengths redder than $\sim$8000\,\AA. We cut the spectra at 8500\,\AA~to reduce fringing effects while trying to maintain a wavelength range as large as possible, since this is important for spectral type determination.

Data reduction was performed using the VIMOS pipeline
provided by ESO, and includes bias subtraction,
flat-field and bad-pixel correction, wavelength and flux calibration, as well as the final extraction of the spectra. The extraction is done by applying the optimal extraction from \citet{horne86}. In total we extracted spectra of 501 objects, distributed within $\sim$15$'$ around the centre of \cluster.

\subsubsection{Spectral analysis}

The VIMOS data have been retrieved from the ESO archive, thus we are not aware of the exact selection procedure for individual objects, however, the vast majority of the selected sources are red in optical-to-infrared colour-magnitude diagrams. 
Because of the fringing, the gravity-sensitive alkali lines typically used for youth (i.e. membership) assessment (Na II doublet at 8183/8195\,\AA~and  Ca II triplet at 8498/8542/8662\,\AA; \citealt{kirkpatrick91,martin99,schlieder12}) are not available. We therefore rely only on the enhanced H$\alpha$ emission as a signature of youth. 

The vast majority of sources with H$\alpha$ in emission show spectral features characteristic for cool dwarfs (late K and M spectral types), which consist of an overall spectral shape which is flat or with a positive slope towards red, 
as well as various molecular (TiO, VO, CaH) and metallic transitions \citep{kirkpatrick91}.
By visual inspection, we identify 327 late-type objects, of which 251 show a discernible H$\alpha$ in emission. Among the remaining 174 spectra of earlier type stars, only 3 show H$\alpha$ in emission. Pronounced H$_\alpha$ emission in stars with SpT$\lesssim$K0 can be taken as a clear sign of accretion, since the chromospheric activity in these stars tends to first fill-in the photospheric absorption lines, unlike in the lower mass stars where the chromospheric activity manifests through H$_\alpha$ in emission  \citep{stauffer97,manara13,kraus14,frasca15,hou16}. Therefore, the 3 early-type objects are directly included in our list of accretors and potential members of the cluster. We note that these objects are also listed as YSO candidates in previous works \citep{bell13,broos13,meng17}. For the late-type H$\alpha$ emitters, additional analysis related to the strength of the H$\alpha$ emission is necessary to identify potential cluster members.

To simultaneously derive the spectral type and extinction, we perform a $\chi$-squared minimisation identical to that used in our previous works (e.g. \citealt{kubiak21}), using a grid of optical spectra of young objects (1-3 Myr) in Cha I, Taurus, $\eta\,$Cha \citep{luhman03a,luhman04a,luhman04b,luhman04c}, and Collinder 69 \citep{bayo11}, with spectral types M1 to M9, with a step of 0.25-0.5 spectral subtypes. To this, we add K5, K7, and M0 field templates, created by averaging a number of available spectra at these sub-types\footnote{From \url{http://www.dwarfarchives.org, http://kellecruz.com/M_standards/}, and \citet{luhman03a, luhman04c}.}. The extinction has been varied between 0 and 5\,mag, with the step of 0.2\,mag, and we applied the extinction law by \citet{cardelli89}, with R$_V$=3.1 \citep{fernandes12}. The derived spectral types and extinction for the analysed objects are given in Table~\ref{tab:spt} of the Appendix.  

\begin{figure}
   \centering
   \includegraphics[width=0.45\textwidth]{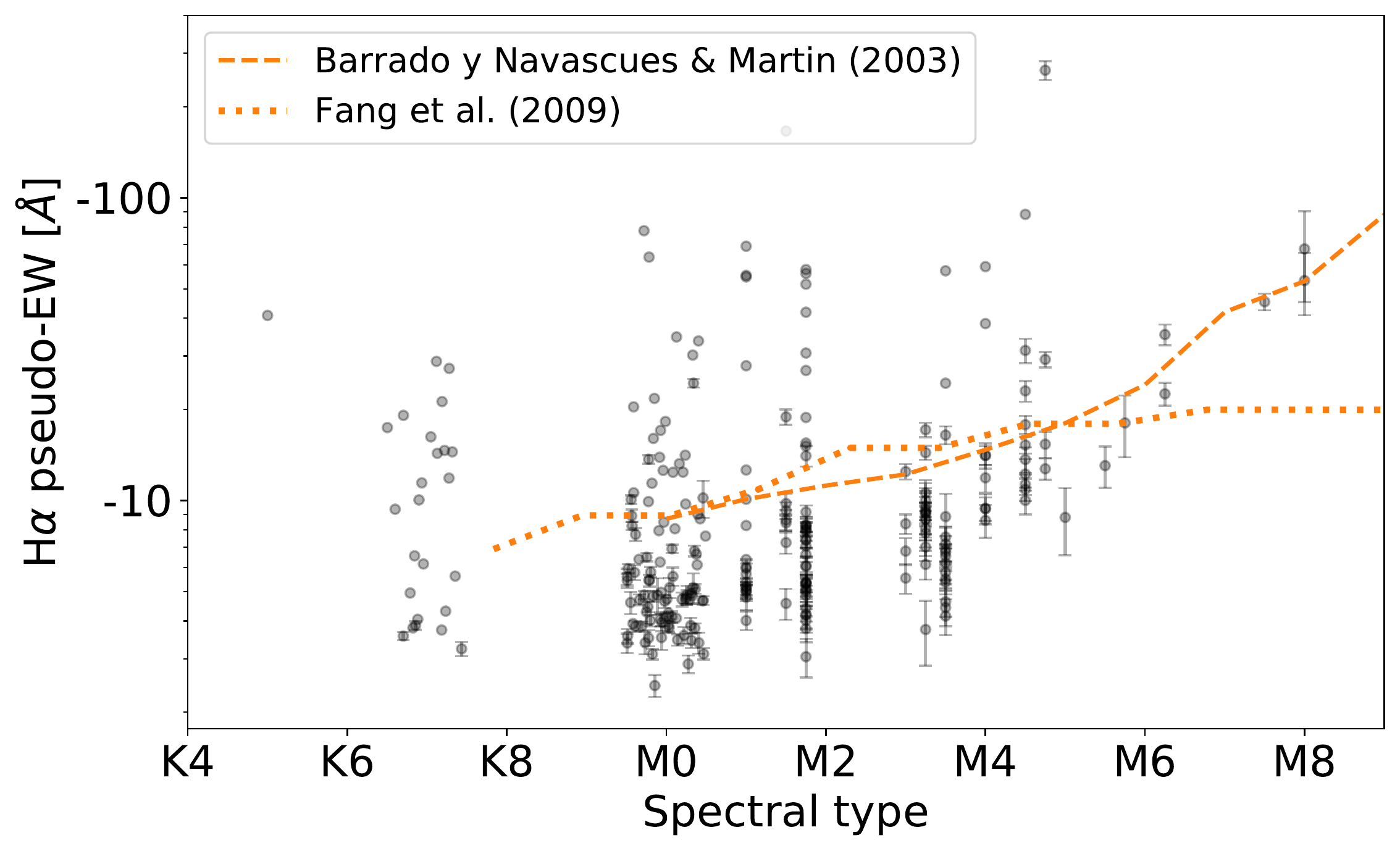}
      \caption{\ha~pseudo-EWs as a function of spectral type obtained from the VIMOS spectra. The dashed and the dotted orange lines represent the accretion thresholds defined by \citet{barrado03} and \citet{fang09}, respectively. The sources with the SpT K7 and M0 have been randomly displaced within (SpT-0.5, SpT+0.5), for clarity. The error bars are shown only when they are larger than the plotting symbols. }
         \label{fig:halpha_ew}
   \end{figure}

The equivalent widths (EWs) of the H$\alpha$ line in sources experiencing accretion are higher when compared to non-accreting pre-main sequence and main-sequence stars, where H$\alpha$ can still be present due to chromospheric activity \citep{white&basri03, barrado03, fang09, alonso15}. The measurements of the pseudo-EW of H$\alpha$ line in emission is shown as a function of the spectral type in Fig.~\ref{fig:halpha_ew}. The error bars are calculated as standard deviations from repeated measurements slightly changing the wavelength range for the pseudo-continuum subtraction, and the range over which the EW is measured, taking into account the spectrum measurement errors.
We also overplot the often-applied criteria to distinguish between accretion and chromospheric activity from \citet{barrado03} and \citet{fang09}, which will be used to select the objects for the training set in the next section. 

\begin{figure*}[htb]
    \center
    \begin{subfigure}{}
        \center
        \includegraphics[width=0.5\linewidth]{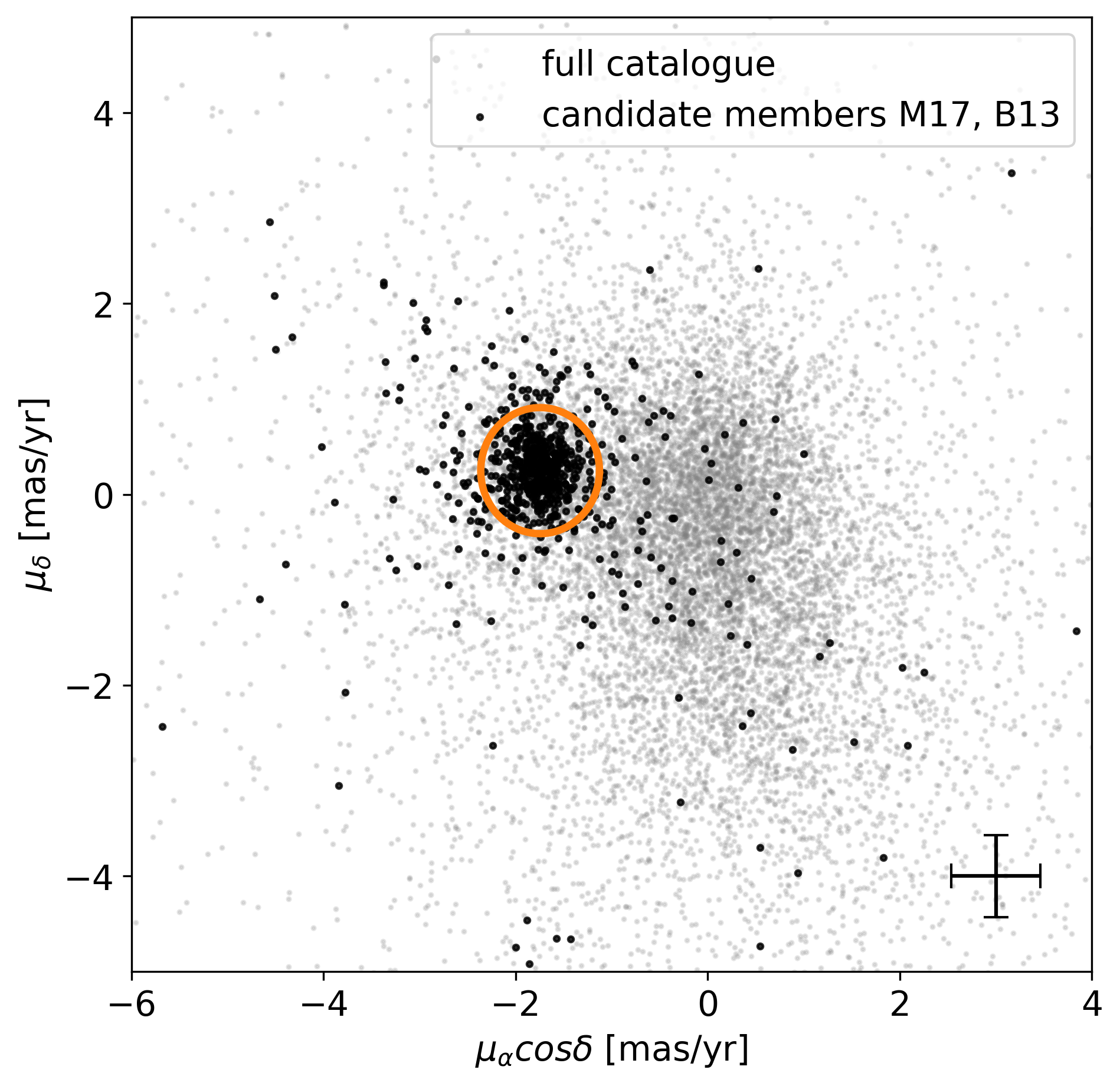}%
        \hfill
        \includegraphics[width=0.5\linewidth]{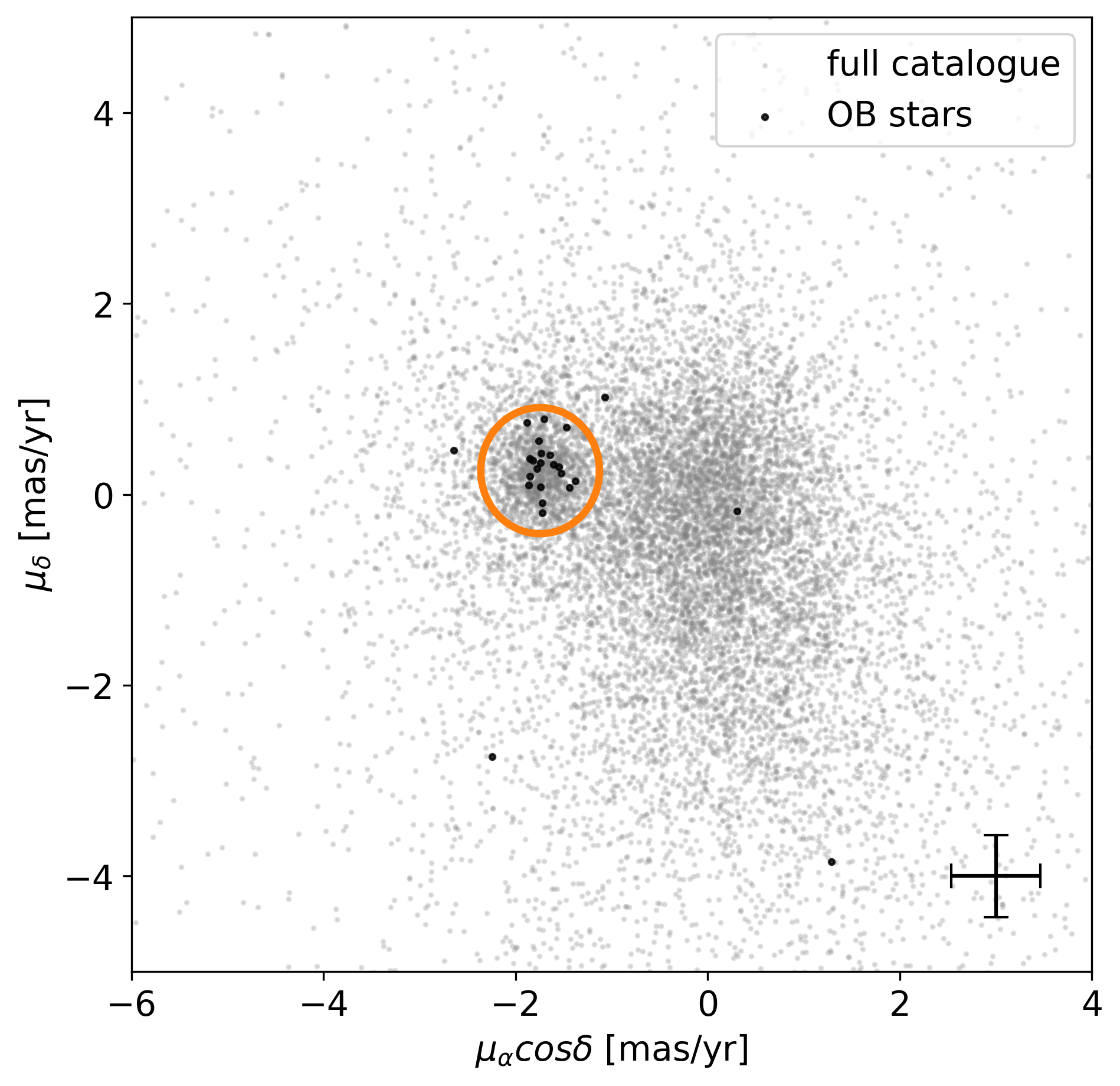}
        \hfill
        \includegraphics[width=0.5\linewidth]{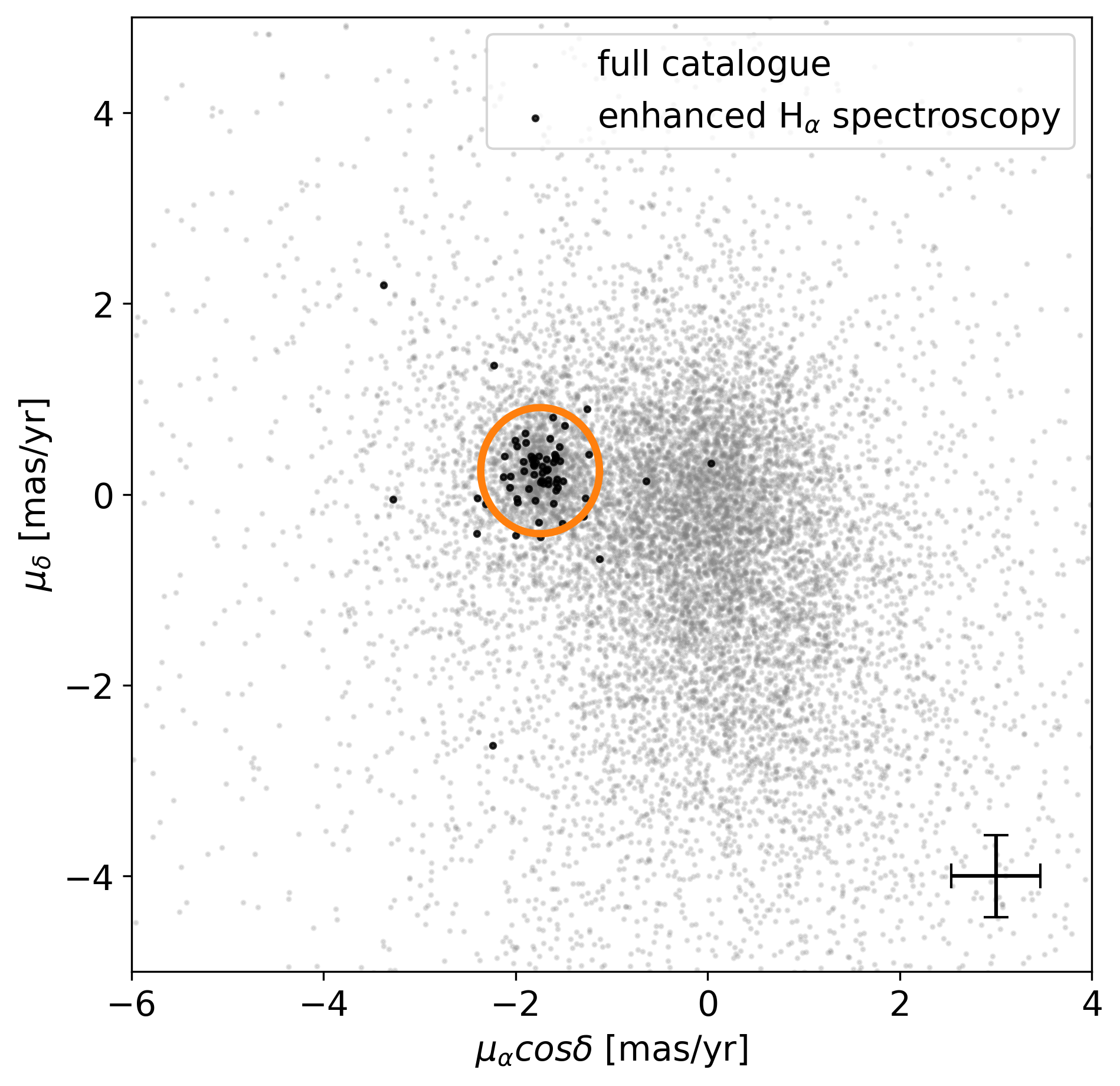}%
        \hfill
        \includegraphics[width=0.5\linewidth]{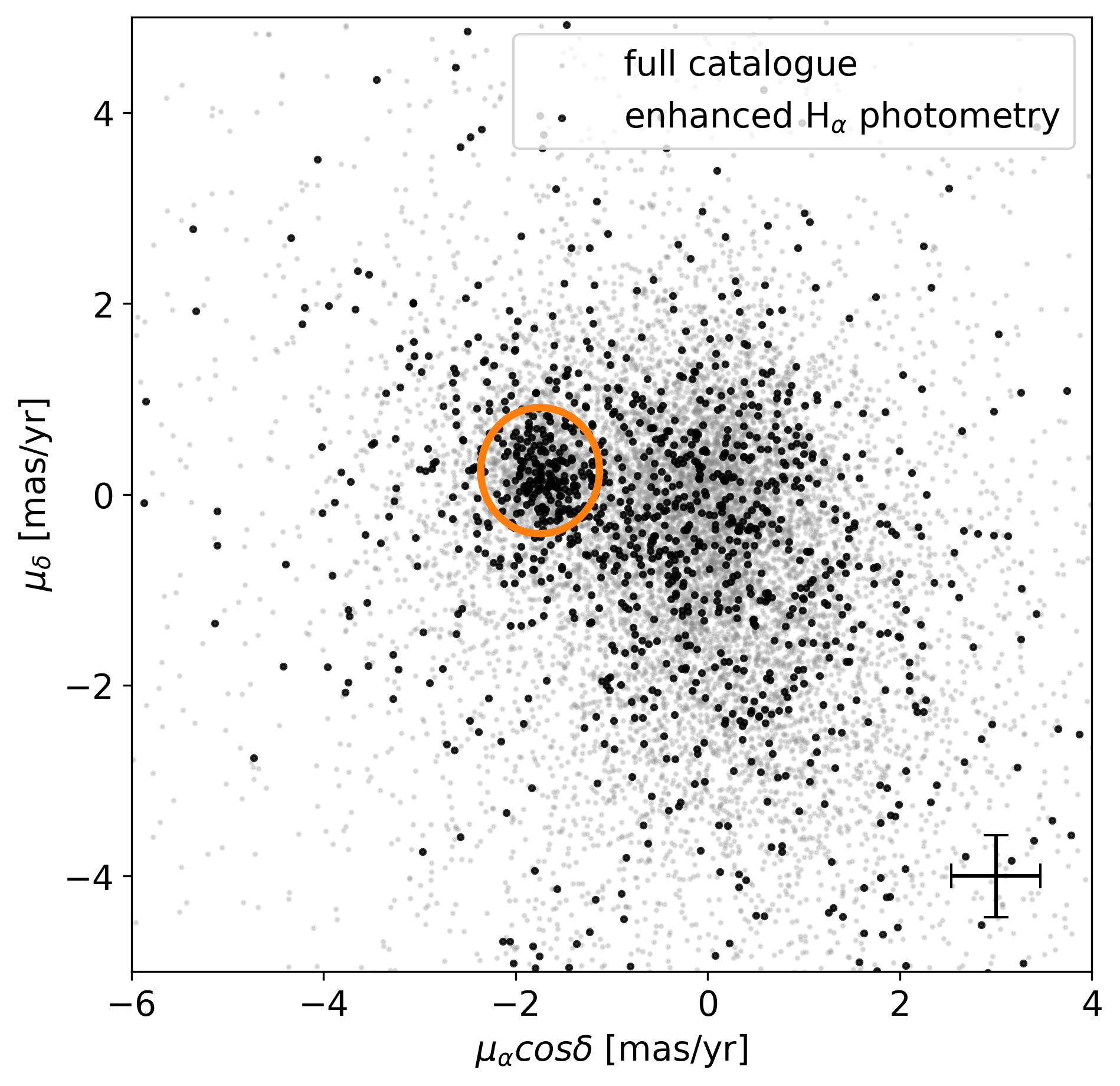}       
        \caption{\gaia~proper motions of bulk of the objects in the direction of \cluster~(grey dots). Black dots mark the proper motions of 688 candidate members from \citet[][B13]{bell13} and \citet[][M17]{meng17} (upper left), OB stars (upper right), strong \ha~emitters from VIMOS spectroscopy (lower left), and objects bright in \ha~from photometry (lower right). The orange ellipse marks the proper motion selection criterion for the construction of the member training set. The mean proper motion uncertainty is shown in the lower right corner of each panel.}
        \label{fig:member_pm}
    \end{subfigure}
\end{figure*}

\begin{figure}
   \centering
   \includegraphics[width=0.45\textwidth]{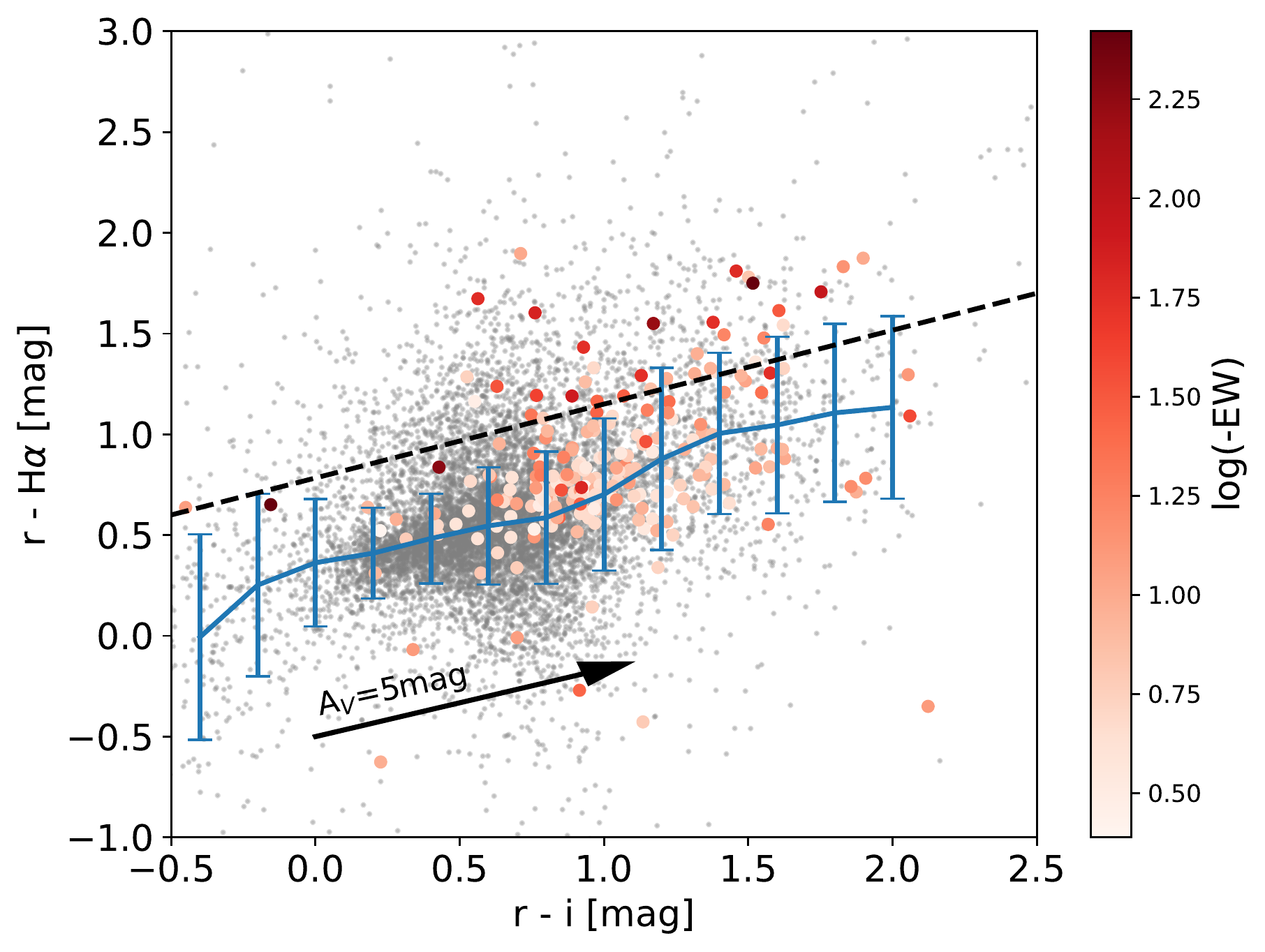}
      \caption{$r-i$, $r-H\alpha$ CMD showing all the sources in the field (grey dots), along with their mean colours and corresponding standard deviations (blue). The red-scale points show the colours of stars with measured pEWs from spectroscopy, coloured by its value. The sources above the black dashed line are selected as potential member candidates due to an enhanced \ha~emission. This selection is further refined using the proper motions and position in CMDs.}
         \label{fig:halpha_phot}
\end{figure}

\begin{figure*}
   \centering
   \includegraphics[width=1.0\textwidth]{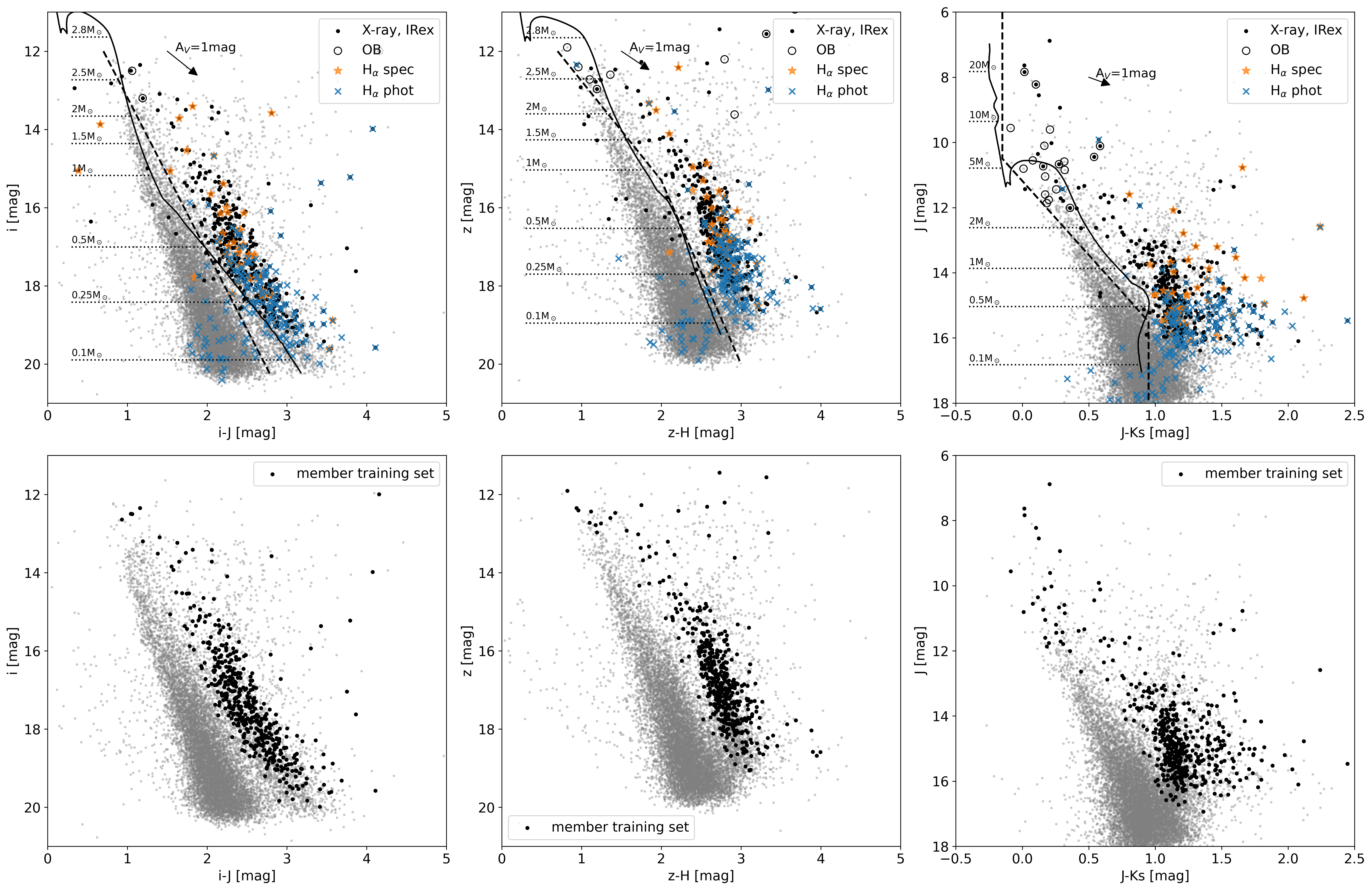}
      \caption{Colour-magnitude diagrams used for the selection of the member training set. The grey dots in all panels show the full catalogue. In the top panels, the proper-motion-selected candidates belonging to various categories (X-ray and IR excess, OB stars, \ha~spectroscopy and \ha~photometry; see text for details) are overplotted. The solid line shows the 2 Myr PARSEC isochrone, and the dashed lines the selection criteria. In the bottom panels, the black points show the 500 finally selected members for the training set, based on the proper motion, colour, and parallax arguments}.
         \label{fig:cmd_sel}
\end{figure*}

\section{Membership using the Probabilistic Random Forest algorithm}
\label{sec:PRF}
\subsection{Construction of the training set}

The Rosette Nebula and \cluster~have been extensively studied in the past, and probable members have been  selected based on spectroscopy, infrared excess, as well as the X-ray emission. Additionally, with the availability of the astrometric measurements from \gaia, these member lists can be further constrained.
The cluster \cluster~is by far the best studied region in the area of interest. Previous works yield a cluster radius in the range $20' - 30'$ \citep{li05,wang08}, and several other detailed studies that will be used in constructing the training set were limited to r$<20'$, where the concentration of the candidate members is highest. For this reason, we limit the area for the training set construction to within the radius of $20'$ of the brightest cluster star HD 46150.

\label{sec:training_set}
\subsubsection{Member list}

 To construct the a list of members to be used for the training, we first compile a catalogue using the following sources in the literature:
\begin{enumerate}
    \item {\bf X-ray and infrared excess candidates.} Both \citet{bell13} and \citet{meng17} provide candidate member lists based on X-ray data \citep{wang08} and infrared excess from \emph{Spitzer} \citep{balog07}. The two lists have a large overlap, and contain 688 unique sources with counterparts in our catalogue. We use this list as the basis for the construction of the member training set, applying further cuts based on the proper motion information from \gaia. In Fig.~\ref{fig:member_pm} we show the proper motions of all the sources in our catalogue (within the plotting limits), along with the 688 candidate members. The 3-sigma-clipped mean proper motions are $\mu_\alpha$cos$\delta$ = $-1.75 \pm 0.41$ mas\,yr$^{-1}$, and $\mu_\delta$= $0.25 \pm 0.44$ mas\,yr$^{-1}$. For our member training set, we chose to keep the sources within the 1.5$\sigma$ proper motion ellipse. We note that this selection is somewhat arbitrary, with an intention to be conservative, rather than complete. Given the size of the uncertainties (indicated by the mean uncertainty in Fig.~\ref{fig:member_pm}), we expect that several of the black dots outside of the orange ellipse are genuine members, which may later be recovered by the algorithm. 449 out of 688 sources pass this proper motion criterion.
  
  \vspace{0.2cm}
    \item {\bf OB stars.} Since the brightest members of the cluster will largely be missed by the selections of \citet{bell13} and \citet{meng17}, we compile a list of candidate OB stars from \citet{romanzunigalada08}, complemented by a list obtained by querying the {\it Simbad} database \citep{simbad}. A total of 25 candidate OB stars are found in our catalogue, of which 20 pass the proper motion criterion defined in point 1. 
    
     \vspace{0.2cm}
    \item {\bf \ha~emitters from spectroscopy.} We select the sources with pseudo-EWs of \ha~consistent with accretion (see Fig.~\ref{fig:halpha_ew}), where we use the criterion by \citet{fang09}. For the earliest spectral types present in our dataset, we extrapolate the \citet{fang09} criterion horizontally. This yields 62 sources, of which 45 pass the proper motion criterion.
    
    \vspace{0.2cm}
    \item {\bf \ha~emitters from photometry.} We search for additional young sources with accretion signatures, with the help of \ha~photometry. In Fig.~\ref{fig:halpha_phot}, we show the $r-i$, $r-H{\alpha}$ diagram of all the sources in the field (grey dots). The blue line and associated error bars show the mean value and standard deviation of $r-$\ha~colour for a set of $r-i$ values, respectively. Coloured points mark the objects with VIMOS spectra, with the colour scale indicating the measured pseudo-EWs of \ha. Majority of pale red points (small absolute pseudo-EWs) follow the blue mean line, while the darker points tend to be located above the mean sequence, as expected for the objects brighter in \ha. This behaviour, however, presents some scatter, with some of the sources with large pseudo-EWs from the VIMOS spectra falling close to the main bulk of sources. This is probably caused by variability in \ha~emission in young stars, due to variable accretion rates, or clumpy disk geometry, as previously reported in multiple works \citep[e.g.][]{jayawardhana06,sousa16,manara21}. We select the stars with enhanced \ha~values by drawing a line (black dashed line in Fig.~\ref{fig:halpha_phot}) running roughly parallel to the mean sequence, above which we find a significant fraction of spectroscopic \ha~emitters. This line is also roughly parallel to the extinction vector,    
    making the selection extinction-independent. According to proper motions (lower right panel in Fig.~\ref{fig:member_pm}), this sample is significantly contaminated at this stage, but it does show a concentration of sources in the area where cluster members tend to reside. The photometric \ha~selection yields 1281 sources, of which 180 pass the proper motion criterion and are retained for the next step in the training set construction.
    
\end{enumerate}

\begin{figure}
   \centering
   \includegraphics[width=0.45\textwidth]{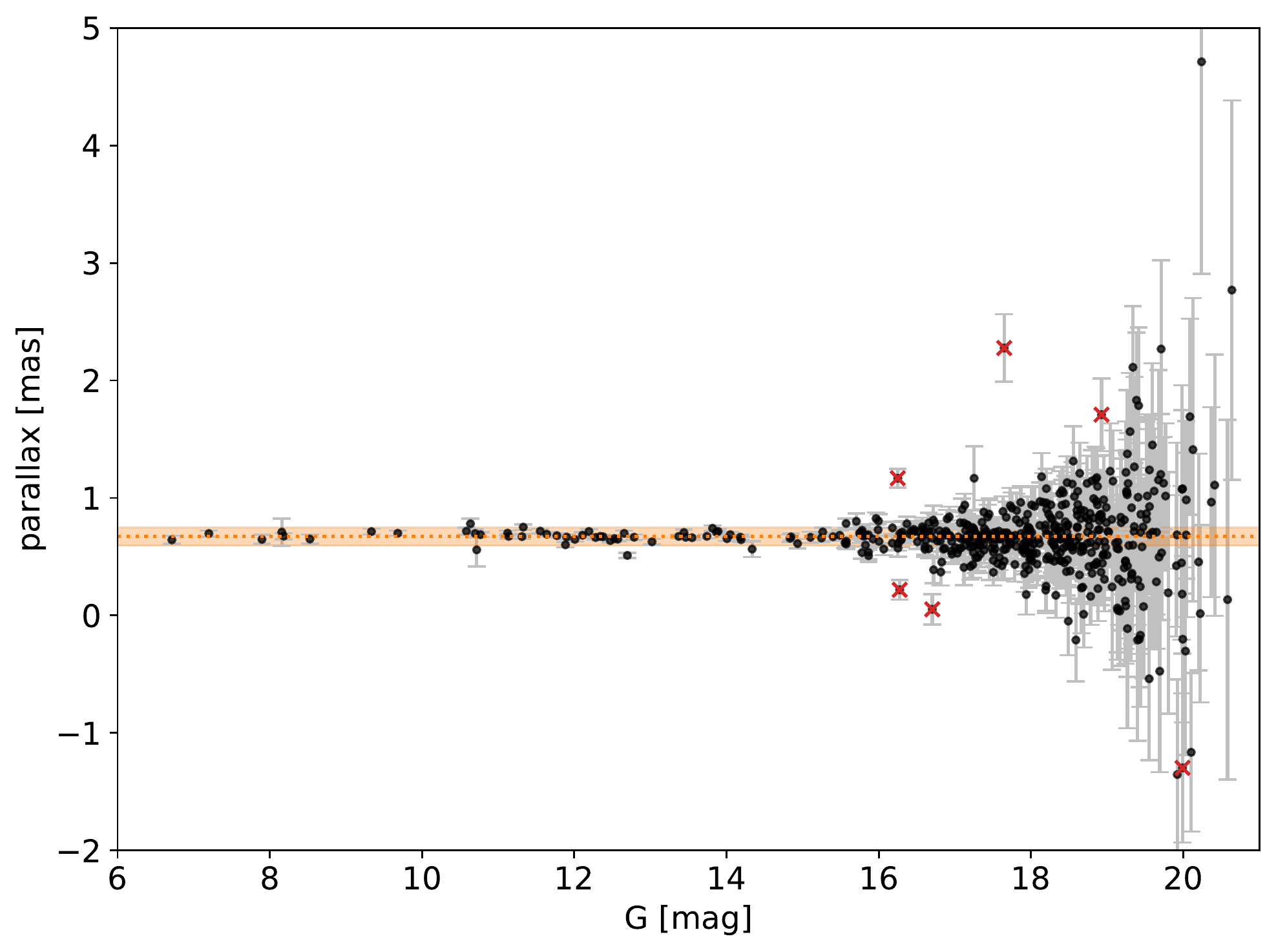}
      \caption{Parallax measurements for the member training set candidates, selected based on proper motions and colours (black dots). The dotted orange line represents the weighted average of the parallaxes, while the shaded area marks $\pm 1\sigma$ range. The red crosses mark the sources that have been excluded from the final member training set (see text for the details of the applied criterion).}
         \label{fig:parallax}
\end{figure}

Joining the four lists of proper motion candidates presented above, and removing the overlapping objects, we obtain a list containing 700 stars, which can further be constrained by their position in a colour-magnitude diagram (CMD; Fig.~\ref{fig:cmd_sel}). As expected, this sample still contains some contamination by the field sources, which is particularly evident in the photometric \ha~sample. Young stars are expected to be located to the right of the main bulk of the field sources due to their pre-main-sequence (PMS) evolutionary status, and the extinction. The solid lines in Fig.~\ref{fig:cmd_sel} show the PARSEC \citep{bressan12,pastorelli2020} isocrone for the age of 2 Myr and a distance of 1500\,pc. We use the isochrones only as a guide for defining the selection lines in the CMD. The OB stars are largely saturated in the optical wavelengths, and therefore only the PMS sources can be selected in the top left and middle panels of Fig.~\ref{fig:cmd_sel}, in contrast to the near-infrared CMD (top right panel), where we can appreciate the transition from the PMS to the main sequence, extending to masses in excess of 20 M$_\odot$. The selection in the $(i, i-J)$ CMD is therefore a single line excluding the candidates that are too blue to be really young, while the selection criterion in the $(J, J-Ks)$ diagram is defined so as to keep most of the bright sources (roughly $\gtrsim$ 1 M$_\odot$) and red PMS stars. Below $J\approx 15.5$\,mag, the sequence of PMS stars merges with that of the field stars (see \citealt{muzic19}), which motivates the vertical selection line in this magnitude range. 
The selection criterion in $(z, z-H)$ has been defined with two lines of slightly different slopes, in order to accommodate the included OB stars, as well as to follow the shape of the PMS sequence. 
The final selection of the member training set consists of the sources that pass the selection in all the three CMDs, in addition to those that pass the selection in any of the diagrams, but are not present in the remaining ones. The latter corresponds exclusively to the bright stars saturated in the $i-$ and $z$- bands. 
After this selection step, we are left with 506 candidate members. The final filtering is then done by looking at the parallaxes (Fig.~\ref{fig:parallax}), and excluding all the sources with the score 
\begin{math}
\zeta = |\varpi-\overline\varpi|/\sqrt{\sigma_\varpi^2 + \sigma^2} >3
\end{math},
where $\varpi$ and $\sigma_\varpi$ represent the parallax measurement and uncertainty for each star, while $\overline\varpi$ and $\sigma$ are the weighted mean and standard deviation of the entire sample, correspondingly.
The six excluded stars are marked with red crosses in Fig.~\ref{fig:parallax}.
The final member list for the training set consists of 500 stars, marked with black points in the lower panels of Fig.~\ref{fig:cmd_sel}.
Of the 500 members in our training set, 420 overlap with the member candidates reported in \citet{bell13} and \citet{meng17}, while the remaining 80 have been added in this work.

A match of our training set to the $Spitzer$ catalogue of the region \citep{balog07} returns 287 common sources, which, according to the YSO classification rules from \citet{fang09} belong either to Class II and transition disk class, or show colours consistent with a naked photosphere. Our training set therefore does not include the more embedded YSOs (Class0/I), which is to be expected  anyways given the main condition of the existence of the $Gaia$ proper motion measurement.

\subsubsection{Non-member list}
The non-members are selected from the annulus with the inner and the outer radii of $15'$ and $25'$, respectively. This way we avoid the densest part of \cluster~and minimise inclusion of potential cluster members, while sampling from a region that has similar extinction properties as the member sample.
To construct the list that would represent the non-member class for the PRF classifier, we use two criteria:
\begin{enumerate}
    \item {\bf Sources with inconsistent proper motions.} This includes all the sources whose individual proper motion within 1$\sigma$ do not intersect the ellipse centred at the mean proper motion of the member list, and with the semi-major axes equal to 3 standard deviations in both directions (Fig.~\ref{fig:member_pm}). 
    
    \item {\bf Blue sources.} This includes the sources located to the left of the selection lines in the CMDs shown in Fig.~\ref{fig:cmd_sel}. Similar criteria have been constructed in other CMDs: $(i,i-Ks)$, and $(z,z-H)$.    
\end{enumerate}
    If a source is found in any of the above-mentioned lists, it is marked as a non-member. The non-member lists contain in total 9679 unique sources, that will be used as a pool for random selection of the sources for the training set. 
    

\subsection{PRF classifiers}
There are several aspects that need to be considered when building a PRF model. These are the following:

\begin{itemize}
\item {\bf Features}. As we have shown in Section~\ref{sec:training_set}, the main features that help distinguish between members and non-members are the position of the stars in various CMDs, as well as their proper motions. Therefore, the main features are the magnitudes ($r$, $i$, $z$, $J$, $H$, $K_s$), colours (constructed from the differences of pairs of the mentioned magnitudes), and proper motions from $Gaia$. Due to the relatively large distance to NGC\,2244, the $Gaia$ parallax tends to have large uncertainties, and is expected to be of limited use for the selection, especially for the fainter objects. We run the PRF algorithm both with and without the parallax as one of the features. In the next section, we discuss the feature importance, i.e. the relevance that each feature has for successful splitting of the two class categories.   

\item {\bf Hyperparameters.} There are several parameters to adjust in the PRF algorithm, although we find that none have a major impact on the results. To test this, we change one of the parameters, while keeping others at a fixed (default) value, and compare the resulting accuracies. In case of the parameters {\it n$\_$estimators} (number of trees), and the parameter {\it max$\_$features} (maximum number of features to be considered at each node split), we find that the accuracies vary within 1$\%$ from the mean value over the tested range ({\it n$\_$estimators}=20-300, {\it max$\_$features}=3-19). We therefore use the default values for these two hyperparameters ({\it n$\_$estimators}=100, {\it max$\_$features}=$\sqrt{N}$, where N is the number of features; N=20 when considering parallax as one of the features, N=19 otherwise). 

\item {\bf Missing data.} Although we do require all the objects our catalogue to have a proper motion measurement, about 5$\%$ lack one or more photometric measurements. Typically, the objects with missing values are either disregarded, or the missing values may be replaced via some of available imputation methods. One of the advantages of the PRF is that no special treatment of the missing data is necessary, a consequence of representing the feature measurements as PDFs. Both at the training and the prediction stages, an object with a missing value at a given feature will have a probability of 0.5 to propagate to either of the two tree nodes (member or non-member). This way the model is
constructed only from the measured features, but at the same time allowing us to predict the class of
objects with missing features without making strong assumptions about the dataset \citep{reis19}. 

\item {\bf Imbalance compensation.} To construct the training set, the member list (500 sources) is paired to a sub-sample of the non-member catalogue (9679). In order to have a more balanced training set, we randomly resample the two classes, which can be done either by over-sampling the member list, under-sampling the non-member list, or a combination of the two. For this task, we use the Python package $imbalanced-learn$ \citep{imbalanced-learn}. The keyword \textit{sampling\_strategy} may be used to control the amount of random over- or under-sampling. For example, \textit{sampling\_strategy}=0.5 will either under-sample the majority class, or over-sample the minority class so that the majority class contains twice as many objects as the minority class. In Table~\ref{tab:sampling_scores} we show different combinations of the \textit{sampling\_strategy} parameters that we used to train the PRF algorithm. We first apply the oversampling to the minority class, followed by the undersampling of the majority class. The choice of the parameters was such to keep the two classes roughly in the same order of magnitude range. 
\end{itemize}

The selected combinations of imbalance compensation parameters, and feature options (parallax included or not) result in a total of 16 different PRF classifiers, whose details are listed in Table~\ref{tab:sampling_scores}.

\subsection{PRF evaluation and comparison}
\label{sec:crossval}
Cross-validation is used to validate the different PRF classifiers, and also to compare their performance. Here, the main training sample (marked A1 to H2) for each classifier is shuffled, and then sub-divided into the test and the train sample, with a proportion 25\%:75\%, respectively. We employ stratified splitting, which ensures the relative class frequencies to remain approximately preserved. For each training sample (A1 to H2), we generate 50 random test and train (25 and 75 division) split samples, and re-run the PRF algorithm for each of them. 

To compare the performance of different PRF classifiers, four different metrics are used. These are the F1 score, the area under the receiver operating characteristic curve (ROC\_AUC), the area under the precision recall curve (PR\_AUC), and the
 Matthews correlation coefficient (MCC; \citealt{yule12,matthews75}). The uncertainties for each score are calculated as the standard deviation of the 50 values obtained through splitting. The results are given in Table~\ref{tab:sampling_scores} and shown in Fig.~\ref{fig:scores} of the Appendix. 
Evidently, several classifiers perform similarly well; for the remainder of the paper we show the results from the run F1, stressing that this choice does not affect the scientific results of the paper when comparing with those obtain by using, for example, the run F2 or C1/C2.

In Fig.~\ref{fig:feat_imp} we show the relative importance of each feature as returned by the classifier in run F1. We note that the feature importance looks similar for all the runs. The uncertainties shown in the plot correspond to the standard deviation of the 50 split values. As may have been expected, the features that are most informative for the classifier are proper motions, in particular the one in right ascension, whereas parallax provides very little information. NIR photometry and colours appear to be more significant than the optical ones. This may partially be due to the fact that the more massive stars do not appear in the optical catalogues (see Fig.~\ref{fig:cmd_sel}).       
\section{Spatial and kinematic properties of candidate members}
\label{sec:kinematics}

\begin{figure}
   \centering
   \includegraphics[width=0.45\textwidth]{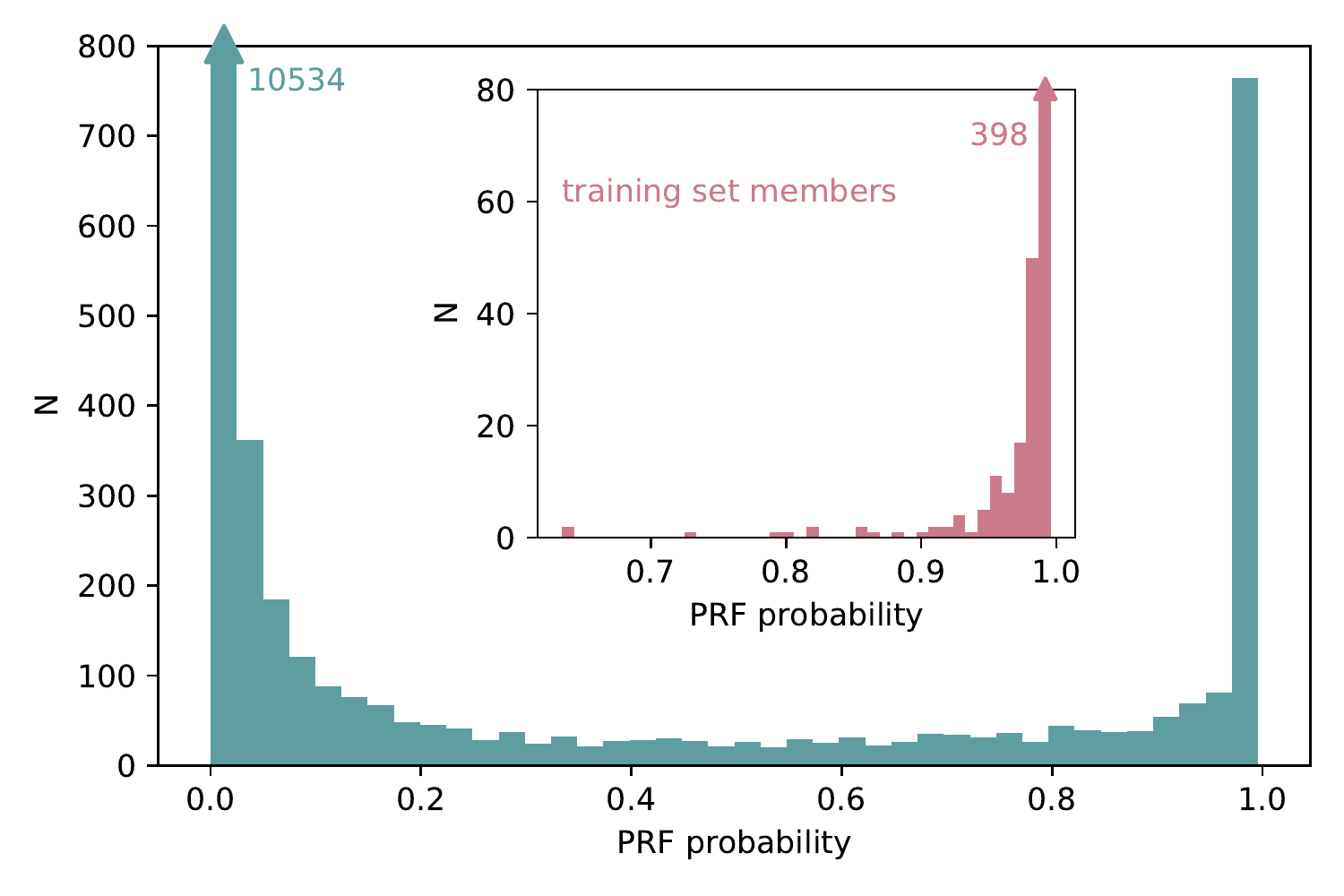}
      \caption{Membership probabilities from the run F1, limited to the objects within 20$'$ radius from the stars HD~46150 for clarity. The inset histogram is showing the probabilities derived for the objects labelled as members in the training set. In both histograms, the values of the edge bins marked with an arrow extend outside of the plotting range; the corresponding bin values are shown on the plot.}
         \label{fig:prf_probs}
\end{figure}
To study different properties of the candidate members of \cluster~and other structures in its vicinity, we focus on the high-probability members from one of our top-ranked runs, F1. 
The classifier F1 has been trained on the entire main training sample and then used to evaluate membership probabilities for unlabelled part of the catalogue.
Fig.~\ref{fig:prf_probs} shows membership probabilities of all the objects in the area within which the training set was selected (circle of 20$'$ radius, centred at HD~46150). The histogram shown in the inset contains the probabilities derived for the objects labelled as members in the training set, with $99\%$ of the objects having probabilities $\geq 0.8$\footnote{The four objects from the member training set that do not pass the 0.8 cut include some of the faintest objects both in the optical and NIR, and two objects with significantly incomplete optical photometry, located close to the selection line in the NIR CMD shown in Fig.~\ref{fig:cmd_sel}.
}. We therefore decide to apply the same cut to select the probable members of the entire region of interest. 
 The distribution of the candidates from the run F1 on the sky is shown in Fig.~\ref{fig:on_sky}, on top of a Planck 857 GHz image \citep{Planck2016}. 
This selection contains 2974 probable members over the entire field. Inside the r=20$'$ circle (black dashed line), we find 1121 candidate members, of which 626 are new, i.e. not found in the training set member list. These numbers will naturally differ for various runs, but we do see a significant overlap, especially for the runs with the best scores (B, C, E, F), where the overlap is on a level of $>95\%$. 

\subsection{Identification of different structures on the sky}

\begin{figure*}
   \centering
   \includegraphics[width=0.9\textwidth]{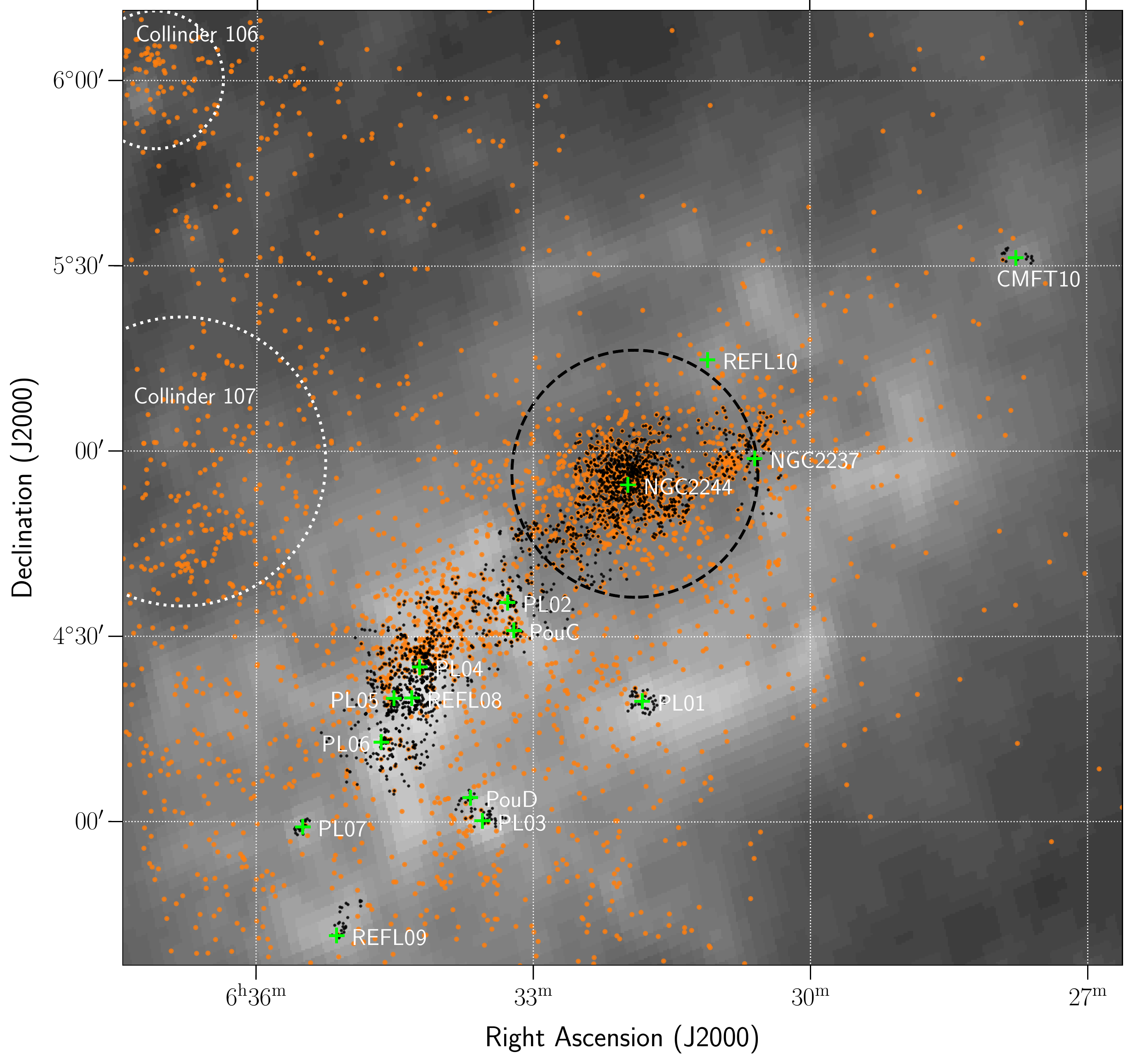}
      \caption{Planck 857 GHz image of the studied region, along with the candidates with membership probabilities $\geq$ 80$\%$ from the run F1 (orange dots), and the mid-infrared and X-ray selected YSOs from \citet{broos13} and \citet[][black dots]{cambresy13}. The black circle marks the 20$'$ ($\sim$8.7\,pc) radius from the brightest star in the field where the training sample has been selected. Green crosses mark the positions of the young clusters and groups identified in \citet{phelps&lada97, romanzuniga06, poulton08, cambresy13}. The dotted white circles are centred at the central positions of the clusters Collinder~106 and 107, with the radii containing half the members, as given in \citet{cantatgaudin20}.  }
         \label{fig:on_sky}
\end{figure*}

Along with the candidates selected here (orange dots), in Fig.~\ref{fig:on_sky}, we show the locations of the previously identified young clusters and groups, marked by the green crosses \citep{phelps&lada97,romanzuniga06,poulton08,romanzunigalada08,cambresy13}. The black dots show the YSOs selected from mid-infared and X-ray data \citep{broos13, cambresy13}.
The central part is clearly dominated by the cluster \cluster, with about 30\% of candidate members being concentrated in a circular region with r=15$'$ around its centre. Slightly to the west of \cluster, we find the young cluster NGC~2237, whose central position as reported by \citet{romanzunigalada08} appears slightly offset from the highest concentration of candidates seen in our selection.
To the south-east of \cluster, we find a more extended region of star formation containing various groups identified in the aforementioned works.
These clusterings appeared as distinct structures in the early selections \citep{phelps&lada97,romanzuniga06, poulton08,romanzunigalada08}, but here their borders are blurred, and it seems that they constitute a rather extended region of star formation, associated with the molecular cloud filaments. This is probably due to the limited depth of the earlier surveys, corroborated by the extended MIR and X-ray population from \citet{broos13} which also appears fairly continuous. We note that their 
selection does not extend over the entire region studied here (e.g. PL01, PouD/PL03, PL07, REFL09 were not included in their study).


To the east, we see objects possibly belonging to open clusters Collinder~106 and 107. The dotted white circles mark their positions and radii within which half of their respective members reside, as determined by \citet{cantatgaudin20}. Both these cluster, as well as NGC~2237 have proper motions, and distances very close to that of \cluster~\citep{cantatgaudin20}.

\subsection{Distance to the region}
\label{sec:distance}
\begin{table}
\begin{center}
\caption{Distances to various structures in the Rosette Nebula} \label{tab:distance}

\begin{tabular}{lcc}\hline\hline
Region     &  d$_1$ [pc]\tablefootmark{a} & d$_2$ [pc]\tablefootmark{b}\\
\hline
all               & $1536 \pm 15$ & $1489 \pm 37$\\
NGC\,2244  & $1483 \pm 24$ & $1440 \pm 32$  \\
NGC\,2237  & $1575 \pm 90$ & $1525 \pm 36$  \\
south-east  & $1487 \pm 43$ & $1447 \pm 35$  \\
\hline
\multicolumn{3}{l}{$^{a}$\footnotesize{from $Gaia$ EDR3 parallaxes}} \\
\multicolumn{3}{l}{$^{b}$\footnotesize{from $Gaia$ EDR3 parallaxes, including the }} \\
\multicolumn{3}{l}{\footnotesize{correction of the parallax bias}}
\end{tabular}    
\end{center}
\end{table}

To estimate the distance to the Rosette Nebula complex, we employ the maximum-likelihood procedure as in \citet{muzic19} and \citet{cantatgaudin18}, using the list of likely members (probability $\geq$0.8) from the run F1. 
We calculate the distances to the entire region, to clusters \cluster~and NGC\,2237 separately, as well as to the south-east concentration of members. 
The NGC\,2237 members were selected from a circle centred at the position given in \citet{romanzunigalada08} with a radius of $10'$ and having the $\Phi$ angle values between 60$^\circ$ and 120$^\circ$ (see Section~\ref{sec:expansion}). The members of ~\cluster~were selected from a circle  centred at ($\alpha$, $\delta$) = (06:31:57.52, +04:54:37.5) and with a radius of $18'$ (see Section~\ref{sec:radial_distr}), excluding the likely NGC\,2237 members. The likely members in the south-east region were selected within a rectangular region with 06:33:00 $<\alpha <$   06:35:00 and  04:15:00 $<\delta <$ 04:40:00.

Based on the observations of quasars, it has been shown that $Gaia$ EDR3 parallaxes may contain systematic offsets from the expected distribution around zero, which depend
in a non-trivial way on the magnitude, colour, and ecliptic latitude of the source \citep{lindegren21}. To estimate the parallax bias, we use the publicly available Python package\footnote{\url{https://gitlab.com/icc-ub/public/gaiadr3_zeropoint}} based on the prescription given in \citet{lindegren21}. The values of the bias show a distribution ranging between -0.05 and 0.02\,mas, with a peak at -0.03\,mas. 
By sampling from the obtained parallax bias distribution, we re-run the maximum-likelihood procedure 100 times, each time subtracting a different value of the bias. The distance and the corresponding uncertainty are then calculated as a weighted mean and standard deviations of the 100 iterations.

The results are given in Table~\ref{tab:distance}. The first column is showing the distance obtained using the maximum likelihood method directly on the $Gaia$ EDR3 parallaxes, while the second column shows distances after applying the parallax bias correction.
It is interesting to note that, while \cluster~and the south-east region appear to be at a similar distance, the cluster NGC\,2237 seems to be located some 90 pc behind. This is significantly larger than the extent of \cluster~in the plane of the sky (diameter $\sim$16\,pc; see Section~\ref{sec:radial_distr}). 

All the distances obtained here are in agreement with those reported in \citet[][1550$^{+100}_{-90}$\,pc]{kuhn19} and \citet[][1400$\pm$100\,pc]{lim21}.
For the remainder of this paper, we assume the distance of 1500\,pc.

\subsection{Radial distribution and the extent of NGC 2244}
\label{sec:radial_distr}

\begin{figure}
   \centering
   \includegraphics[width=.45\textwidth]{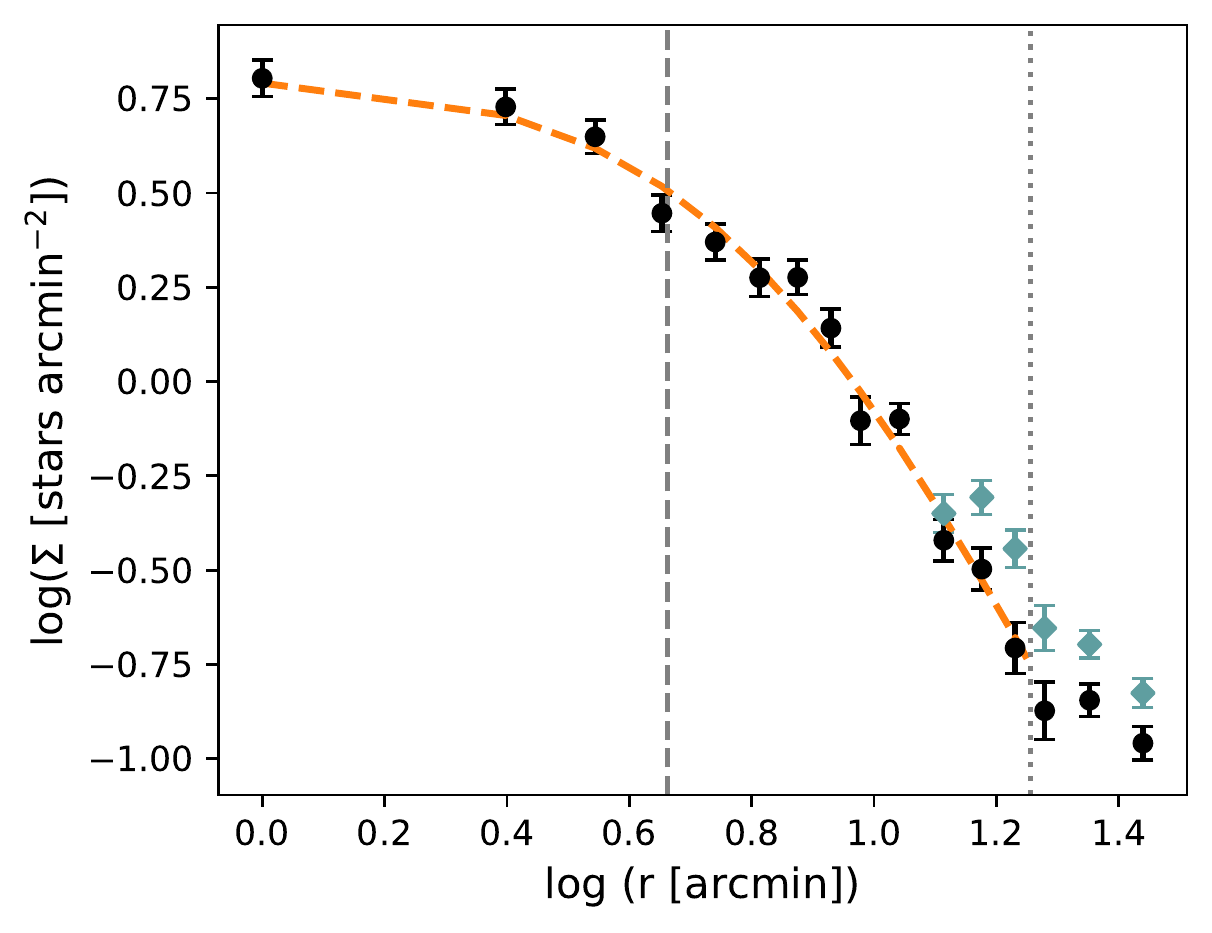}
      \caption{Radial profile of the cluster \cluster~in logarithmic scale. The profile (black filled circles) has been calculated by averaging the number of objects contained inside annuli defined around the \cluster~centre, where both the region occupied by members of NGC~2237, and the overdensity of young stars to the south-east of \cluster~have been masked. The blue diamonds at larger radii show the values without any masking. The dashed orange line shows the best fit EFF profile \citep[][see text for details]{elson87}. The last three points, where the profile starts to flatten out have not been included in the fit. The vertical dashed line represents the core radius derived from the profile fitting, and the dotted one marks the approximate outer radius of the cluster (r=$18'$).}
         \label{fig:radial_profile}
\end{figure}

To determine the radial profile of \cluster, we first calculate a stellar surface density distribution using a two-dimensional Gaussian kernel density estimator (KDE). The maximum of this distribution is taken as the centre of the cluster, and is located at the position ($\alpha$, $\delta$) = (06:31:57.52, +04:54:37.5). The radial profile, shown in Fig.~\ref{fig:radial_profile}, is calculated from the number of stars within the concentric annuli around the central position, divided by the area of the respective annulus. At radii r$\gtrsim 12'$, stars belonging to NGC~2237 and the south-east overdensity start to affect the radial distribution, and should be excluded from the calculation. Filled black circles in Fig.~\ref{fig:radial_profile} show the radial profile with these two regions masked, while the blue diamonds consider all the stars. Outwards from r=18$'$ the density distribution flattens out, which we take as an indication of reaching the cluster outer radius. This radius corresponds to $\sim$8\,pc at a distance of 1500\,pc, and is marked with the vertical dotted line.      
To fit the radial distribution shown in Fig.~\ref{fig:radial_profile}, we take a generalised profile known as Elson-Fall-Freeman (EFF) profile \citep{elson87} in the following form:

\begin{equation}
    \Sigma(r) = \Sigma_0\Big[1+\Big(\frac{r}{a}\Big)^2\Big]^{-\gamma/2},
\end{equation}
where $r$ stands for the projected distance from the centre of the cluster, $\Sigma_0$ is the surface density at its centre, and $a$ is a scale parameter. The core radius, r$_c$ of the King profile \citep{king66} is then given by
\begin{equation}
    r_c=a\,\Big(2^{2/\gamma} -1\Big)^{1/2}
.\end{equation}
The best fit profile, obtained from the black points excluding the rightmost three, is shown as an orange dashed line in Fig.~\ref{fig:radial_profile}, and its parameters are: 
$\Sigma_0 = 6.4 \pm 0.7$ stars arcmin$^{-2}$, 
$a = 6.3 \pm 1.1'$,
$\gamma = 3.2 \pm 0.5$. From this, we obtain the core radius of the cluster, r$_c$ = $2.00 \pm 0.35$\,pc. The scale parameter $a$ and the slope are in agreement (within errors) with those obtained in \citet{lim21}. The slope $\gamma$ is between that of a modified Hubble model (an approximation of an isothermal sphere; $\gamma$=2) and a Plummer model \citep[$\gamma=4$;][]{binneytremaine,portegieszwart10}, in line with observations of a substantial number of other young clusters, with typical values between 2 and 4. Some examples are NGC\,6231 \citep{kuhn17}, RCW\,38, the Flame Nebula Cluster, M17 \citep{kuhn14}, IC\,4665 \citep{miret19}, Ruprecht\,147 \citep{olivares19}, NGC\,3603 and Westerlund I \citep{portegieszwart10}.
The mean surface density of the cluster, $\Sigma_{mean}$, is $\sim$\,2.2\,stars\,arcmin$^{-2}$, equivalent to 12 \,stars\,pc$^{-2}$ at a distance of 1500 pc.

 Table~\ref{tab:parameters} contains a summary of various physical parameters of the cluster \cluster, including those derived in this section. 
 
\subsection{Stellar kinematics}

\begin{figure*}
   \centering
   \includegraphics[width=1\textwidth]{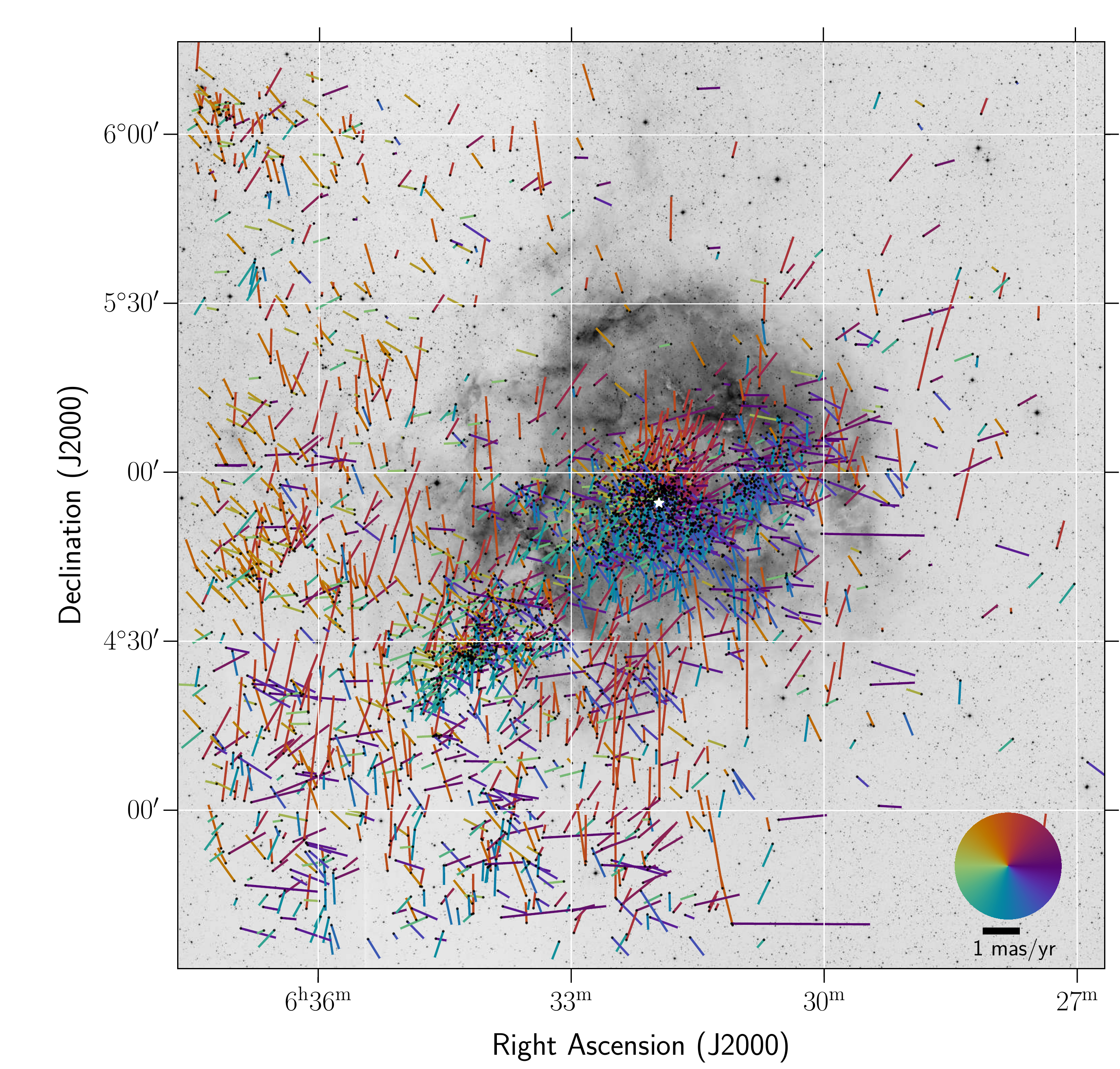}
      \caption{Proper motion vectors relative to the mean proper motion of \cluster, on top of a DSS red image shown in grey scale. Stellar positions marked by black points are the point of origin for the bars showing proper motions. The orientations of the
      bars and their colours indicate their direction of motion. The colour segregation in the central part of the image indicated that many stars in the cluster \cluster~move away from its centre. The white star indicates the position with the highest surface density of stars, which we adopt as the centre of \cluster. For a zoom into some particular regions of this plot, see Figs.~\ref{fig:on_sky_pm_zoom1} and \ref{fig:on_sky_pm_zoom2}.}
         \label{fig:on_sky_pm}
\end{figure*}

In this section we study the two-dimensional internal motions of the stars, relative to the mean motion of the cluster \cluster. 
The error-weighted mean proper motion was calculated using the sources within the 12$'$ radius from the centre of \cluster, a conservative radius that ensures most of the selected sources would indeed be members of the cluster. The mean proper motion was subtracted from all the stars in the region. In this way we can study the internal kinematics of \cluster, as well as relative motions of other young stellar groups with respect to \cluster. 
We then correct for the so-called perspective expansion (or contraction), an effect that is caused by the projection of the cluster's radial motion at the different positions of cluster members (i.e. the further away a star is from the centre of the projection - in this case the centre of the cluster - the larger is the fraction of the radial velocity of the cluster that is projected into the observed proper motion components;
\citealt{vanleeuwen09,helmi18,kuhn19}). The adopted radial velocity is v$_r$=12.8 km\,s$^{-1}$, derived by \citet{lim21} from high-resolution spectroscopy of a large number of YSOs in \cluster. For the centre of the stellar distribution, we take the position ($\alpha$, $\delta$) = (06:31:57.52, +04:54:37.5), obtained from the maximum of the two-dimensional stellar density distribution, and calculated using a two-dimensional KDE.  
We find that the effect is generally small, $\sim$0.03\,mas\,yr$^{-1}$ at the distance of 1$^\circ$ from the centre, and $\sim$0.05\,mas\,yr$^{-1}$ at the largest relative distances present in our dataset.

In Fig.~\ref{fig:on_sky_pm}, we show the relative proper motions of all the selected sources, with respect to the mean motion of \cluster. The positions of stars are marked by black dots, with a bar extending from each dot showing the direction of the relative motion. The bars are also colour-coded according to the direction on the sky. In the region occupied by \cluster~(see Fig.~\ref{fig:on_sky_pm_zoom1} for a zoomed-in version), one can clearly notice a circular colour segregated pattern, caused by a large number of source moving away from the cluster centre, indicating expansion. A similar behaviour has previously been reported by \citet{kuhn19} and \citet{lim21}. 
To the west of \cluster, a large number of sources shows a bulk motion in the southward direction. These stars are likely the members of NGC~2237, and separate from \cluster. The zoomed-in version of the plot concentrated on NGC\,2237 is shown in the right panel of Fig.~\ref{fig:on_sky_pm_zoom2}. 
Towards the east edges of Fig.~\ref{fig:on_sky_pm}, we note a region in the lower left corner where relative motions appear to have random orientations, as well as the region occupied by members of Collinder 106 and 107, whose members seem to have a largely coherent motion pointing north-east. 

Taking a closer look at the kinematics of the south-east concentration connecting the mid-infrared clusters (Fig.~\ref{fig:on_sky_pm_zoom2} left), we see that the relative proper motions appear randomly distributed, except for the stars around the position 06:34:35, 04:40:21 (corresponding roughly to the group PL05 shown in Fig.~\ref{fig:on_sky}), which do seem to have a common preferred motion towards the south-east.

\begin{figure*}
   \centering
   \includegraphics[width=0.9\textwidth]{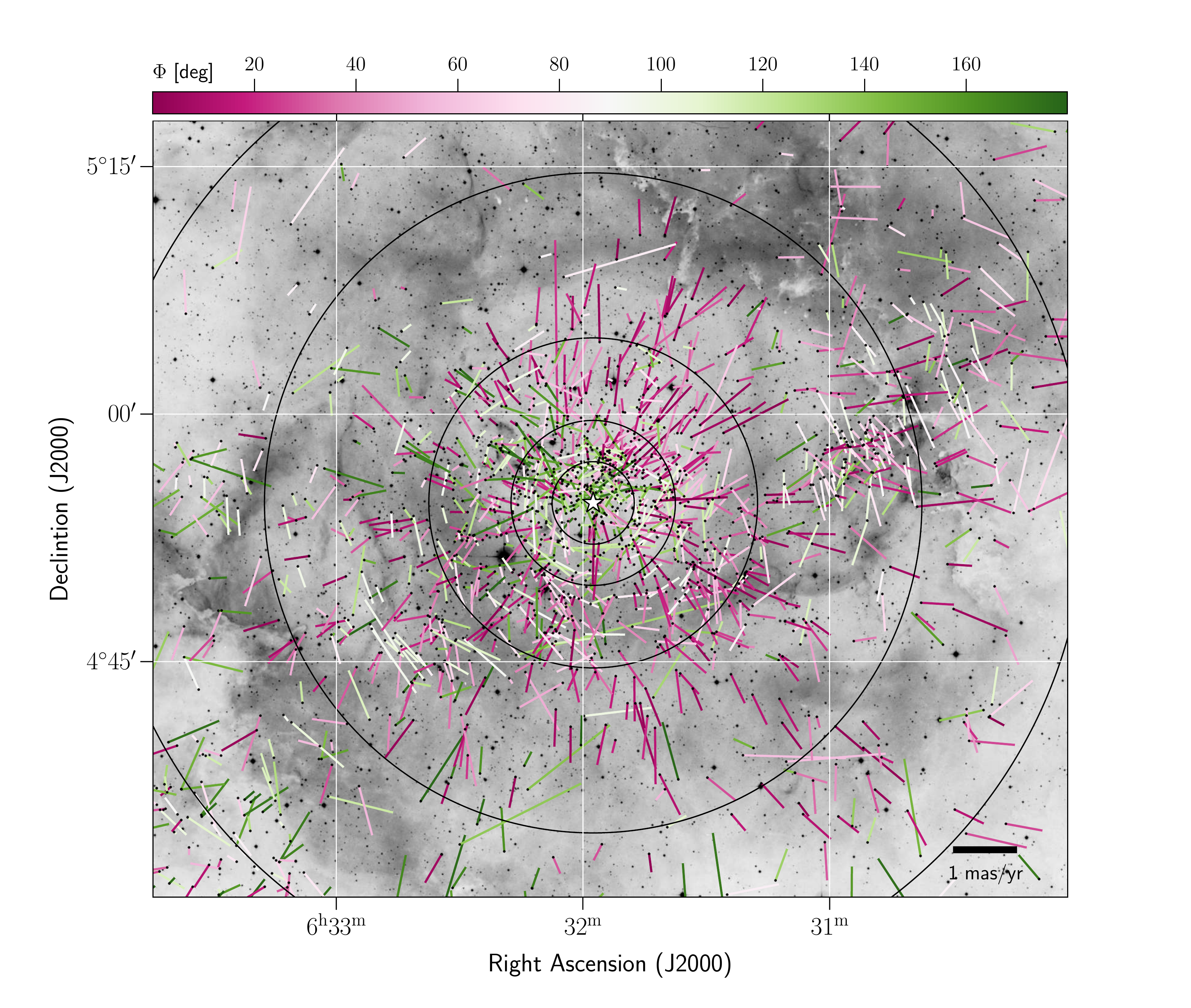}
      \caption{Same as Fig.~\ref{fig:on_sky_pm_zoom1}, but with the relative proper motions coloured according to the angle between star's relative proper motion vector and the line that connects it with the centre of the cluster. The purple hues mark the stars with a predominant motion away from the cluster centre, and the green towards it. We note that while near-white sources move close to perpendicularly to the line that connects them with the centre of the cluster, they may be moving in two opposite directions with respect to each other. The concentric circles correspond to the radii of (2.5, 5, 10, 20, 30)$'$ from the cluster centre (white star), used to plot the distribution of angle $\Phi$ in Fig.~\ref{fig:phi_kde}.}
         \label{fig:on_sky_phi}
\end{figure*}

\begin{figure}
   \centering
   \includegraphics[width=0.45\textwidth]{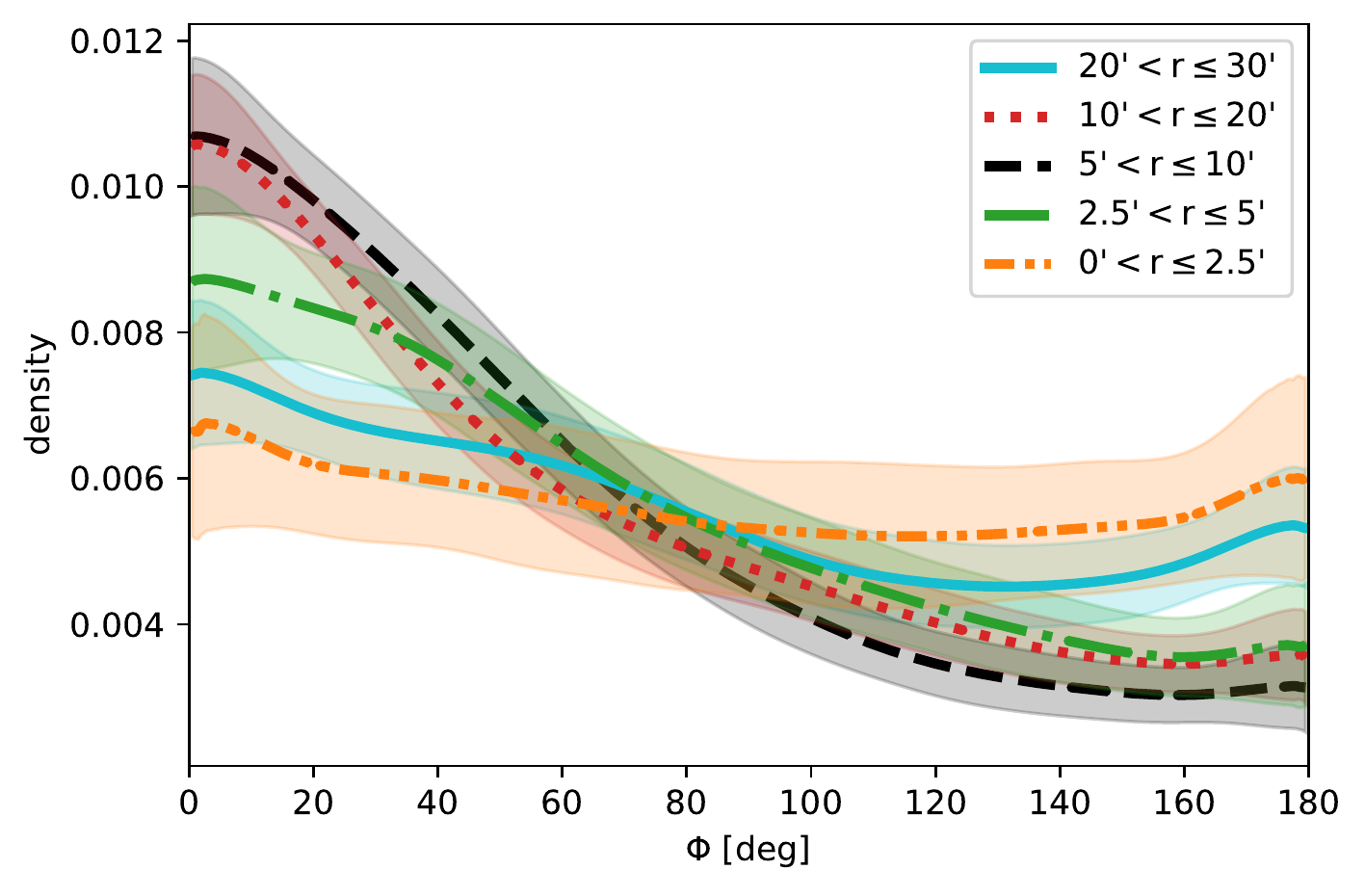}
      \caption{KDE distributions of angle $\Phi$, for various radii from the centre of \cluster~shown on sky in Fig.~\ref{fig:on_sky_phi}. The confidence intervals correspond to 1$\sigma$.}
         \label{fig:phi_kde}
\end{figure}

\subsection{Expansion of \cluster}
\label{sec:expansion}

\begin{figure}
   \centering
   \includegraphics[width=0.45\textwidth]{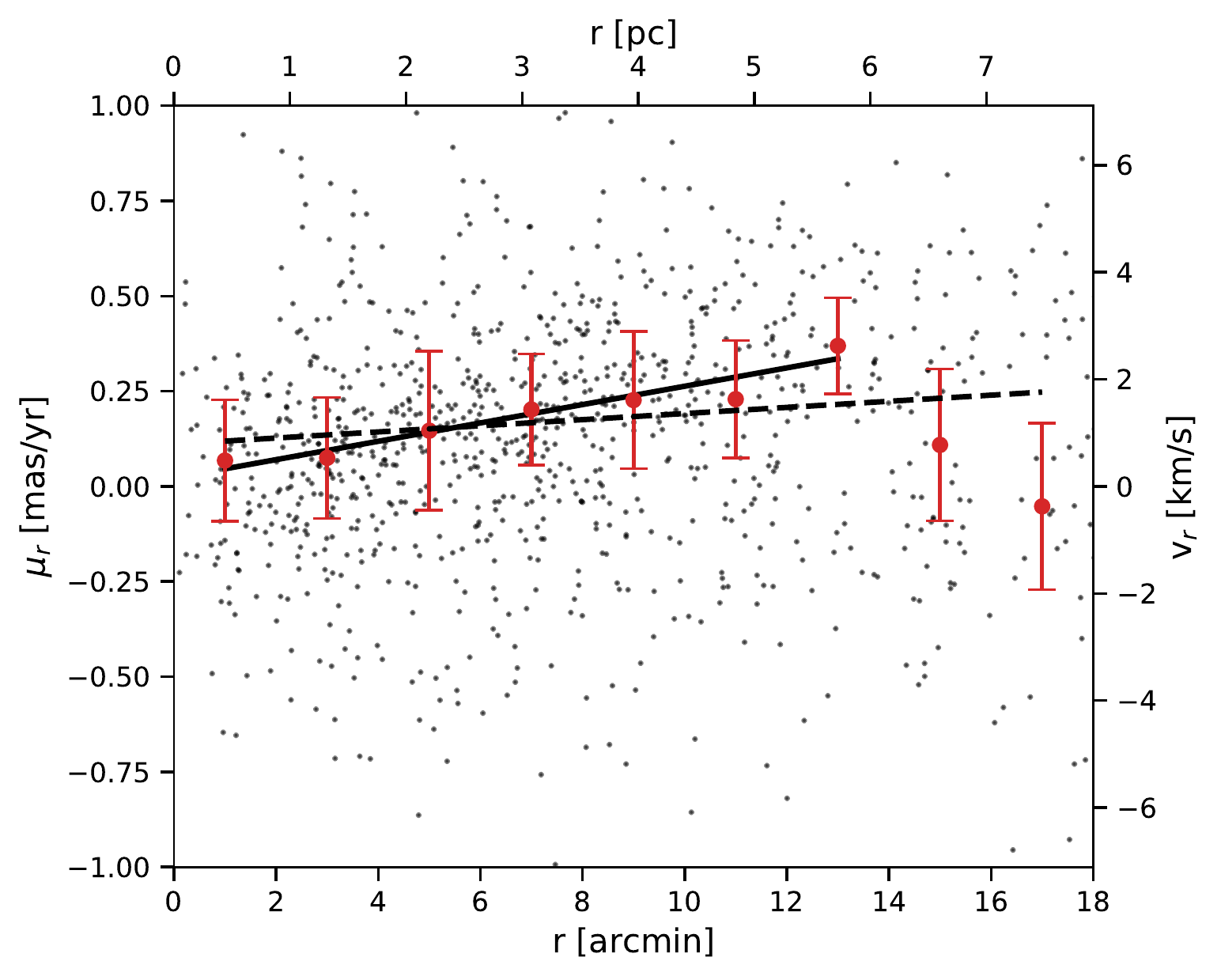}
      \caption{Radial component of the relative proper motions as a function of distance from the centre of \cluster. Stars are binned by radial distance with the bin size of 2$'$, and the red points and error-bars indicate the weighted mean and the standard deviation for each bin. The top and the right axes assume a distance of 1500 pc. The positive values of the velocity indicate expansion. The linear fit over the entire range (black dashed line) has a slope of $0.008 \pm 0.009$ mas\,yr$^{-1}$\,arcmin$^{-1}$ and intercept of $0.11 \pm 0.09$\,mas\,yr$^{-1}$. The fit for r$<14'$ (solid black line), has a slope of $0.024 \pm 0.003$ mas\,yr$^{-1}$\,arcmin$^{-1}$ and intercept of $0.022 \pm 0.028$\,mas\,yr$^{-1}$.}
         \label{fig:pm_radius}

\end{figure}

\begin{figure}
   \centering
   \includegraphics[width=0.4\textwidth]{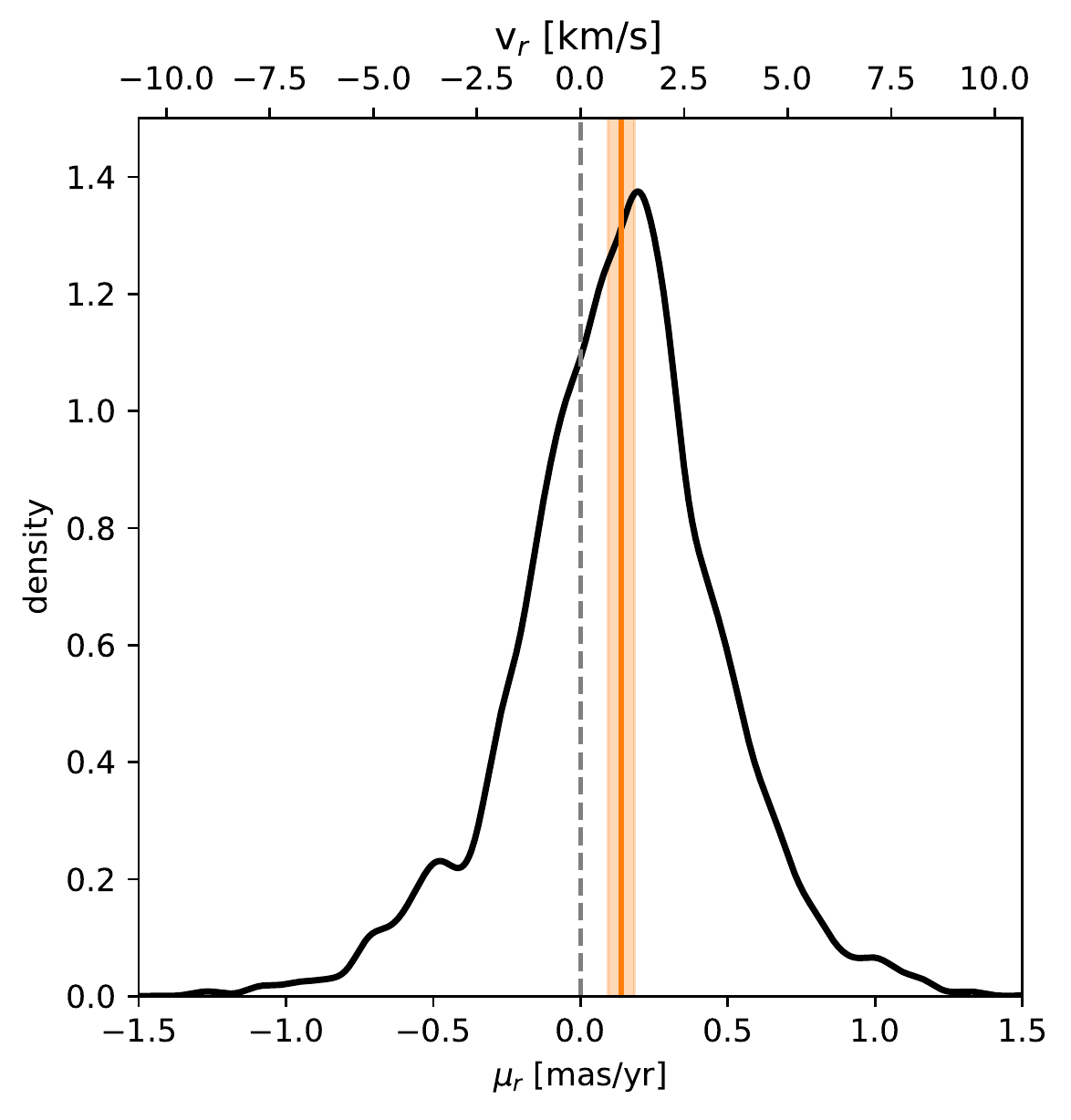}
      \caption{Distribution of the radial component of the relative proper motions for the stars within the 18$'$ radius from the centre of \cluster~(shown as a KDE plot). The grey dashed line indicates the zero velocity, the solid orange line the median of the distribution, and the shaded orange region indicates the 3$\sigma$ uncertainty on the median. The values on the top axis assume a distance of 1500\,pc.}
         \label{fig:pm_distribution}
\end{figure}

To study in more details the internal kinematics of \cluster~and its immediate surroundings, we calculate the angle $\Phi$, defined as an angle between the star's relative proper motion vector and the line that connects it with the centre of the cluster (see \citealt{lim21}). $\Phi = 0^\circ$ corresponds to proper motion vectors pointing away from the cluster centre, and $\Phi=180^\circ$ towards it. Fig.~\ref{fig:on_sky_phi} shows the relative proper motion vectors on top of an optical image of the region. The vectors are colour-coded according to the value of the angle $\Phi$, with purple and green hues marking the stars moving radially away from the centre, or towards it, respectively. In this representation, the outward radial motion of a large number of members of \cluster~can easily be appreciated. In Fig.~\ref{fig:phi_kde}, we show the density distribution of the angle $\Phi$, for several angular distance bins from the cluster centre, as indicated by black circles in Fig.~\ref{fig:on_sky_phi}. The density has been calculated with the KDE, using a Gaussian kernel with bandwidth = 10. The shaded areas show the 1$\sigma$ confidence interval, obtained through a Monte Carlo (MC) simulation with 1000 iterations, in which the angle $\Phi$ was varied within the uncertainties for each object. At all radii, except for the innermost and the outermost part which are fairly flat, we see a pronounced peak at low $\Phi$ values, corresponding to the purple-hued sources in Fig.~\ref{fig:on_sky_phi}. 

Here we must note that in the colour representation of Fig.~\ref{fig:on_sky_phi}, the white colour represents the motion perpendicular to the line that is connecting a star to the centre of \cluster, however, not in a vectorial form, meaning that diametrically opposite motions may be represented by similar colours. This is not the case for NGC~2237, where we see close-to-white shading only for the sources moving southwards. The other apparent group within the 20$'$-30$'$ annulus, contains sources moving in both directions (see also Fig.~\ref{fig:on_sky_pm_zoom1}), and therefore may be just a chance projection. 

In Fig.~\ref{fig:pm_radius} we show the radial component of the relative proper motion vector ($\mu_r$) as a function of the distance from the cluster centre.
 $\mu_r$ has been calculated as a projection of the relative proper motion vector onto the vector connecting the centre of the cluster with the star in question. 

Positive value of $\mu_r$ means that the radial component of the proper motion is directed away from the cluster centre, while the negative value signals the opposite behaviour. Stars were binned by radial
distance with the bin size of 2$'$, and the weighted mean and the corresponding standard deviations of $\mu_r$ for each bin are shown in red. We can clearly see that $\mu_r$ steadily increases as we move further out from the cluster centre up to the projected distance of $\sim 14'$, providing evidence of radially dependent expansion velocity. The trend, however does not hold at the edges of the cluster. We note that this is not caused by the members of NGC 2237, which have been masked (see Section~\ref{sec:radial_distr}) for this part of the analysis. We perform two weighted linear regressions to the red points in Fig.~\ref{fig:pm_radius}, one including only the stars within r=$14'$ (6.1\,pc at 1500\,pc), and the other over the entire extent of the cluster (r$<18'$; 7.8\,pc). For the two fits, we obtain the following slopes:
\begin{itemize}
    \item 
    $\rm{r}<14'$: $0.024\pm0.003$\,mas\,yr$^{-1}$\,arcmin$^{-1}$ = $0.39 \pm 0.05$\,km\,s$^{-1}$\,pc$^{-1}$,
    \item $\rm{r}<18'$: $0.008\pm0.009$\,mas\,yr$^{-1}$\,arcmin$^{-1}$ = $0.13 \pm 0.15$\,km\,s$^{-1}$\,pc$^{-1}$.
\end{itemize}

Previously, \citet{kuhn19} have also detected a similar trend of increase of the expansion velocity with radius, out to the radius of $11'$, deriving the slope of $0.6 \pm 0.1$\,km\,s$^{-1}$\,pc$^{-1}$, which is somewhat larger than the one derived here, and only marginally consistent. We note that, when repeating the fit using the same radius as in \citet{kuhn19}, the value remains essentially unchanged with respect to the one obtained for the radius of $14'$.

The median of the distribution of $\mu_r$ out to r=18$'$ (Fig.\,\ref{fig:pm_distribution}) peaks at $0.138\pm 0.015$\,mas\,yr$^{-1}$, which is equivalent to v$_r=1.0\pm 0.1$\,km\,s$^{-1}$ at a distance of 1500\,pc. This value is in agreement with the result $1.23\pm0.17$\,km\,s$^{-1}$ reported by \citet{kuhn19} for a distance of 1550\,pc (or $1.19\pm0.16$\,km\,s$^{-1}$ when scaled to the distance considered here).

\section{Physical characterisation of candidate members}
\label{sec:physical_character}

\subsection{SED fitting}
\label{sec:seds}
The spectral energy distributions (SEDs) were constructed using the optical to mid-infrared photometry,
and the SED fitting was performed with the help of the Virtual Observatory SED Analyzer (VOSA; \citealt{bayo08}). VOSA compares the observed fluxes with the synthetic photometry at some assumed distance for each object (fixed to 1500 pc), looking for the best fit effective temperature ($T_{\mathrm{eff}}$), extinction ($A_{\mathrm{V}}$), and surface gravity (log($g$)) combination. For the objects showing excess in the WISE photometry, the SED fitting is performed over the optical and near-infrared portions of the SED, otherwise the full available range is included in the fit. The metallicity was fixed to the Solar value. We use the BT-Settl models \citep{allard11}, and probe $T_{\mathrm{eff}}$ over the range offered in VOSA (400 - 70,000 K), with a step of 100 K, and $A_{\mathrm{V}}$ between 0 and 5\,mag, with a step of 0.25\,mag (automatically determined by VOSA). The log($g$) was varied between 3.5 and 5.0, which is suitable for young cool very low-mass stars, and the field stars.
We note, however, that the SED fitting procedure is largely insensitive to log($g$), resulting in flat probability density distributions over the inspected range (see e.g. \citealt{bayo17}).
In Figure~\ref{fig:seds} of the Appendix, we show a subset of SED fits, for a wide range of fitted temperatures, with (bottom panels) or without (top panels) infrared excess. 


\subsection{Luminosity function}

\begin{figure}
   \centering
   \includegraphics[width=0.45\textwidth]{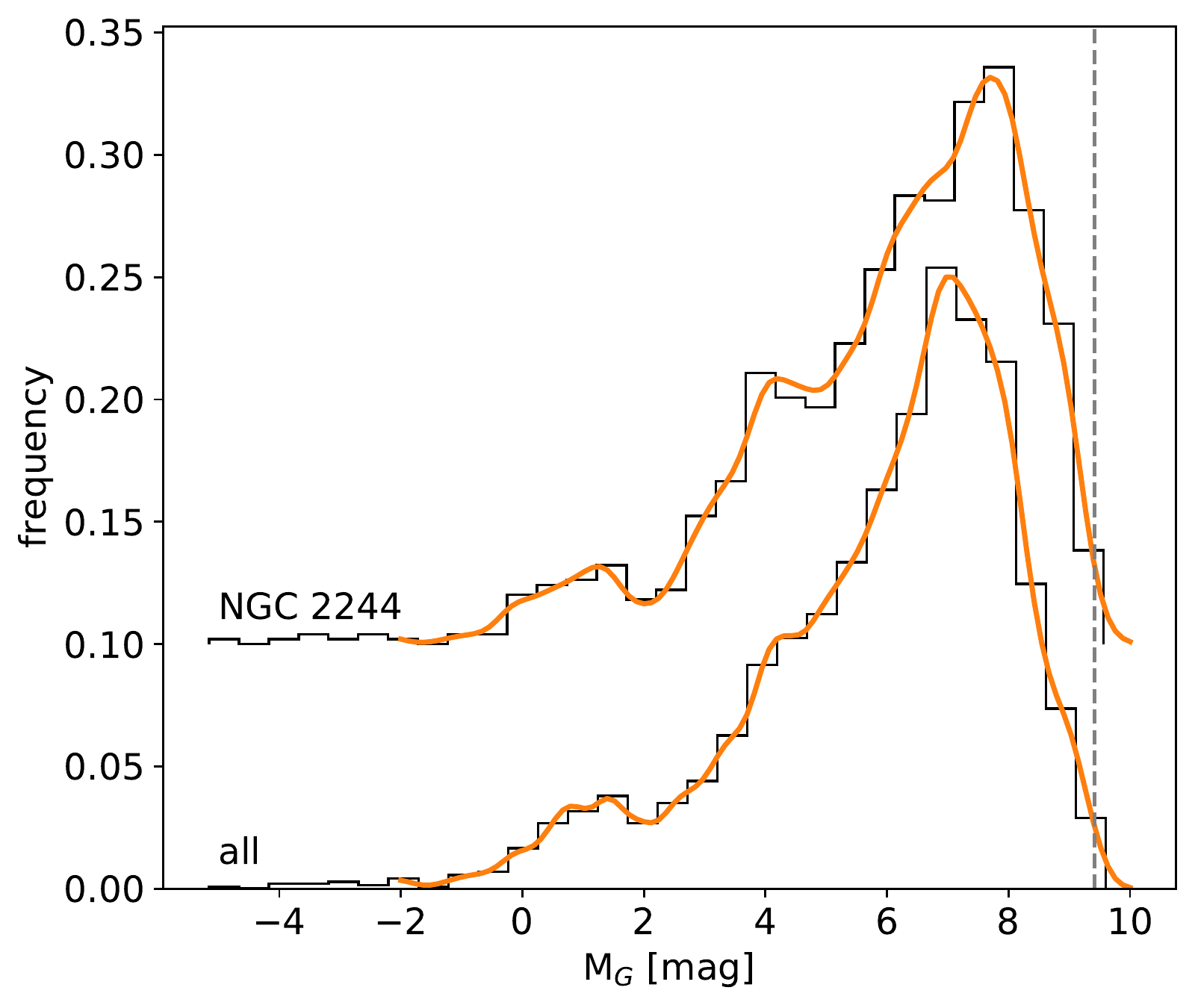}
      \caption{Luminosity function in $Gaia$ G-band for the probable members in the entire studied region (run F1; bottom histogram), and only for \cluster~(within the cluster radius of r$=18'$, excluding the probable members of NGC\,2237; top histogram). The orange lines represent KDE distributions calculated with Silverman rule bandwidth. The upper distribution is shifted vertically by 0.1 for clarity. The completeness limit of the $Gaia$ $G$-band with the distance modulus subtracted is shown as a vertical dashed line.}
         \label{fig:lumf}
\end{figure}
Fig.~\ref{fig:lumf} exhibits the luminosity function in form of a histogram of absolute magnitudes in $Gaia$ $G$-band, and a one-dimensional Gaussian KDE with Silverman bandwidth \citep{silverman86}. The lower histogram shows high probability members over the entire region, while the upper histogram contains only those belonging to \cluster. Here we include only the sources within our estimated cluster radius of r=$18'$ (Section~\ref{sec:radial_distr}), while at the same time excluding possible members of NGC\,2237 (as described in Section~\ref{sec:distance}). To obtain the absolute magnitudes, we correct the apparent one by the distance modulus corresponding to the distance of 1500\,pc, and the extinction derived in Section~\ref{sec:seds}, assuming A$_G$ = 0.789 A$_V$ \citep{wang_chen19}. The luminosity function of \cluster~shows a pronounced dip at M$_G\approx 5$\,mag, probably corresponding to the Wielen dip, a feature often observed in luminosity functions of open clusters, and possibly caused by the changes in the atmospheric physics with mass \citep{wielen74,kroupa90,jeffries01,naylor02}. As discussed in \citet{guo21}, the Wielen dip appears at M$_G$=5\,mag for an age of 1 Myr, but by 10 Myr it shifts to M$_G$=7\,mag, where it stabilises. Observations of the Wielen dip at M$_G\approx 5$\,mag speaks in favour of the young age of the Rosette population, which will be further discussed in the following sections.

On the other hand, the age dating using the so-called H-peak \citep{guo21} yields a somewhat older age of $\sim$5 Myr, assuming that the smaller peak seen at M$_G\approx$1.3\,mag corresponds to this feature. The H-peak was first reported in \citet{piskunov96}, and is thought to be caused by the change of the mass-luminosity relation at the transition from the pre-main sequence to the main sequence, which creates a local peak in the stellar number density at a given luminosity. As the mass of the stars arriving at the main sequence gets lower with time, so the H-peak also evolves towards fainter magnitudes.

\subsection{Mass and age derivation from the HR diagram}
\label{sec:massage}
\begin{figure*}
   \centering
   \includegraphics[width=\textwidth]{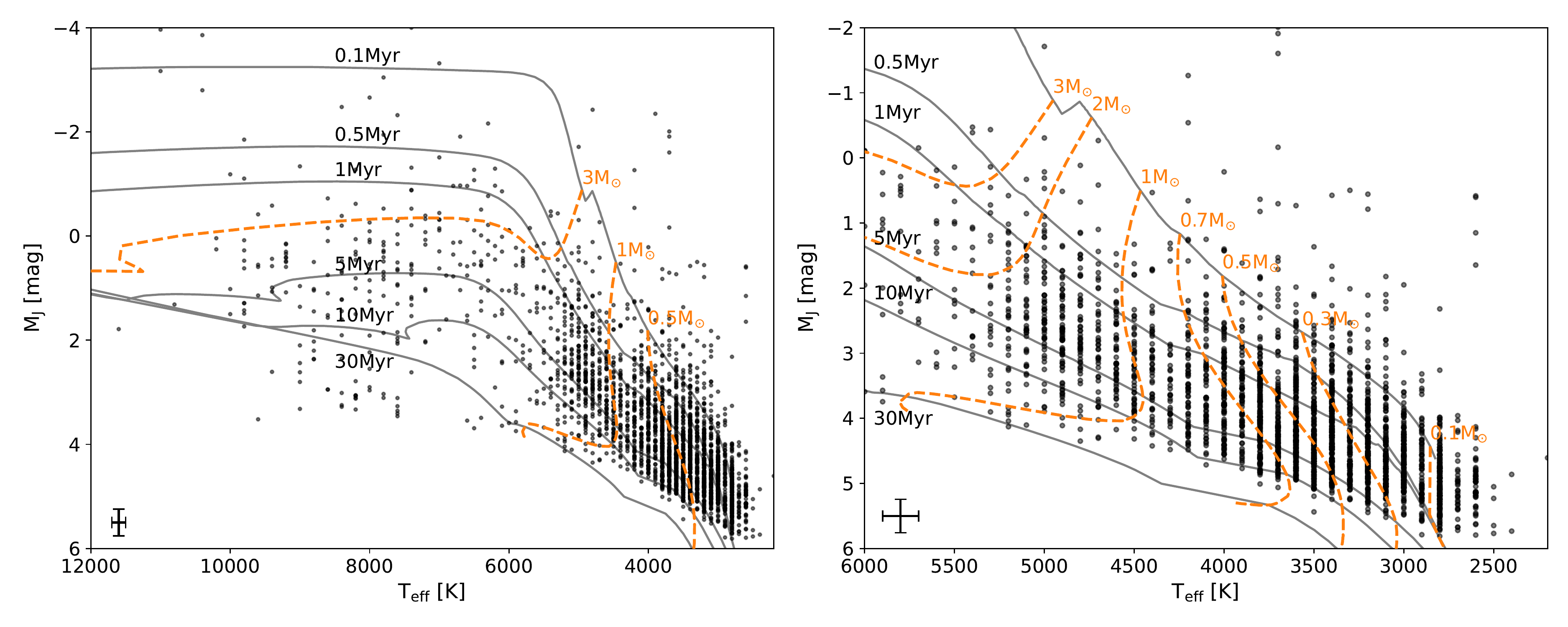}
      \caption{Hertzsprung-Russell diagram showing the candidates with membership probabilities $\geq$ 80$\%$ from the run F1 (black dots). The right panel is the zoomed-in version of the one to the left. The isochrones (grey solid lines) and the lines of constant mass (dashed orange lines) are from the PARSEC series. A typical errorbar is shown in the lower left corner.}
         \label{fig:hrd}
\end{figure*}


\begin{figure}
   \centering
   \includegraphics[width=0.5\textwidth]{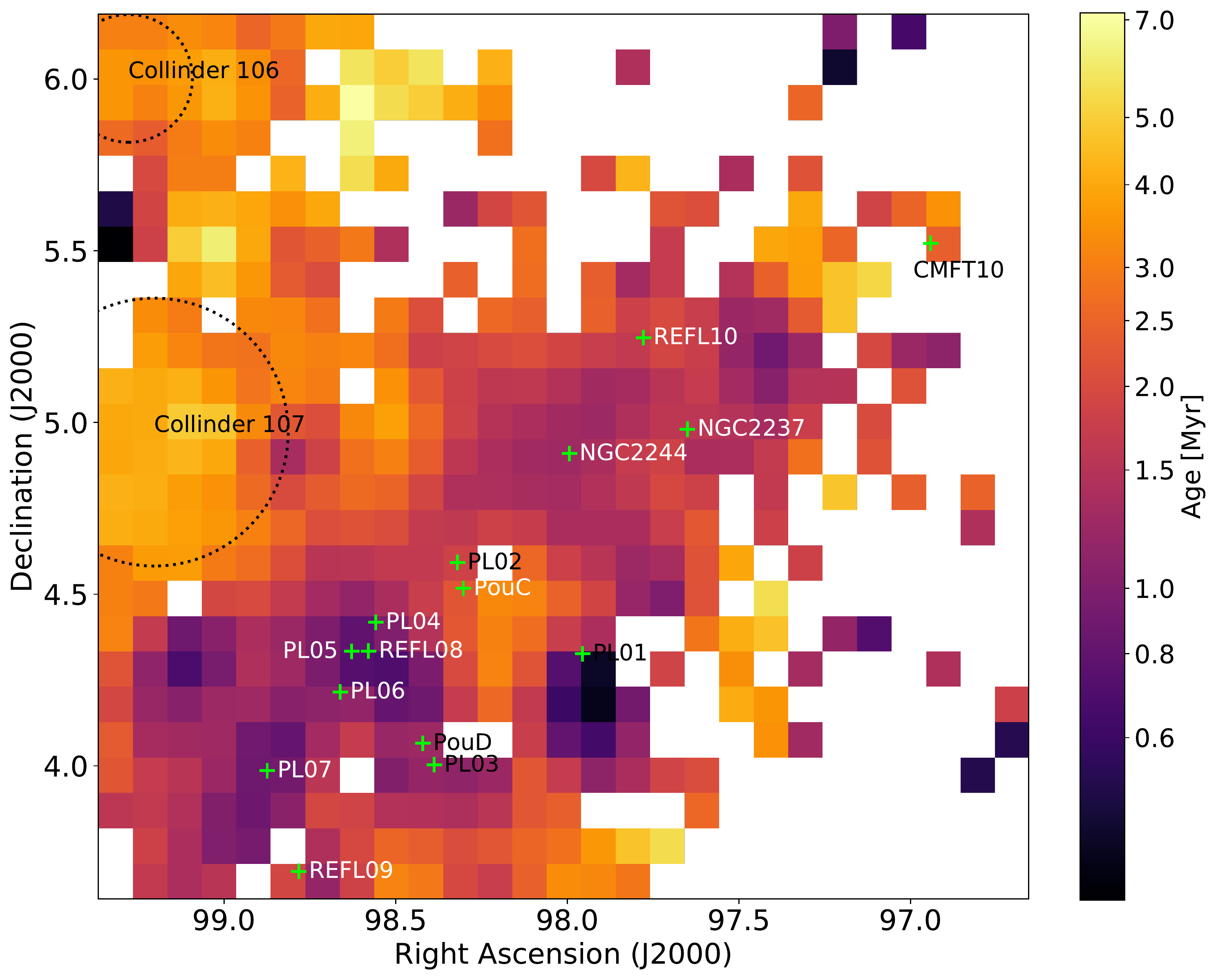}
      \caption{Map of mean ages as a function of the position on the sky, with he bin size of $6'\times6'$. To facilitate a comparison with other figures in the paper, we mark various known structures, identical to those in Fig.~\ref{fig:on_sky}}.
         \label{fig:ages_onsky}
\end{figure}

Using the $T_{\mathrm{eff}}$ and $A_{\mathrm{V}}$ obtained in Section~\ref{sec:seds}, and a distance of 1500 pc, we can construct a HR diagram of the region (Fig.~\ref{fig:hrd}), for the objects with membership probabilities $\geq 80\%$ (run F1). We overplot the PARSEC evolutionary tracks \citep{bressan12,pastorelli2020} for the ages between 0.1 and 30 Myrs. 
The majority of the selected objects is consistent with ages below 10 Myrs, speaking in favour of the robustness of our selection method. The hotter stars appear, on average, older than the cooler stars, an issue that has been repeatedly reported in the past (e.g. \citealt{pecaut12,prisinzano19}). Thus, stars with $T_{\mathrm{eff}} > 6000 K$ are excluded from the calculation of the mean age of the region.

To derive the age and the mass of the objects shown in the HR diagram, we use interpolation on the PARSEC isochrones with ages between 0.1 Myr and 100 Myr, with a step of 0.05 dex. To assess the uncertainties of the derived parameters, we perform a MC simulation by varying the $T_{\mathrm{eff}}$ and $M_{\mathrm{J}}$ within their respective uncertainties, assumed to be 100K for the $T_{\mathrm{eff}}$ (SED fitting step). The uncertainties for $M_{\mathrm{J}}$ include the photometric errors, as well as the uncertainty of 0.25 mag in $A_{\mathrm{V}}$, added in quadrature. The distance uncertainty is not included in this calculation. The derived mass and age distributions for each object typically do not follow a normal distribution, and may be highly asymmetric. In case of the mass, we save the derived distributions (see an example in the Appendix  Fig.~\ref{fig:mass_distr_ind}), which are later used to draw random samples for the derivation of the Initial Mass Function (IMF). In case of the age, we save the medians of 100 MC values for each object.

\subsection{Ages of various structures in the region}

As is commonly seen in HR diagrams of star-forming regions, the objects in the Rosette Nebula show a significant span in ages, from below 0.5 Myr to about 10 Myr. 
To search for potential structure in the age distribution of the studied region, in Fig.~\ref{fig:ages_onsky} we plot the ages of
all the probable members with $T_{\mathrm{eff}} < 6000 K$ in the form of 
a two-dimensional, binned and Gaussian-kernel-smoothed on-sky map, colour-coded according to the mean age in each bin, and with the bin size of $6'\times6'$. 

The mean age of the entire region is $1.6\pm 0.5$ Myr, derived as the average, and the standard deviations of the ages of all the probable members with $T_{\mathrm{eff}} < 6000 K$. Here, it is important to keep in mind that the quoted uncertainty is statistical, and does not include potential systematics related to the use of a particular set of models, nor the uncertainties inherent to models at these young ages. 
The mean age of \cluster~(within the radius of 18$'$) is $1.3\pm 0.4$\,Myr, which is close to the typically assumed age of the region of $\sim2$\,Myr \citep{hensberge00,park02,dias02,bell13,wareing18}, although slightly younger, and at odds with estimates by \citet[][$\sim 5\,$Myr]{kharchenko13}, and \citet[][$\sim 12\,$Myr]{cantatgaudin20b}. As discussed in Appendix~\ref{sec:CG20}, a comparison of our selection with that used in the latter work shows that their member list may suffer a significant degree of contamination, which in turn can make the cluster appear older than it really is.

From Fig.~\ref{fig:ages_onsky}, it appears that the region to the south-east of \cluster~contains on average the youngest stars in the entire Rosette Nebula. For example, the region centred at REFL08 with a radius of 10$'$ has the mean age of $1.0\pm 0.4$\,Myr.
This is in agreement to findings of \citet{poulton08} and \citet{cambresy13}, who report a lower concentration of Class I sources associated with the cluster \cluster~than in the regions associated with the molecular cloud (south-east of \cluster), which should therefore be younger. 

The mean ages of our high probability candidates potentially associated with the clusters Collinder~106 and 107, calculated for the sources within the dotted circles in Fig~\ref{fig:ages_onsky} are $3.3\pm 0.7$\,Myr and $4.0\pm 0.9$\,Myr, respectively. However, the derived ages should not be taken as representative for these two clusters, because our algorithm picks up only a fraction of their reddest members, which can be clearly seen when compared to the selection of \citet{cantatgaudin20b}, who report ages of 18\,Myr and 24\,Myr for Collinder~107 and 106, respectively.
According to \citet{kharchenko13}, Collinder~106 has an age of $\sim$3-5 Myr, while the Collinder~107 appears slightly older ($\sim$15 Myr). On the other hand, \citet{dias02} report an age of 10 Myr for Collinder~107, and an old age inconsistent with the open cluster nature for Collinder~106. 

\section{Properties of the cluster \cluster}
\label{sec:ngc2244_properties}

In Sections~\ref{sec:radial_distr} and \ref{sec:expansion}, we determined the radial extent of \cluster, and studied its internal kinematics, indicating expansion. In Section~\ref{sec:massage}, we derived masses and ages for all the probable members across the Rosette Nebula complex. This section aims at further characterisation of \cluster, by the derivation of its mass function, assessment of mass segregation, and finally discussion on its dynamical state, including comparison to numerical simulations.

\subsection{Initial mass function}
\label{sec:imf}
\begin{figure}[htb]
    \center
    \begin{subfigure}{}
        \center
        \includegraphics[width=0.9\linewidth]{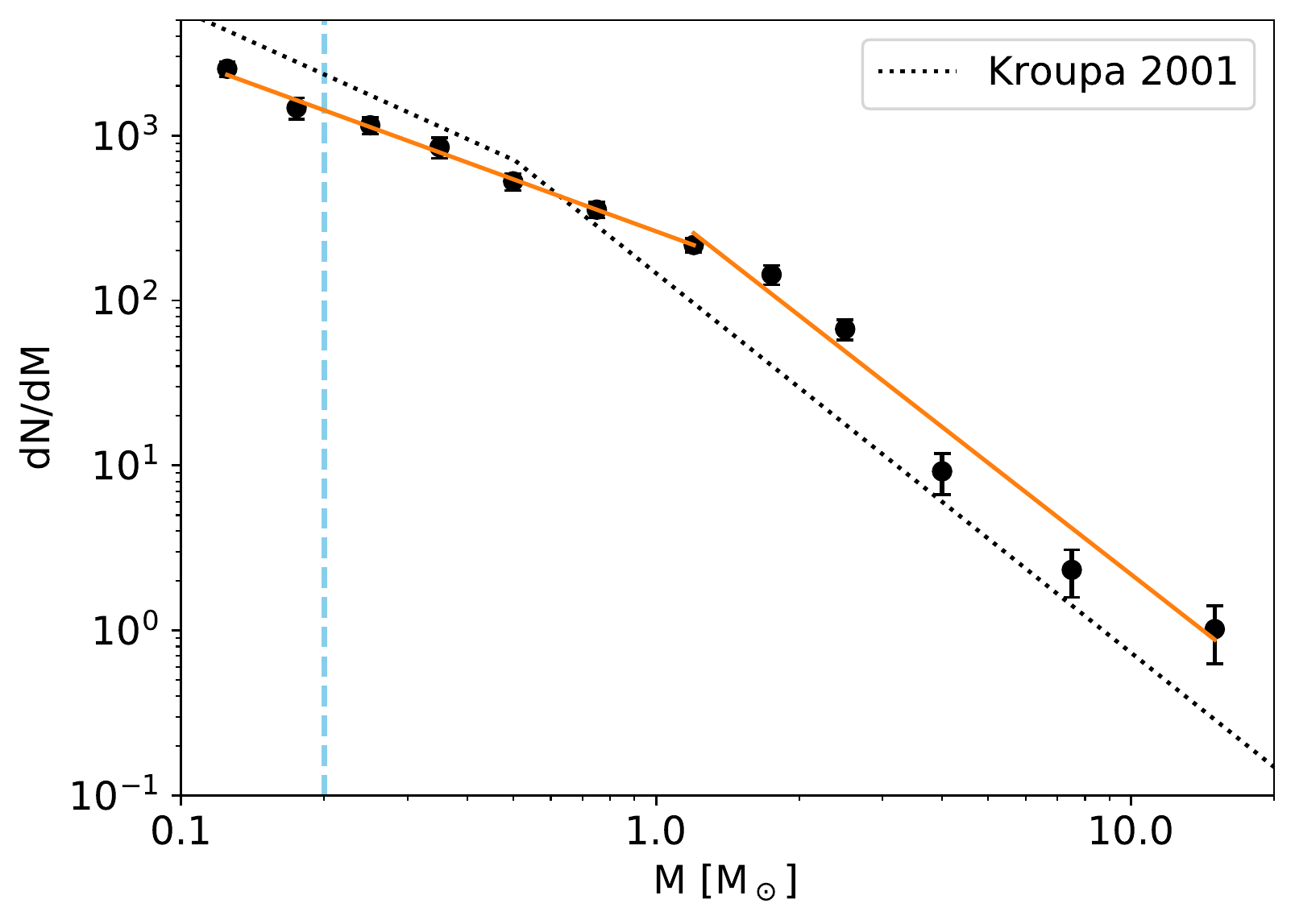}%
        \hfill
        \includegraphics[width=0.9\linewidth]{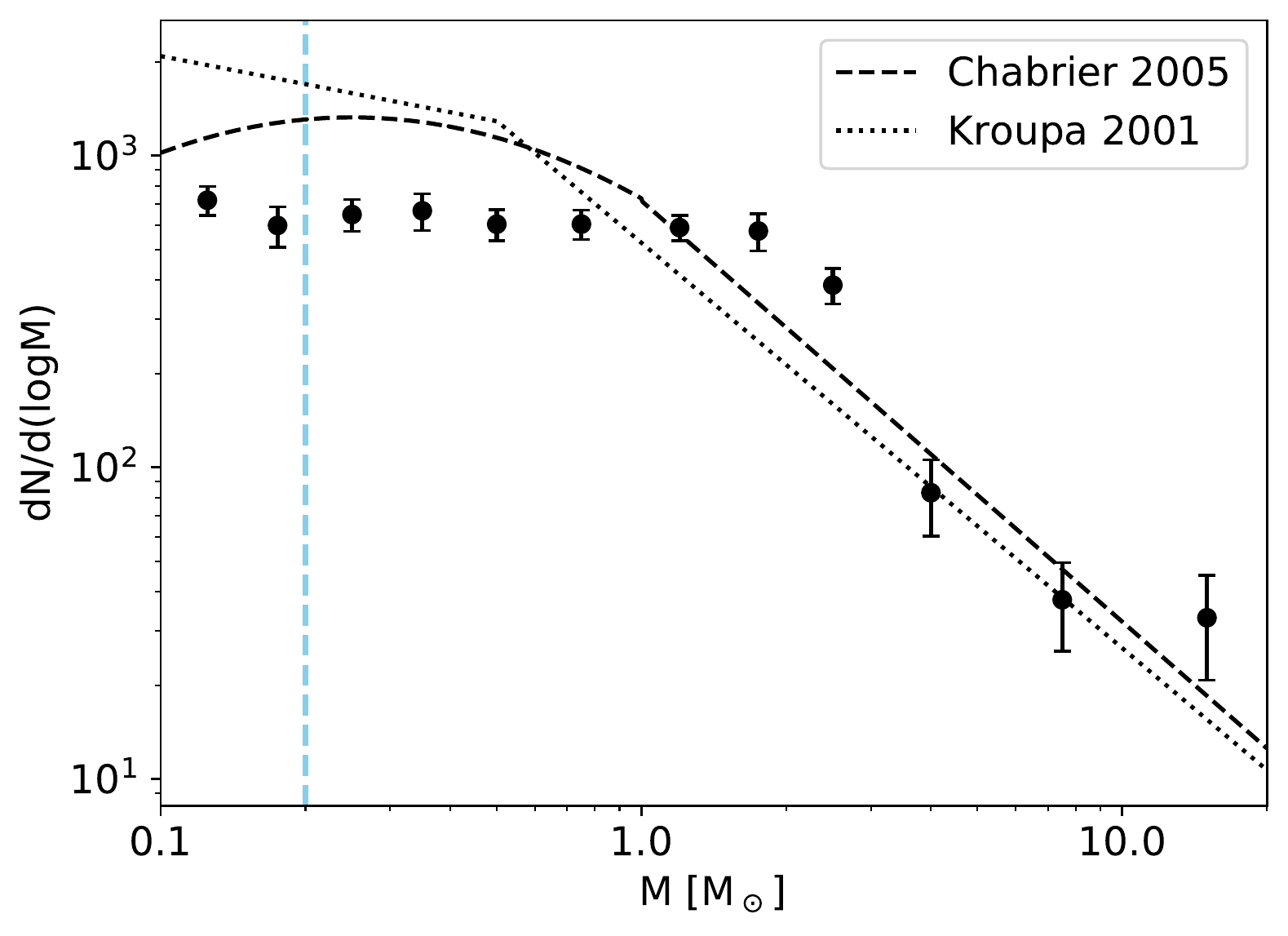}  
        \caption{Initial mass function of the cluster \cluster, in two different representations: $dN/dM$ (top panel) and $dN/d(logM)$ (bottom panel). The solid orange lines in the top
panel show a power law fit with a break $\sim 1.5$M$_\odot$, with slopes $\alpha = 1.05\pm 0.02$ at the low-mass side, and $\alpha = 2.26 \pm 0.25$ for the massive part. Two standard mass functions are shown as dotted \citep{kroupa01}, and dashed \citep{chabrier05} lines, normalised to match the total number of objects that entered the IMF derivation. The vertical blue dashed line marks the completeness limit.}
        \label{fig:imf}
    \end{subfigure}
\end{figure}

In this section we derive the IMF of \cluster~(within a circle of $18'$ from the cluster centre), for
masses in the range from 0.2 M$_\odot$ to 20 M$_\odot$.
NGC\,2237 members have been removed as described in Section~\ref{sec:distance}.
The mass of each star is represented by a distribution derived in Section~\ref{sec:massage}. Each of these distributions is smoothed by a Gaussian KDE (using the Silverman rule to determine the kernel), and used to extract a random sample of N$_1$ values per object, constructing thus N$_1$ realisations of the mass distributions for the entire sample. For each of these realisations, we perform N$_2$ bootstraps (random samplings with replacement), resulting in N$_1\times$N$_2$ mass distributions. Each distribution is then binned onto the same grid, and the final value and its uncertainty for each bin is calculated as mean and standard deviation of N$_1\times$N$_2$ values. The bin sizes have been selected as to contain a high number of elements (typically 70-120), except for the last three high mass bins which are more sparsely populated (10 - 20 stars). We set N$_1$=N$_2$=100. 

Fig.~\ref{fig:imf} shows the derived IMF in two commonly used forms, $dN/dM$ (upper panel), and $dN/d(logM)$ (lower panel), along with the standard IMFs from \citet{kroupa01} and \citet{chabrier05}.
We fitted the IMF shown in the upper panel with two power-laws in the form $dN/dM \propto M^{-\alpha}$, with
a break at $\sim$1.5\,M$_{\sun}$ (solid orange lines), and obtain the slopes 
 $\alpha = 1.05\pm 0.02$, and $\alpha =2.26 \pm 0.25$ for the low-mass and the massive part, respectively. 
 The high-mass slope is very close to the standard Salpeter slope ($\alpha = 2.35$; \citealt{salpeter55}), and in agreement with the slope $\alpha \sim 2.1$ derived for X-ray-selected members with masses $> 0.5$M$_\odot$ \citep{wang08}, and $\alpha = 2.4 \pm 0.1$ derived by \citet{lim21} for the masses above 1M$_\odot$. 
At the low-mass side, we can compare the slope $\alpha$ with the results in nearby star-forming regions and other young clusters that have been derived below 1\,M$_\odot$: NGC\,1333 (0.9-1) and IC\,348 (0.7-0.8; \citealt{scholz13}), Chamaeleon-I and Lupus 3 ($0.8 \pm 0.1$; \citealt{muzic15}), RCW 38 ($0.8\pm0.1$; \citealt{muzic17}). The slope derived here for \cluster~is in agreement with these different regions, especially considering that all these results include systematics that are difficult to take into account, such as the age uncertainties, choice of models and the extinction law, or mass range to calculate the fit.
Finally, both slopes of the IMF are in agreement with our earlier results which focussed only on a narrow region close to the centre of \cluster~\citep{muzic19}.

In the $dN/d(logM)$ representation, the IMF of \cluster~is fairly flat over the mass range 0.2 - 1.5\,M$_{\sun}$, which has already be noted in \citet{muzic19}, and also reported in \citet{damian21}, over the mass range 0.1 - 1\,M$_{\sun}$. A similar behaviour has also been observed in the
cluster IC\,4665 over the range 0.1 - 1\,M$_{\odot}$ \citep{miret19}, and NGC\,2362 for the masses $\sim 0.06 -1$\,M$_{\odot}$ \citep{damian21}.


\subsection{Mass segregation}
\label{sec:mass_seg}
\begin{figure*}[htb]
    \center
    \begin{subfigure}{}
        \center
   \includegraphics[width=0.4\textwidth]{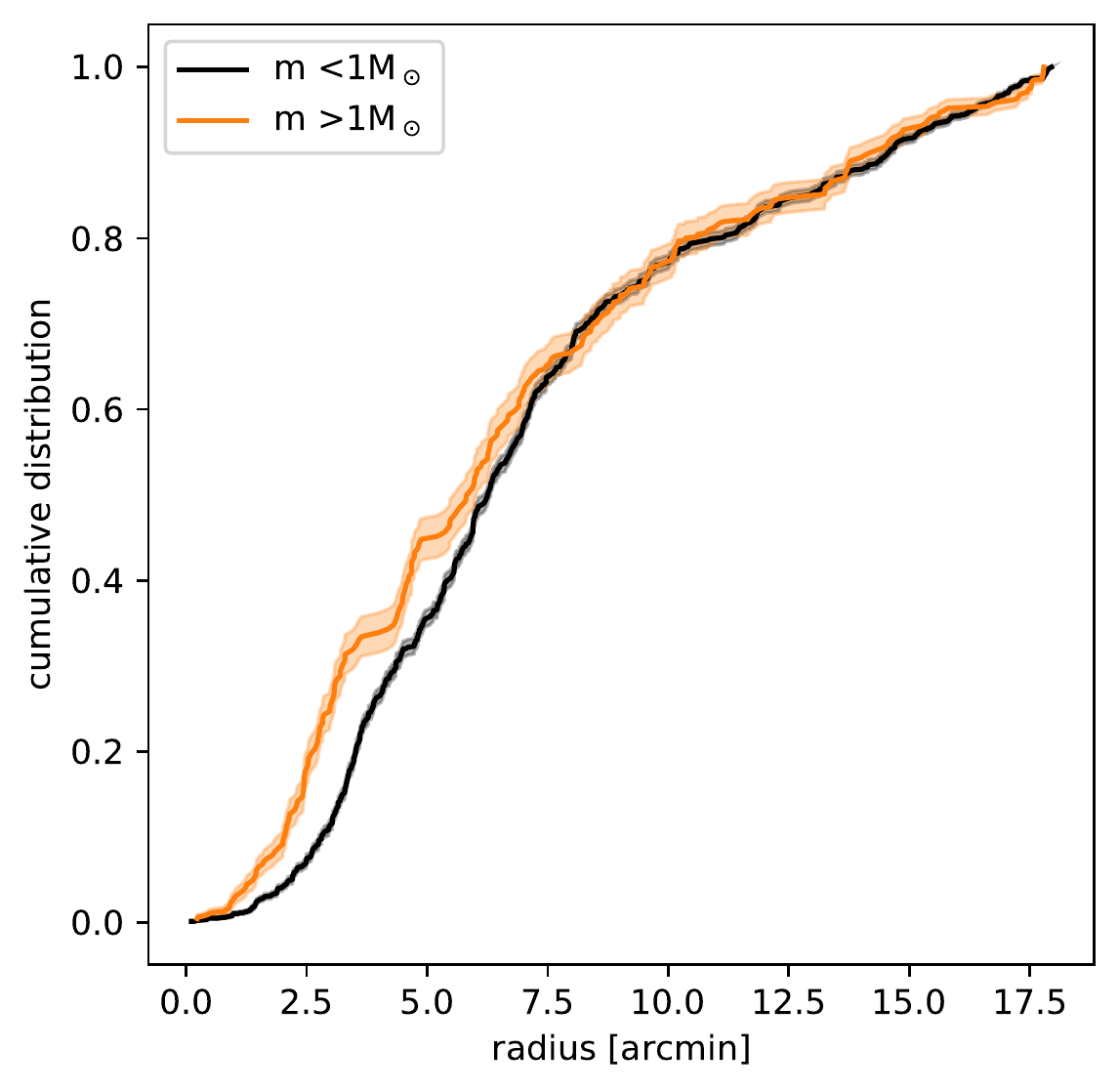}
   \hfill
      \includegraphics[width=0.5\textwidth]{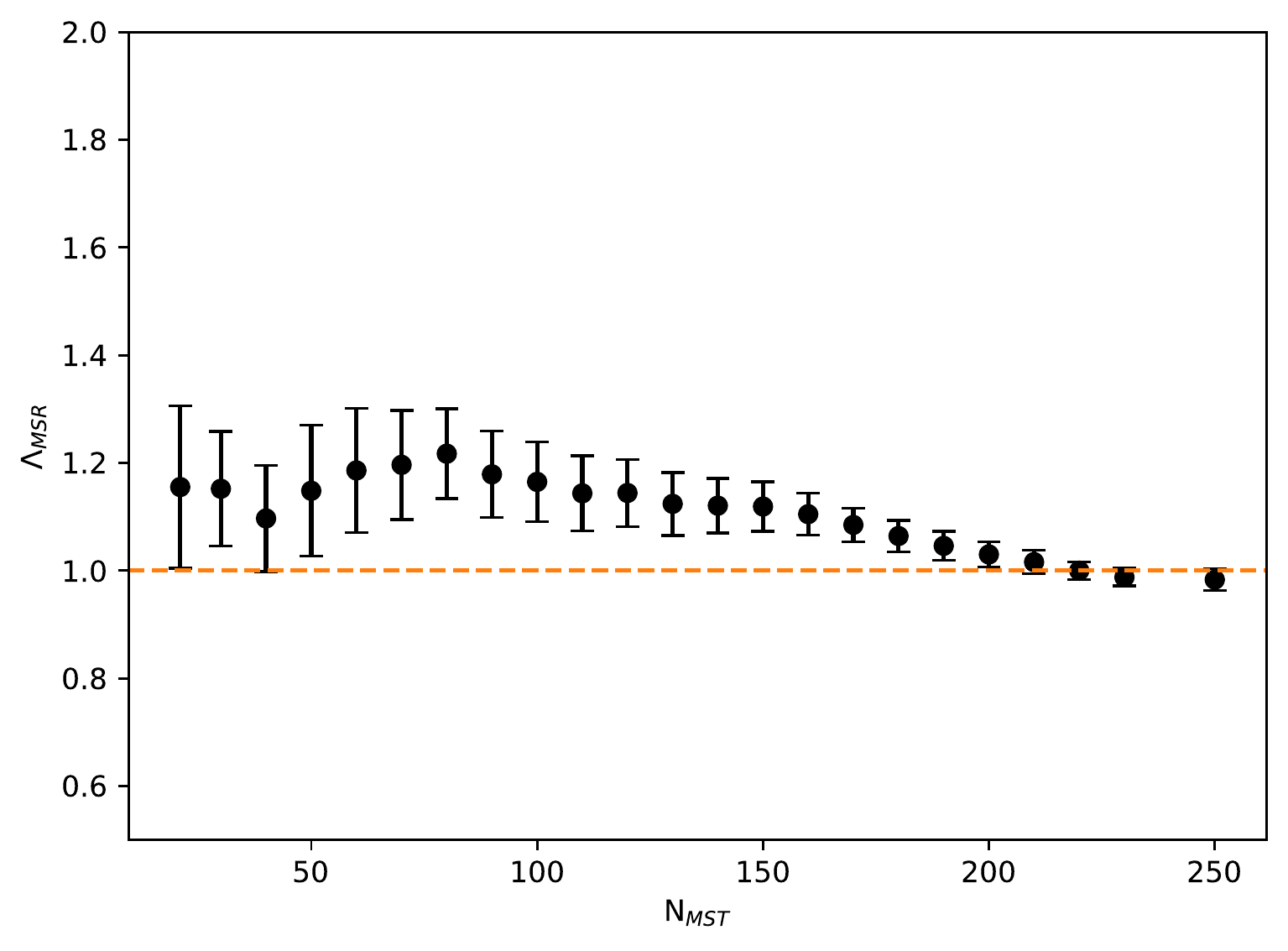}
      \caption{
      Assessment of mass segregation in \cluster.
      \textbf{Left:} Cumulative distributions of mass as a function of the projected distance from the \cluster~centre. The masses are divided at 1\,M$_\odot$. The confidence intervals correspond to 1$\sigma$. \textbf{Right:} Mass segregation ratio, $\Lambda_{MSR}$ as a function of the N$_{MST}$ stars. The horizontal dashed line indicates $\Lambda_{MSR}$ = 1, corresponding to no mass segregation. 
     }
         \label{fig:mass_seg}
 \end{subfigure}
\end{figure*}

To inspect potential mass segregation in \cluster, we perform an MC analysis similar to that in Section~\ref{sec:imf}, sampling the masses from individual mass distributions. At each step, a cumulative distribution of mass as a function of the projected distance from the cluster centre is saved, and finally a mean value and standard deviations are calculated for each value of projected distance. Fig.~\ref{fig:mass_seg} (left panel) contains the results of this procedure for two mass bins: stars above 1\,M$_\odot$ are represented by the orange line, and those below by the black one. We see clearly that the more massive stars are preferentially found closer to the cluster centre, indicating mass segregation. The Anderson-Darling two-sample test \citep{andersondarling} returns p-values lower than 0.001 for the means of the two distributions, suggesting that they cannot be drawn from the same parent population.

Another way to probe the mass segregation in a cluster is by using the $\Lambda_{MSR}$ parameter \citep{allison09, maschberger11}. It is determined with the help of the minimum spanning tree (MST), a graph constructed by drawing lines between pairs of points in a distribution in a way to connect all the points with the shortest possible total path length and no closed loops. 
$\Lambda_{MSR}$ is obtained by dividing the median edge length of the MST of randomly chosen N$_{MST}$ stars, with the median edge length of N$_{MST}$ most massive stars. A region without any mass segregation would have $\Lambda_{MSR} \sim 1$, as opposed to $\Lambda_{MSR} > 1$ which indicates mass segregation, or $\Lambda_{MSR} < 1$, signalling an inverse effect with respect to mass. We run an MC simulation similar to that from Section~\ref{sec:imf}, including resampling and bootstrapping to obtain 10$^4$ mass distributions. For each mass distribution, we vary N$_{MST}$ between 20 and 250 (the latter roughly corresponds to 1\,M$_\odot$), which allows us to calculate the mean $\Lambda_{MSR}$ and the corresponding standard deviation. The MSTs were constructed using the Python package MISTree \citep{naidoo19}. 
In the right panel of Fig.~\ref{fig:mass_seg} we see that  $\Lambda_{MSR}$ has values consistent with, or just slightly higher than unity. Since it has been shown that even some simulated random distributions of stars can yield $|\Lambda_{MSR}|$>1 \citep{parker&goodwin15}, values of $|\Lambda_{MSR}|$>1.5 - 2 are typically required in the literature to unambiguously detect mass segregation \citep[e.g.][]{dib18,parker18}.
At face value, this would  indicate that no significant mass segregation is present in \cluster, contrary to what is seen in the left panel of Fig.~\ref{fig:mass_seg}.
The reason for this may lie in the fact that, even if the more massive stars are centrally concentrated, so are the other stars in our cluster, and therefore a random choice of stars will favour short branch distributions.
It has in fact been demonstrated by \citet{parker&goodwin15} that when the location of the most massive stars coincides with the regions of the highest surface density, $\Lambda_{MSR}$ tends to only marginally signal mass segregation. Furthermore, although the massive stars may be more centrally concentrated, some of them are still present in the outer parts of the cluster, leading to longer branches, increasing the average value for the massive stars, and thus pushing the ratio closer to unity. 

Mass segregation in \cluster~has previously been reported by \citet{chen07}, who observe a similar effect separating the stars according to their apparent magnitude. They divide the sample at m$_B$=13\,mag, which roughly corresponds to 2-3M\,$_\odot$ at the distance of 1500\,pc and the age of 2\,Myr (and additionally depends on the extinction). At the same time, they do not observe any velocity - mass correlation, speaking in favour of primordial mass segregation. On the other hand, \citet{wang08}, report no evidence of mass segregation, by comparing the cumulative distributions of stars with estimated masses $<2$M$_\odot$ with those with masses $>8$M$_\odot$. Clearly, the three works use different mass intervals to determine whether stellar masses may be segregated. Changing the limit mass to 2 or 3 M$_\odot$ as in \citet{chen07}, we still observe the same behaviour in the cumulative mass distributions. However, the comparison with \citet{wang08} is more difficult as the low number of high mass stars, along with the uncertainties in their mass determination result in very wide confidence intervals.  

To summarise, the two indicators of mass segregation applied in this work return somewhat contradictory results. The same is found by comparing different sources in the literature. We conclude that, while there may be some degree of mass segregation in \cluster, it does not seem to be very pronounced.

\subsection{Dynamical state of \cluster}
\label{sec:dynamics}

\subsubsection{Velocity dispersion}
\label{sec:veldisp}

One-dimensional velocity dispersion $\sigma_{1D}$ can be obtained from 
\begin{equation}
    \sigma_{1D}^2=\frac{\sigma_{\mu\alpha}^2 + \sigma_{\mu\delta}^2}{2},
\end{equation}
where $\sigma_{\mu\alpha}$ and $\sigma_{\mu\delta}$ mark the velocity dispersion in right ascension and declination, respectively. The latter can be calculated as weighted standard deviations of the measured proper motions. In order to obtain a robust measurement of the velocity dispersion, the proper motions are resampled $10^4$ times, by taking into account the proper motion uncertainties and their correlations, as well as bootstrapping.  
For \cluster~($r<18'$) we obtain $\sigma_{1D}$ = $0.201\pm 0.016$\,mas\,yr$^{-1}$, equivalent to $1.43 \pm 0.11$\,km\,s$^{-1}$. This estimate is in between two previous determinations of one-dimensional velocity dispersion from the $Gaia$ DR2 data, $\sigma_{1D} = 1.94 \pm 0.10$\,km\,s$^{-1}$ \citep{kuhn19}, and $\sigma_{1D} = 1.1 \pm 0.9$\,km\,s$^{-1}$ (obtained from proper motions) or $\sigma_{1D} = 1.4 \pm 1.2$\,km\,s$^{-1}$ (with radial velocity included) from \citep{lim21}. In both cases, the cited results have been re-scaled to account for the difference in assumed distances. To facilitate the comparison with the numerical simulations, which will be presented in Section~\ref{sec:sim_comp}, we also calculate the inter-quartile range of the velocity dispersion ($\sigma_{IQR}$; equation 8 in \citealt{parker14}). The obtained value is almost identical to the value of $\sigma_{1D}$ (see Table~\ref{tab:parameters}). 

The total stellar mass of \cluster, estimated from an MC simulation equivalent to the one used to derive the IMF, is $1000\pm70$ M$_\odot$\footnote{The mass estimated here is a lower limit on the total mass of the cluster due to incompleteness at the low-mass end, and the limited extent of the sample (r$<18'$), however, it is not expected to be much larger. For example, a total contribution of missing brown dwarfs is expected to be at most a few percent, taking into account their frequencies relative to stars in the center of the cluster \citep{muzic19}}, and the half-mass radius r$_h$=$7.9 \pm 0.5'$, equivalent to $3.4 \pm 0.2$\,pc at a distance of 1500\,pc.
 The velocity dispersion $\sigma_{vir}$ of a virialised cluster with the mass M and half-mass radius r$_h$ can be obtained from
\begin{equation}
    \sigma_{vir}^2 = \frac{M\cdot G}{\eta r_h},
\end{equation}
where $\eta$ is a constant that depends on the radial profile of the cluster. For $\eta=10$, corresponding to a Plummer model, this yields $\sigma_{vir} = 0.36 \pm 0.02 $\,kms$^{-1}$. Taking $\eta=9$, which may be more appropriate for a cluster with $\gamma=3$  \citep{portegieszwart10}, we obtain $\sigma_{vir} = 0.37 \pm 0.02 $\,kms$^{-1}$. Even for very small values ($\eta$=5), the virial velocity dispersion remains significantly smaller than $\sigma_{1D}$. 
Placing it in the context of other young clusters in Figure 17 of \citet{kuhn19}, which shows the dependence of the virial velocity dispersion vs. the observed one for several Milky Way's young clusters, \cluster~appears well above the  line indicating zero total energy, suggesting that it may be unbound. The pattern of expansion seen within the half-mass radius (Fig.~\ref{fig:pm_radius}), in principle supports this conclusion. However, numerical simulations following dynamical evolution of young star-forming regions by \citet{parker16}, show that the measured velocity
dispersion may not be a good diagnostic of the current virial state of these systems, as most clusters in their simulations appear supervirial after a short time, independently of the initial conditions. On the other hand, regions that are significantly supervirial from the start display a velocity dispersion that is well in excess of the virial one ($\sigma_{1D}$/$\sigma_{vir} \sim 4$ in their simulations) at very early times ($<$2\,Myr). This may be the case of \cluster, and will be further discussed below. The number of sources in the simulation (1500) by \citet{parker16} is comparable to \cluster~($\sim1000$). The current density of \cluster~is significantly lower than that of the simulated clusters, which have a radius of only 1\,pc. However, given that the cluster crossing time seems to be comparable to its age, it may as well have expanded to its current size.

\subsubsection{Relaxation timescale}
\label{sec:reltime}
A relaxation timescale of a cluster due to two-body interactions can be estimated as
\begin{equation}
    t_{relax} \approx \frac{0.1 N}{ln N}\times t_{cross},
\end{equation}
where N is the total number of stars in the cluster, and t$_{cross} \approx$ r$_{h}$/$\sigma_{1D}$ is the cluster crossing time \citep{binneytremaine}. Using the half-mass radius and one-dimensional velocity dispersion calculated above, along with N=1\,055 (high probability members within the 18$'$ radius), we obtain t$_{cross} \approx 2.3 \pm 0.2$\,Myr, and t$_{relax} \approx 35 \pm 3\,$Myr. The relaxation time is an order of magnitude longer than the estimated age of the cluster, signalling that the potential mass segregation (see Section~\ref{sec:mass_seg}) cannot be a consequence of dynamical relaxation.

\subsubsection{Tidal stability}
\label{sec:tidal}
A static tidal field enforces an upper limit on the size of a stellar system, by stripping the stars outside of its Jacobi (or Hill) radius \citep{binneytremaine}. If a cluster is to survive the Milky Way's tidal field, it should be smaller than its Jacobi radius:

\begin{equation}
    r_J=\Big[\frac{GM}{4A(A-B)}\Big]^{1/3}, 
\end{equation}
where M is the total mass of the system, and A and B are the Oort's constants at its Galactic position. Using the relations given in \citet{piskunov07}, we obtain A$\approx$12\,km\,s$^{-1}$\,kpc$^{-1}$ and B$\approx$-10.5\,km\,s$^{-1}$\,kpc$^{-1}$. For the total mass calculated above, we obtain the Jacobi radius r$_J\approx$15.9\,pc, which is approximately twice the outer cluster radius determined in Section~\ref{sec:radial_distr}, indicating that the cluster is not significantly influenced by the Galactic potential.

\subsubsection{Structural analysis}
\label{sec:sim_comp}

\begin{table*}
\centering
\caption{Summary of various kinematic and structural parameters of \cluster. }
\label{tab:parameters}
\begin{tabular}{llcc}\hline\hline
Parameter   & Value   & Value at 1500 pc & Section\\
\hline
$\sigma_{1D}$ &  0.201$\pm$0.016\,mas\,yr$^{-1}$  &  1.43$\pm$ 0.11\,km\,s$^{-1}$ & \ref{sec:veldisp}\\
$\sigma_{IQR,1D}$ &  0.198$\pm$0.045\,mas\,yr$^{-1}$  &  1.41$\pm$ 0.32\,km\,s$^{-1}$ & \ref{sec:veldisp} \\
$\sigma_{vir}$ &  0.050$\pm$0.003\,mas\,yr$^{-1}$  &  0.36$\pm$ 0.02\,km\,s$^{-1}$ & \ref{sec:veldisp}\\
M$_{tot}$ & 1000$\pm$70 M$_\odot$ & &  \ref{sec:veldisp}\\
Age entire region &  $1.6\pm0.5$\,Myr& & \ref{sec:massage} \\
Age \cluster &  $1.3\pm0.4$\,Myr& & \ref{sec:massage} \\
r$_{c}$ & $4.6\pm0.8'$ & $2.00\pm 0.35$\,pc & \ref{sec:radial_distr} \\
r$_h$ & 7.9$\pm0.5'$ & $3.4\pm0.2$\,pc & \ref{sec:veldisp} \\
r$_J$ & $\approx36'$ & $\approx15.9 $\,pc & \ref{sec:tidal}\\
t$_{cross}$ & $2.3\pm0.2$\,Myr & & \ref{sec:reltime}\\
t$_{relax}$ & $35\pm3$\,Myr & & \ref{sec:reltime}\\
$\overline{m}$ & 0.57 & & \ref{sec:sim_comp}\\
$\overline{s}$ & 0.66 &  &  \ref{sec:sim_comp}\\
Q & 0.86 & &  \ref{sec:sim_comp}\\
$\Sigma_{0}$ &  $6.4\pm0.7$\,stars\,arcmin$^{-2}$ & $34 \pm 4$\,stars\,pc$^{-2}$ &  \ref{sec:radial_distr}\\
$\Sigma_{mean}$ & $\sim$2.2\, stars\,arcmin$^{-2}$ & $\sim$12 stars\,pc$^{-2}$  &  \ref{sec:radial_distr}\\
$\Sigma_{LDR,10}$ & $1.06\pm 0.40$ & &\ref{sec:sim_comp} \\
$\sigma_{VDR, 10}$ & $0.50\pm0.18$ & &\ref{sec:sim_comp} \\
$\sigma_{VDR, 50}$ & $0.85\pm0.12$ & &  \ref{sec:sim_comp}\\
\hline
\end{tabular}
\end{table*}
 
In Section~\ref{sec:veldisp}, we compared the velocity dispersion in \cluster~to the simulations by \citet{parker16}, and in Section~\ref{sec:mass_seg} we derived the mass segregation ratio $\Lambda_{MSR}$. In this section, we derive several additional structural parameters that can be used for a comparison to the set of simulations presented in various papers of the same group \citep{parker14, parker16, wright&parker19}, and thus try to set some constraints on the formation and early evolution of \cluster. For this analysis, we limit the cluster extent to 2 half-mass radii ($\sim 15'$), in order to avoid including the members of NGC\,2237.

\begin{itemize}

\item Q-parameter. First introduced by \citet{cartwright04}, the scale-free Q-parameter distinguishes between a smooth large-scale radial density gradient and multiscale fractal subclustering.   
Q-parameter is determined as ratio of the normalised mean length of the MST ($\overline{m}$) and the normalised mean edge length of the complete graph ($\overline{s}$). For \cluster, we obtain $\overline{m}$=0.57, $\overline{s}$=0.66, and Q=0.86. This rather moderate value of Q indicates neither a substructured nor a very centrally concentrated distribution. It may correspond to a dynamically young structure (Q roughly constant from the beginning), or may have started off with a smaller value of Q, followed by an episode of violent relaxation \citep{parker14,parker21}. However, as will be shown below, the latter can probably be excluded as it presumes low to moderate values of the virial ratio.   

\item 
The local density ratio $\Sigma_{LDR}$  is the ratio between
the median surface density of the N most massive stars, and the median surface
density of all the stars in the region \citep{parker14,kuepper11}. The surface density of each star is defined as $\Sigma$ = (N-1)/($\pi$ r$_N^2$), where
r$_N$ is the projected separation from the Nth closest neighbour. The value of $\Sigma$ is fairly insensitive to the choice of N; we set N=10. We obtain $\Sigma_{LDR}$ =  $1.06\pm 0.40$. 
This result indicates that no mass segregation is present, which is consistent with the result using $\Lambda_{MSR}$ (Section~\ref{sec:mass_seg}).

\item Velocity dispersion ratio $\sigma_{VDR}$ is defined as the ratio between the velocity dispersion of the N most massive stars and the velocity dispersion of all stars in the sample \citep{wright&parker19}. One-dimensional velocity dispersion have been determined as described in Section~\ref{sec:veldisp}. We measure $\sigma_{VDR}$ for values of N between 10 and 50, in steps of 5. We observe that $\sigma_{VDR}$ generally increases with N, but the values are always maintained below 1. For N=10, the value of the ratio is $\sigma_{VDR}$=$0.50\pm0.18$, whereas for N=50 we obtain $\sigma_{VDR}$=$0.85\pm0.12$.
\end{itemize}

We can now try to compare the above obtained parameters with the N-body simulations of star clusters from  \citet{parker14, parker16, wright&parker19}. In the series of papers, the authors use the same simulations of clusters with 1500 members distributed within a sphere with an initial radius of 1\,pc, starting with several values of the initial virial ratio (subvirial, virial, supervirial), and the degree of substructure (low, moderate, high). We can compare the calculated parameters for \cluster, $\sim2\,$Myr old cluster, with predictions of numerical simulations. 
\begin{itemize}
    \item Evolution of Q with time. Subvirial initial conditions are clearly excluded as they result in much larger values of Q at 2\,Myr. Both virial and supervirial initial conditions may be compatible with the derived value of Q, with the real value of the virial ratio lying somewhere between the extremes used in the simulations. For the virial initial condition, there is a preference towards higher fractal dimensions, i.e. no presence of significant substructure.   
    \item Evolution of $\Lambda_{MSR}$ with time. Similar to the previous item, the subvirial initial conditions are clearly excluded, while the supervirial ones are preferred, independently of the degree of substructure. 
    \item Evolution of $\Sigma_{LDR}$ with time. Both a high degree of substructure, and subvirial initial conditions can be excluded. The calculated values are best represented by the case of high virial ratios and high fractal dimensions (no substructure).
    \item Q - $\Lambda_{MSR}$ and Q - $\Sigma_{LDR}$. As expected from the behaviour of the three mentioned parameters with time, the two relations show preference towards high virial ratios, together with moderate-to-high fractal dimensions. 
    \item Q - $\sigma_{IQR}$ and $\Sigma_{LDR}$ - $\sigma_{IQR}$. Comparison in the planes containing velocity dispersion is slightly complicated by the fact that the initial velocity dispersion in simulations increases with the total number of stars in the simulation \citep{parker16}. However, the value of $\sigma_{IQR}$ obtained for \cluster~is generally higher than that of the simulations (but still consistent within the errors), although the number of stars in the cluster is $\sim 30\%$ lower than that in the simulations. Thus, we can basically make similar observations as when looking at the evolution of Q and $\Sigma_{LDR}$ with time, and conclude that the best match seems to be provided by the supervirial case, with large fractal dimension.
    \item $\sigma_{VDR}$. Independently of the value of N, we obtain $\sigma_{VDR} <1$, i.e. the most massive stars move on average slower than the low-mass stars. This excludes a region that was initially highly substructured, independently of the virial state.     
\end{itemize}

In conclusion, the numerical simulations by Parker et al. indicate that \cluster~may have undergone a hot collapse, i.e. formation under supervirial conditions, with a low level of substructure which has not been significantly altered since the formation phase.



\section{Summary and conclusions}
\label{sec:summary}

In this work, we study the $2.8\degr \times 2.6\degr$ region in the Rosette Nebula, centred in the young cluster \cluster.
 Starting from a catalogue containing optical to mid-infrared photometry from multiple sources, as
   well as accurate positions and proper motions from $Gaia$\,EDR3, we applied the PRF algorithm to derive the membership probability for each source within our field of view. Based on the list of almost 3000 probable members, of which about a third are concentrated within the radius of 20$'$ from the centre of \cluster, we identified various clusters and concentrations of young stars. The masses, extinction, and ages were derived by fitting the SED to the atmosphere and evolutionary models, and the internal dynamics was assessed via proper motions relative to the mean proper motion of \cluster. Here we summarise the main results of the paper:
  \begin{itemize}
      \item Our distance estimate to the overall Rosette Nebula stellar population is $1489\pm37$\,pc, and to \cluster~ $1440\pm32$\,pc, derived from $Gaia$\,EDR3 parallaxes.
      While the region of active star formation associated with the Rosette molecular cloud, located to the south-east of \cluster, appears to be located at a similar distance as the cluster, another, smaller cluster in the region, NGC\,2237, seems to be located $\sim$90\,pc behind it.

      \item The region is clearly young, with the mean age derived from the HR diagram (1.6 $\pm$ 0.5\,Myr) being consistent with previous estimates of $\sim 2$\,Myr. The young age is further supported by the position of the Wielen dip in the luminosity function at M$_G\approx5\,$mag. 
      A region to the south-east of \cluster, associated with the molecular cloud, contains -- on average -- the youngest stars in the Rosette Nebula.
      
      \item   The radial profile of \cluster~was fitted with the EFF profile \citep{elson87}, which returns the peak stellar surface density $\Sigma_0$ = $6.4\pm0.7$\,stars\,arcmin$^{-2}$ , and the (King profile) core radius r$_c$ = $2.0 \pm 0.4\,$pc. We estimate that the cluster extends out to a radius of $\sim 18\,$arcmin, which is equivalent to $\sim$8\,pc at a distance of 1500\,pc.
      
      \item   \cluster~is showing a clear expansion pattern, with an expansion velocity that increases with radius, in agreement with previous studies \citep{kuhn19}. The median of the distribution of the radial component of the relative proper motion vector ($\mu_r$) out to r$=18'$ is at $0.138\pm0.015\,$mas\,yr$^{-1}$, which is equivalent to v$_r$ = $1.0 \pm 0.1$\,km\,s$^{-1}$ at a distance of 1500\,pc. 
      
      \item  The IMF is well represented by two power laws ($dN/dM\propto M^{-\alpha}$), with slopes $\alpha = 1.05 \pm 0.02$ for the mass range 0.2 - 1.5\,M$_\odot$, and $\alpha = 2.3 \pm 0.3$ for the mass range 1.5 - 20\,M$_\odot$, and it is in agreement with slopes detected in other star-forming regions. 
      
      \item The velocity dispersion of \cluster~is well above the virial velocity dispersion derived from the total mass ($1000\pm70$\,M$_\odot$) and half-mass radius ($3.4\pm0.2$\,pc). 
      
      \item The relaxation timescale is an order of magnitude larger than the estimated age of \cluster, suggesting that the cluster is still far from being dynamically relaxed.  
  \end{itemize}
  
 The numerical simulations by \citet{parker14}, \citet{parker16}, and \citet{wright&parker19} make predictions about the evolution of various structural and kinematic parameters with time, from which one can try to deduce the initial conditions of cluster formation. This comparison suggests that \cluster~may be unbound and that it may have possibly even formed in a super-virial state. Recently, \citet{bonilla22} have compared numerical simulations of clusters forming in turbulence-dominated environments over multiple free-fall timescales, with those forming by a global hierarchical collapse over a much shorter timescale. They predict that stars in clusters that formed via the former mechanism should exhibit an inverse mass segregated velocity dispersion (massive stars have a larger velocity dispersion), as opposed to a flat distribution of velocity dispersion with mass in the latter scenario. \citet{bonilla22} use
 $Gaia$ EDR3 to show that the stars in the Orion Nebula Cluster exhibit a constant velocity dispersion as a
function of mass, similar to what we found for \cluster. This 
suggests that the two clusters may have been formed by hierarchical collapse within one free-fall time of its parental cloud, rather than in a turbulence-dominated environment.

\begin{acknowledgements}
      We thank the anonymous referee for constructive comments, and Nick Wright for useful discussions on the topic during the Cool Stars 21 conference.
      KM, KK, and VAA acknowledge funding by the Science and Technology Foundation of Portugal (FCT), grants No. IF/00194/2015, PTDC/FIS-AST/28731/2017, UIDB/00099/2020, and SFRH/BD/143433/2019. KPR acknowledges support by ANID FONDECYT Iniciaci\'on 11201161.
      
      This research has received funding from the European Research Council (ERC) under the European Union’s Horizon 2020 research and innovation programme (grant agreement No 682903, P.I. H. Bouy), and from the French State in the framework of the ”Investments for the future” Program, IdEx Bordeaux, reference ANR-10-IDEX-03-02. 
      
      We gratefully acknowledge the support of NVIDIA Corporation with the donation of one of the Titan Xp GPUs used for this research.
    Based on observations made with the INT operated on the island of La Palma by the Isaac Newton Group in the Spanish Observatorio del Roque de los Muchachos of the Instituto de Astrofísica de Canarias. Based on data obtained from the ESO Science Archive Facility and with ESO Telescopes at the La Silla Paranal Observatory under programme ID 0104.C-0369, 080.C-0697, 67.C-0042, 68.C-0048.This paper makes use of data obtained from the Isaac Newton Group Archive which is maintained as part of the CASU Astronomical Data Centre at the Institute of Astronomy, Cambridge.This research is based on data obtained from the Astro Data Archive at NSF’s NOIRLab. NOIRLab is managed by the Association of Universities for Research in Astronomy (AURA) under a cooperative agreement with the National Science Foundation.
  This research used the facilities of the Canadian Astronomy Data Centre operated by the National Research Council of Canada with the support of the Canadian Space Agency.
      Based on observations obtained with WIRCam, a joint project of CFHT, Taiwan, Korea, Canada, France, at the Canada-France-Hawaii Telescope (CFHT) which is operated by the National Research Council (NRC) of Canada, the Institut National des Sciences de l'Univers of the Centre National de la Recherche Scientifique of France, and the University of Hawaii. Based on observations obtained with MegaPrime/MegaCam, a joint project of CFHT and CEA/DAPNIA, at the Canada-France-Hawaii Telescope (CFHT) which is operated by the National Research Council (NRC) of Canada, the Institut National des Science de l'Univers of the Centre National de la Recherche Scientifique (CNRS) of France, and the University of Hawaii. The observations at the Canada-France-Hawaii Telescope were performed with care and respect from the summit of Maunakea which is a significant cultural and historic site.
      This work has made use of data from the European Space Agency (ESA) mission {\it Gaia} (\url{https://www.cosmos.esa.int/gaia}), processed by the {\it Gaia} Data Processing and Analysis Consortium (DPAC, \url{https://www.cosmos.esa.int/web/gaia/dpac/consortium}). Funding for the DPAC has been provided by national institutions, in particular the institutions participating in the {\it Gaia} Multilateral Agreement.  This publication makes use of data products from the Two Micron All Sky Survey, which is a joint project of the University of Massachusetts and the Infrared Processing and Analysis Center/California Institute of Technology, funded by the National Aeronautics and Space Administration and the National Science Foundation.The Pan-STARRS1 Surveys (PS1) have been made possible through contributions of the Institute for Astronomy, the University of Hawaii, the Pan-STARRS Project Office, the Max-Planck Society and its participating institutes, the Max Planck Institute for Astronomy, Heidelberg and the Max Planck Institute for Extraterrestrial Physics, Garching, The Johns Hopkins University, Durham University, the University of Edinburgh, Queen’s University Belfast, the Harvard-Smithsonian Center for Astrophysics, the Las Cumbres Observatory Global Telescope Network Incorporated, the National Central University of Taiwan, the Space Telescope Science Institute, the National Aeronautics and Space Administration under Grant No. NNX08AR22G issued through the Planetary Science Division of the NASA Science Mission Directorate, the National Science Foundation under Grant No. AST-1238877, the University of Maryland, and Eotvos Lorand University (ELTE) and the Los Alamos National Laboratory.

\end{acknowledgements}

%
%

\bibliographystyle{aa} 
\bibliography{rosette}

\begin{appendix}

\section{Derived spectral types and extinction}

\begin{table}
	\caption{Positions, spectral types, extinction and pseudo-EW of H$_\alpha$ in emission measured from VIMOS spectra.$^{*}$ }
	\begin{center}
	\footnotesize
		\begin{tabular}{ccccc}\hline\hline
			 RA & Dec & SpT & A$_V$ & EW \\
			    &     &     & [mag] & [\AA] \\
			\hline
			06:31:18.23 & 04:50:20.8 & K7.0 & 2.6 & -6.17$\pm$ 0.11 \\
			06:31:18.24 & 04:52:23.6 & M0.0 & 0.4 & -5.11$\pm$ 0.18 \\
			06:31:21.43 & 04:54:05.7 & M0.0 & 1.6 & -4.13$\pm$ 0.13 \\
			06:31:22.86 & 04:49:19.8 & M0.0 & 2.6 & -3.45$\pm$ 0.19 \\
			06:31:24.22 & 04:52:37.5 & M1.0 & 0.4 & -12.61$\pm$ 0.39 \\
			06:31:25.79 & 04:58:14.3 & K7.0 & 1.8 & -3.80$\pm$ 0.11 \\
			06:31:27.57 & 04:51:53.1 & M3.5 & 0.0 & -24.40$\pm$ 0.72 \\
			06:31:27.65 & 04:54:02.7 & M1.75 & 1.0 & -5.33$\pm$ 0.36 \\
			06:31:28.84 & 04:52:13.2 & M0.0 & 0.6 & -11.40$\pm$ 0.20 \\
			06:31:29.41 & 04:50:34.8 & M1.75 & 2.8 & -5.29$\pm$ 0.50 \\
			\hline
		\end{tabular}
	
	\end{center}
$^{*}$ The full table is available at the CDS.
	\label{tab:spt}
\end{table}

In Table~\ref{tab:spt} we list the spectral types and extinction derived from the VIMOS spectra described in Section~\ref{sec:vimos}.

 \section{PRF scores}
\begin{figure*}
   \centering
   \includegraphics[width=0.8\textwidth]{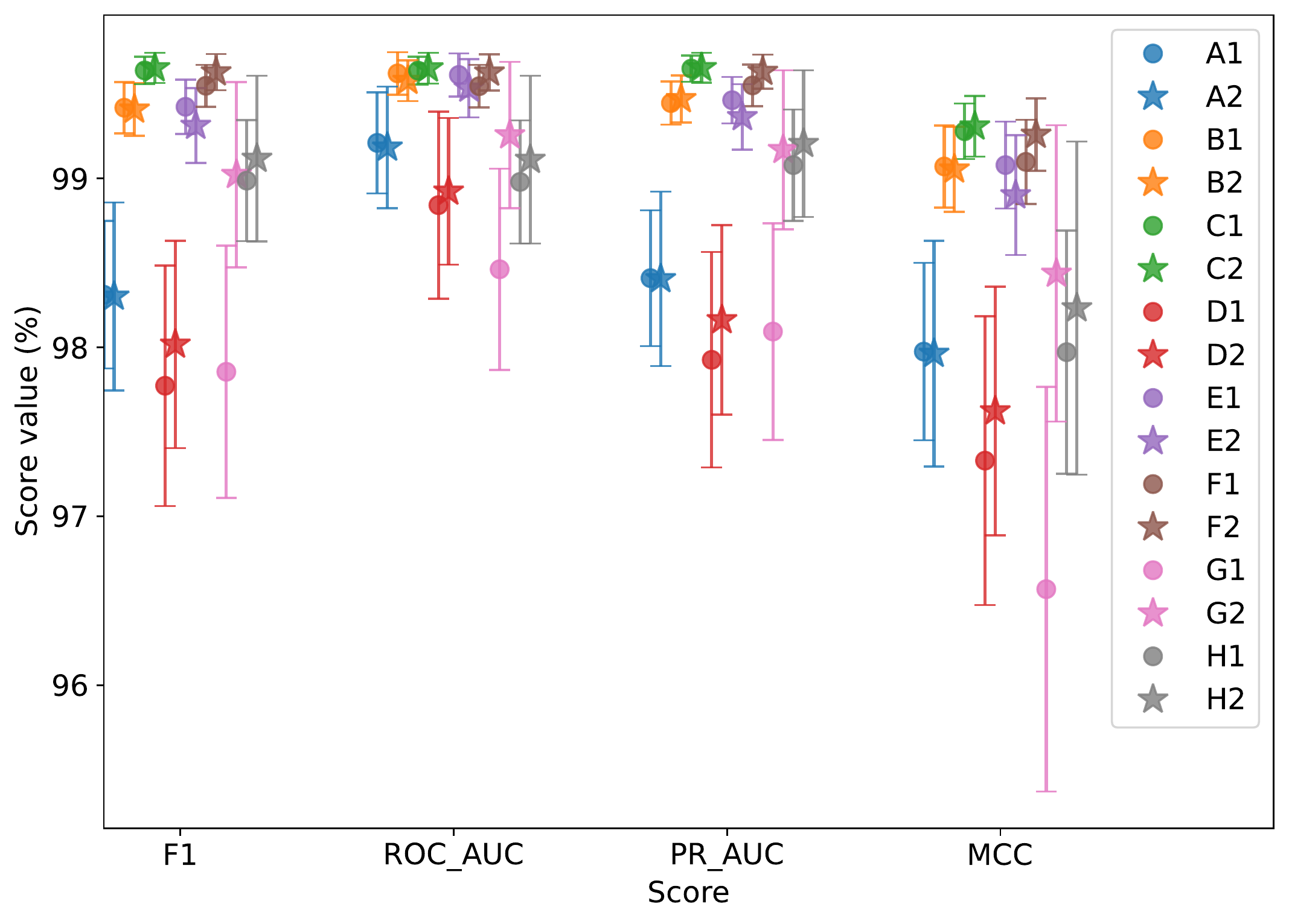}
      \caption{Performance metrics for each of the classifier runs (see Table~\ref{tab:sampling_scores} for IDs, run details and the exact score values). Points are offset on the x-axis for clarity.}
         \label{fig:scores}
\end{figure*}

 In Table~\ref{tab:sampling_scores} we show the details of the 16 PRF runs, named A1-H2 (1 and 2 mark the inclusion or exclusion, respectively, of the parallax in the list of features), together with the scores obtained through cross-validation (see Section~\ref{sec:crossval}). The scores are also shown in Fig.~\ref{fig:scores}.
  \begin{table*}
        \caption{Parameters for the construction of the training set and the corresponding PRF scores.}
        \begin{center}
                \begin{tabular}{c c c | c c c | c c c c}
                \hline \hline
                         & \multicolumn{2}{c|}{{\it sampling strategy}} & & & & \multicolumn{4}{c}{{\it performance metrics}} \\
                                        ID & under$\_$sample & over$\_$sample & plx & N$_\text{{memb}}$ & N$_\text{{non-memb}}$ & F1($\%$) & ROC\_AUC($\%$) & PR\_AUC($\%$) & MCC($\%$) \\
                        \hline
            A1 & 0.06 & 0.20 & y   & 1666 & 8333 & 98.31$\pm$0.44 & 99.21$\pm$0.30 & 98.41$\pm$0.40 & 97.97$\pm$0.53 \\
                        A2 & 0.06 & 0.20 & n   & 1666 & 8333 & 98.30$\pm$0.56 & 99.18$\pm$0.36 & 98.41$\pm$0.52 & 97.96$\pm$0.67 \\
                        B1 & 0.06 & 0.60 & y   & 4999 & 8333 & 99.42$\pm$0.15 & 99.62$\pm$0.13 & 99.45$\pm$0.13 & 99.07$\pm$0.24 \\
                        B2 & 0.06 & 0.60 & n   & 4999 & 8333 & 99.41$\pm$0.16 & 99.58$\pm$0.12 & 99.47$\pm$0.14 & 99.06$\pm$0.25 \\
                        C1 & 0.06 & 1.00 & y   & 8333 & 8333 & 99.64$\pm$0.08 & 99.64$\pm$0.08 & 99.65$\pm$0.08 & 99.28$\pm$0.16 \\
                        C2 & 0.06 & 1.00 & n   & 8333 & 8333 & 99.65$\pm$0.09 & 99.65$\pm$0.09 & 99.66$\pm$0.09 & 99.31$\pm$0.18 \\
                        D1 & 0.10 & 0.20 & y   & 1000 & 5000 & 97.77$\pm$0.71 & 98.84$\pm$0.55 & 97.93$\pm$0.64 & 97.33$\pm$0.85 \\
                        D2 & 0.10 & 0.20 & n   & 1000 & 5000 & 98.02$\pm$0.61 & 98.92$\pm$0.43 & 98.16$\pm$0.56 & 97.62$\pm$0.74 \\
                        E1 & 0.10 & 0.60 & y   & 3000 & 5000 & 99.42$\pm$0.16 & 99.61$\pm$0.13 & 99.46$\pm$0.14 & 99.08$\pm$0.26 \\
                        E2 & 0.10 & 0.60 & n   & 3000 & 5000 & 99.31$\pm$0.22 & 99.53$\pm$0.17 & 99.36$\pm$0.19 & 98.90$\pm$0.36 \\
                        F1 & 0.10& 1.00 & y   & 5000 & 5000 & 99.55$\pm$0.12 & 99.55$\pm$0.13 & 99.55$\pm$0.12 & 99.10$\pm$0.25 \\
                        F2 & 0.10& 1.00 & n   & 5000 & 5000 & 99.63$\pm$0.11 & 99.63$\pm$0.11 & 99.63$\pm$0.10 & 99.26$\pm$0.21 \\
                        G1 & 0.50 & 0.60 & y   & 600  & 1000 & 97.86$\pm$0.75 & 98.46$\pm$0.60 & 98.09$\pm$0.64 & 96.57$\pm$1.20 \\
                        G2 & 0.50 & 0.60 & n   & 600  & 1000 & 99.02$\pm$0.55 & 99.26$\pm$0.43 & 99.17$\pm$0.47 & 98.44$\pm$0.88 \\
                        H1 & 0.50 & 1.00 & y   & 1000 & 1000 & 98.99$\pm$0.36 & 98.98$\pm$0.36 & 99.08$\pm$0.33 & 97.97$\pm$0.72 \\
                        H2 & 0.50 & 1.00 & n   & 1000 & 1000 & 99.12$\pm$0.49 & 99.11$\pm$0.50 & 99.21$\pm$0.43 & 98.23$\pm$0.99 \\
                        \hline
                \end{tabular}
        \end{center}
        \label{tab:sampling_scores}
\end{table*}

\section{Feature importance}

\begin{figure*}
   \centering
   \includegraphics[width=0.8\textwidth]{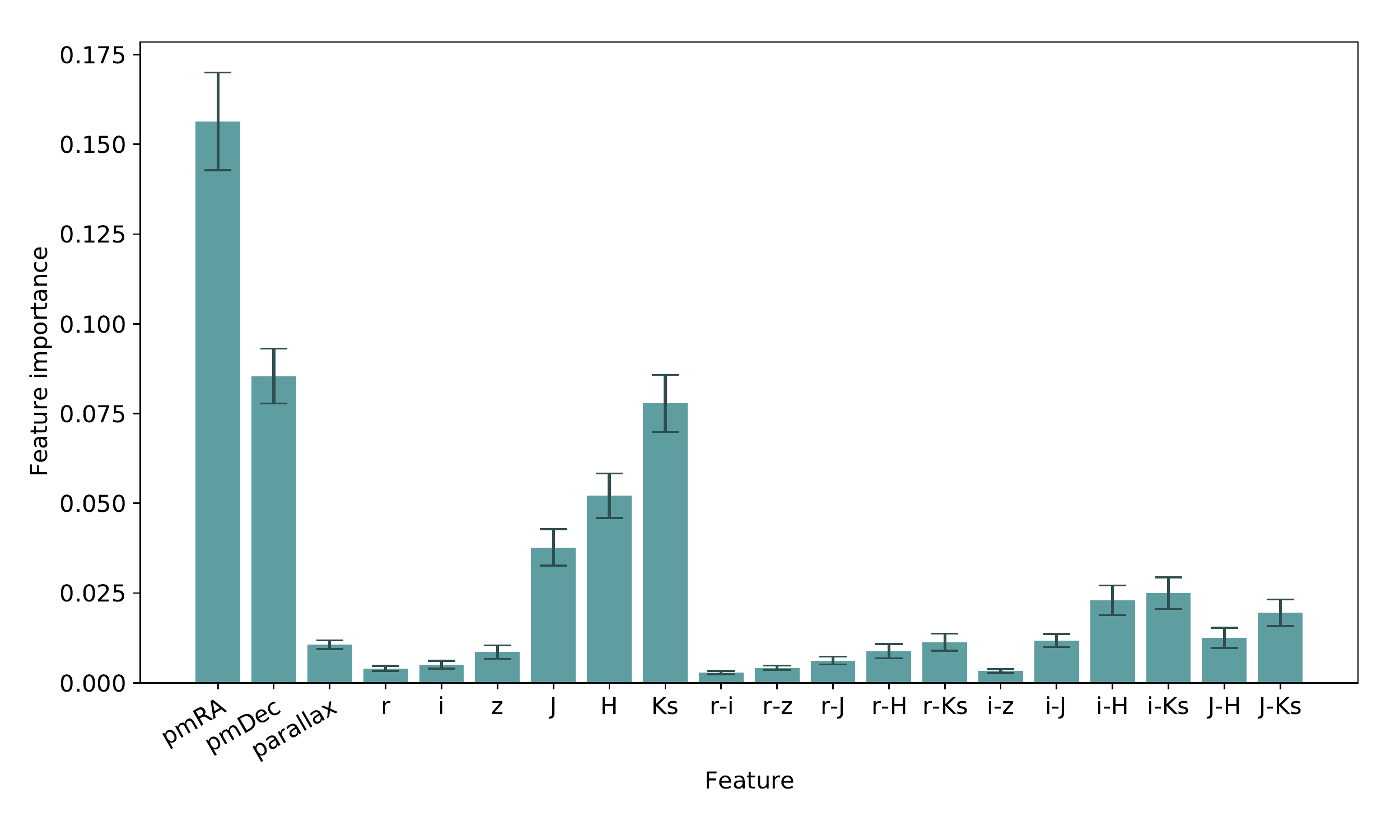}
      \caption{Feature importance plot for the run F1,  showing the relative values of importance for each feature, as returned by the classifier.}
         \label{fig:feat_imp}
\end{figure*}

In Fig.~\ref{fig:feat_imp} we show the relative importance of each feature as returned by the classifier in run F1. We note that the feature importance looks similar for all the runs. The uncertainties shown in the plot correspond to the standard deviation of the 50 split values.

\section{Relative proper motions}

\begin{figure*}
   \centering
   \includegraphics[width=1\textwidth]{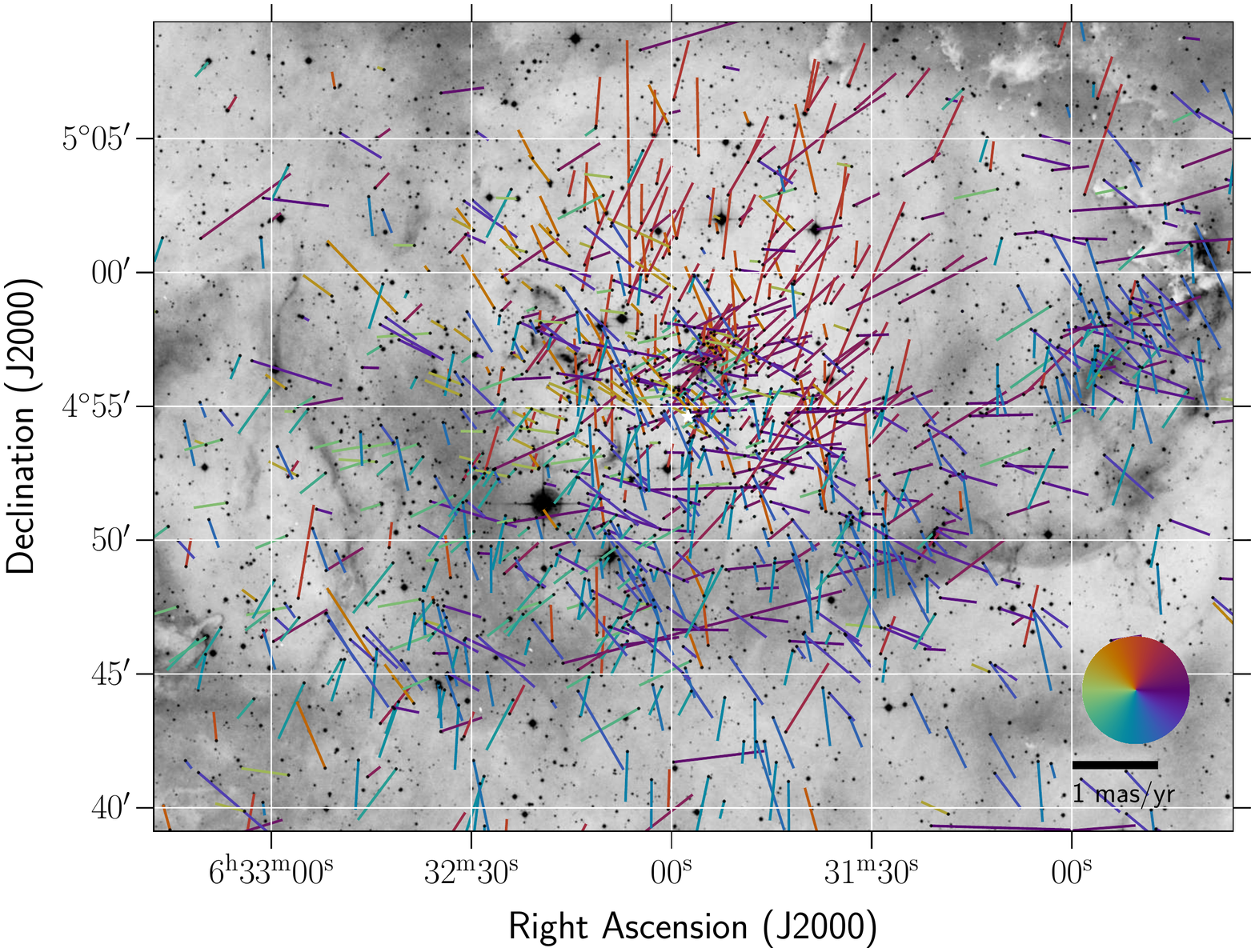}
      \caption{Same as Fig.~\ref{fig:on_sky_pm}, but focussed on \cluster.}
         \label{fig:on_sky_pm_zoom1}
\end{figure*}

\begin{figure*}[htb]
   \center
   \begin{subfigure}{}
        \center
        \includegraphics[width=0.5\linewidth]{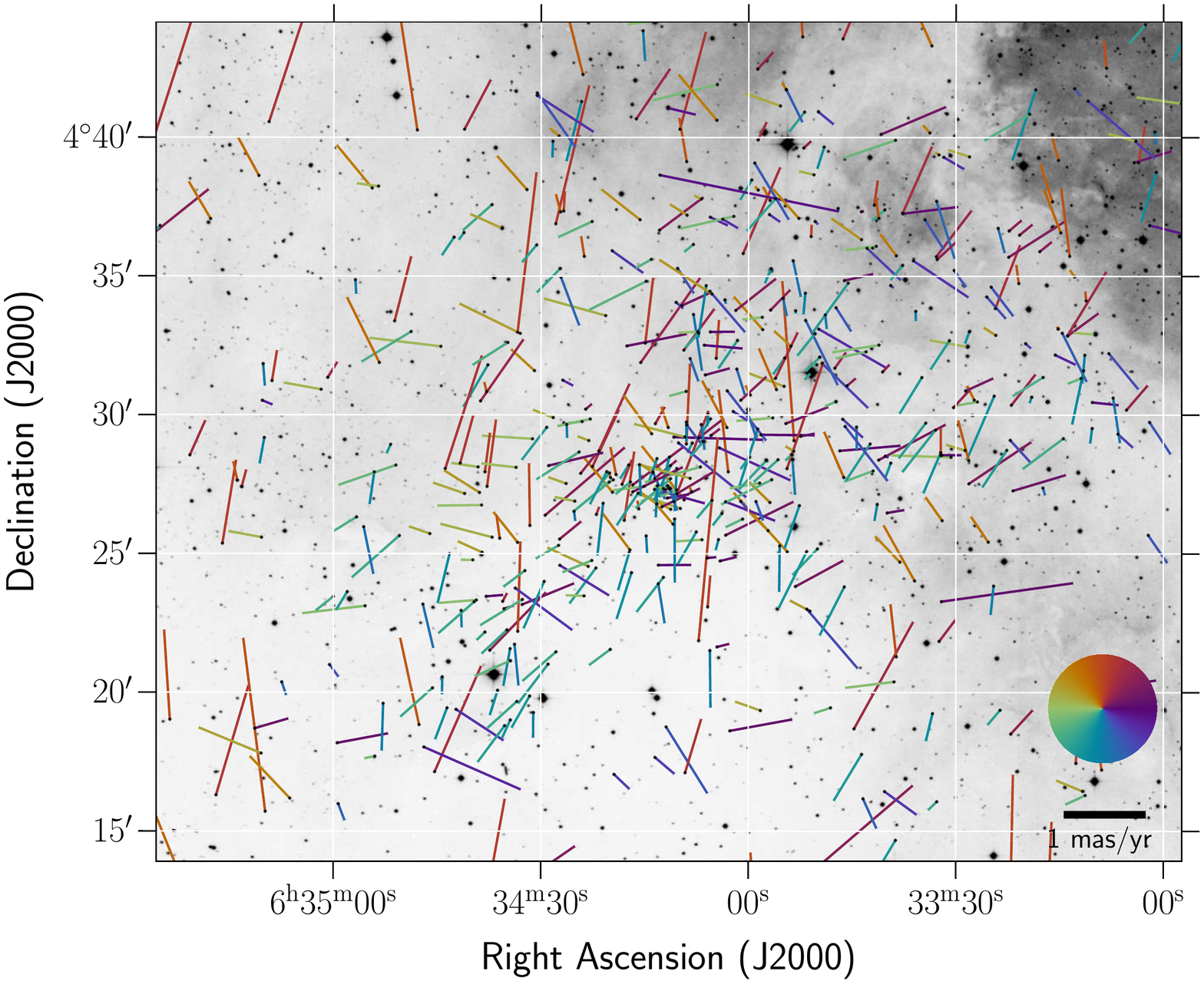}%
        \hfill
        \includegraphics[width=0.5\linewidth]{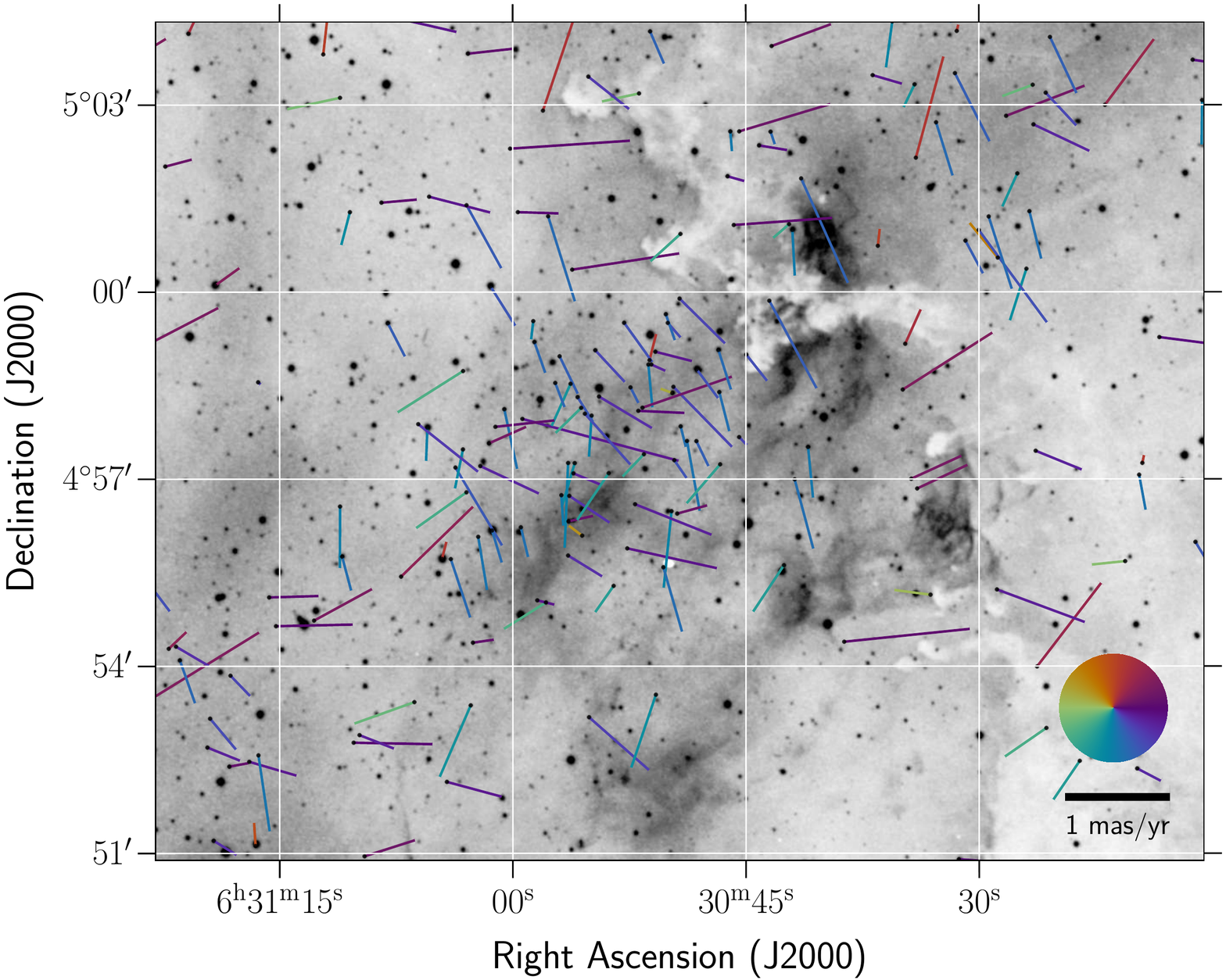}
      \caption{Same as Fig.~\ref{fig:on_sky_pm}, but focussed on the region rich in young stars to the south-east of \cluster~(left), and NGC~2237 (right).}
         \label{fig:on_sky_pm_zoom2}
    \end{subfigure}
\end{figure*}

In Figs.~\ref{fig:on_sky_pm_zoom1} and \ref{fig:on_sky_pm_zoom2}, we show the zoom-in versions of Fig.~\ref{fig:on_sky_pm}, which allows appreciation of details in relative proper motions for the NGC\,2244 and NGC\,2237 regions, as well as the south-east concentration connecting the mid-infrared clusters.

\section{SED fitting and mass distribution examples}

\begin{figure*}
   \centering
   \includegraphics[width=1\textwidth]{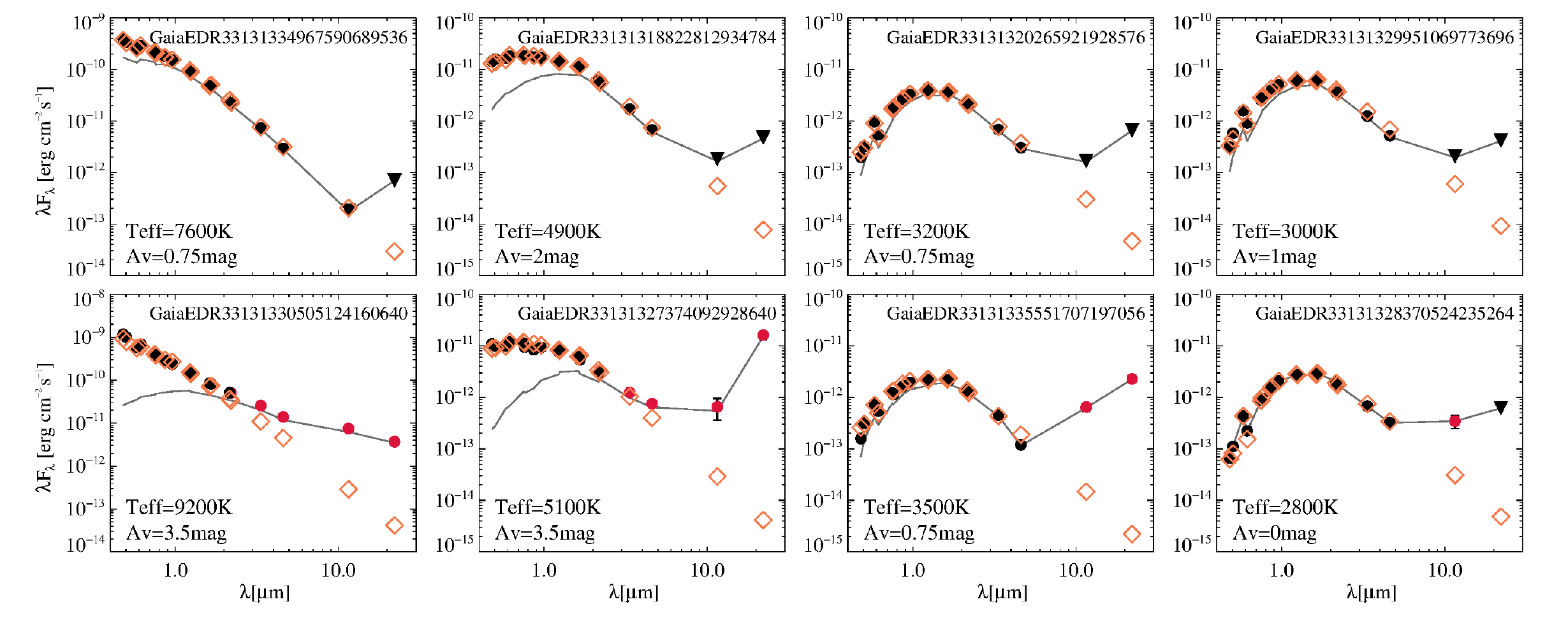}
      \caption{Spectral energy distribution for a subset of candidate members in \cluster. The grey line shows the observed photometry. The filled circles correspond
to the observed photometry corrected by the best-fit value of the extinction, where the black circles are those used for the fitting, and the red ones were ignored due to excess emission at these wavelengths. The black triangles mark the upper limits. The orange diamonds display the best-fitting BT-Settl model value at each fitted wavelength.}
         \label{fig:seds}
\end{figure*}

\begin{figure}
   \centering
   \includegraphics[width=0.45\textwidth]{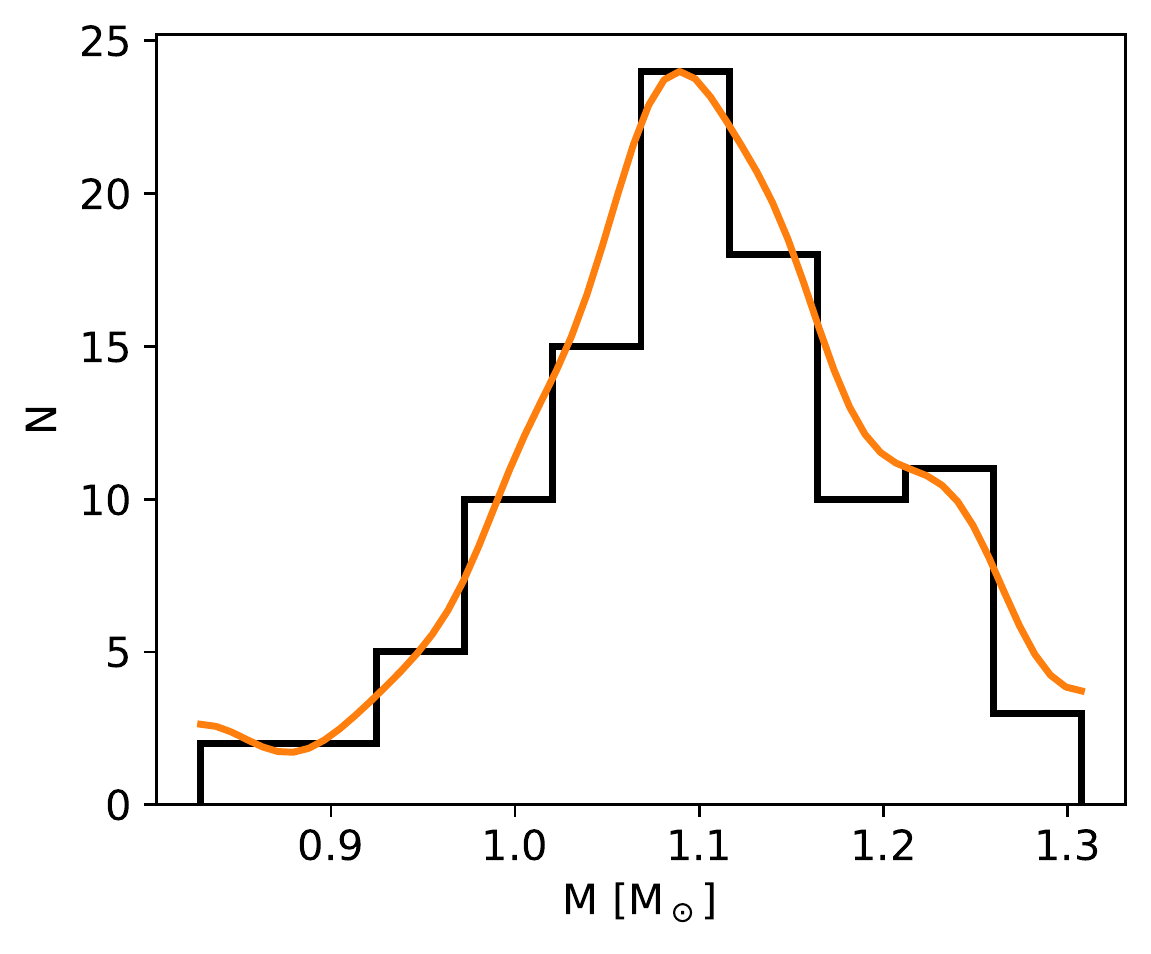}
      \caption{Distribution of masses derived from the HRD (Section\ref{sec:massage}) for the object Gaia EDR3 3131334868807128704, shown as a histogram (black) and a KDE (orange).}
         \label{fig:mass_distr_ind}
\end{figure}

In Fig.~\ref{fig:seds} we show the spectral energy distribution for a subset of candidate members in \cluster. The observed photometry is shown with a grey line, while the fitted photometry (corrected by the best-fit extinction value) is shown by the black dots. Red dots mark the points not used in the fitting process as they correspond to the excess emission. The upper limits are represented by black triangles, and the best-fitting model values by orange diamonds. Fig.~\ref{fig:mass_distr_ind} contains the distribution of masses for a single objects from our dataset (Gaia EDR3 3131334868807128704) as derived from the HRD as described in Section~\ref{sec:massage}.

\section{Comparison to \citet{cantatgaudin20}}
\label{sec:CG20}

\cluster~and its surroundings have been extensively studied in the past, which we took advantage of when constructing the training set. In this section we compare our selection with the result of a recent study based on the $Gaia$ DR2 data.
\citet{cantatgaudin20} employed the UPMASK method \citep{upmask} to select members of almost 1500 star clusters. In this implementation, UPMASK groups stars according to their parallax and proper motions, verifying that their distribution on the sky is more concentrated than what can be expected from random fluctuations in a uniform distribution. The procedure is repeated several times, and each star is attributed a membership probability related to the number of times it has been selected as a member in each iteration.   

\begin{figure*}
   \centering
   \includegraphics[width=\textwidth]{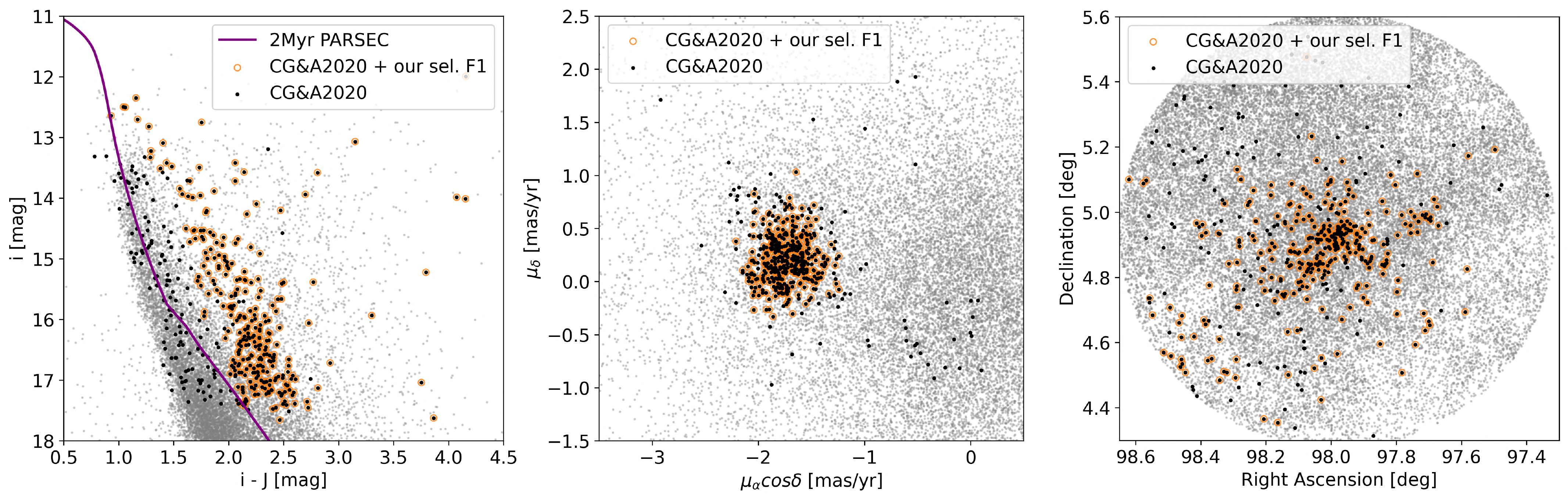}
      \caption{Colour-magnitude diagram (left), proper motions (middle), and the positions on the sky (right) for all the sources featuring in our initial catalogue, located within 45$'$ from the star HD~46150 (small grey dots). Candidates with membership probability $>70 \%$ from \citet{cantatgaudin20} are shown as black dots, and those that pass also our selection (F1, membership probability $>80\%$) as orange open circles. We estimate that there may be as much as 35$\%$ of contaminants in the selection sample of \citet{cantatgaudin20}. }
         \label{fig:cg2020}
\end{figure*}

In Fig.~\ref{fig:cg2020} we show the CMD, proper motion plot, and the positions on the sky for all the sources featuring in our initial catalogue, located within 45$'$ from the star HD~46150 (small grey dots). Candidates with membership probability $>70 \%$ from \citet{cantatgaudin20} are shown as black dots, and those that pass also our selection (F1, membership probability $>80\%$) as orange open circles. The $70 \%$ cut is the same used in \citet{cantatgaudin20} to plot probable members of various clusters. Their selection is based exclusively on astrometry, and, consequently, several sources that are too blue to be in a pre-main sequence stage pass their selection. From the 460 high-probability members of \citet{cantatgaudin20}, 301 pass our selection (F1). We therefore estimate that there may be as much as 35$\%$ of contaminants in their selection sample. This number depends on the membership probability that is adopted for the final selection of probable members, and drops to $\sim 30\%$ raising the cut on the \citet{cantatgaudin20} selection to $80\%$ or $90\%$. Raising the cut in our dataset to $90\%$ has a negligible effect on this number.

The reason for the contamination in the \citet{cantatgaudin20} sample probably lies in the fact that the cluster is relatively far away, and the individual stellar parallaxes come with significant uncertainties. Moreover, their selection is purely astrometric, meaning that they do not use the colour information, and in particular the infrared bands, which significantly help in our classification. Things may be further complicated by the variable extinction present in the region, which affects the number counts and appearance of the cluster on the sky. Recently, the selection of \citet{cantatgaudin20} cut at $50\%$ membership probability has been used as a training set by \citet{mahmudunnobe21}, in their application of the Random Forest algorithm to select members in nine galactic open clusters. Their training set in \cluster~likely included a significant fraction of contaminants (we estimate $\sim 50\%$ applying the same analysis as above), resulting in a precision that is among the lowest achieved in their work (88$\%$). Consequently, a significant fraction of $\sim$3000 new members found by \citet{mahmudunnobe21} are likely to be contaminants, which is quite clear when looking at their CMD and proper motion plots. In conclusion, we caution against blind usage of unsupervised members selections as training sets for supervised ML algorithms, particularly in clusters at large distances and with non-negligible amounts of extinction. A rigorous vetting of members is indispensable to obtain reliable results.

\end{appendix}

\end{document}